\documentclass[usenatbib,useAMS]{mnras}
\usepackage{graphicx}
\usepackage[fleqn]{amsmath}
\usepackage{txfonts}
\usepackage[T1]{fontenc}
\usepackage{amssymb,amsfonts,hyperref,color,subfig}
\hypersetup{linkcolor=red,citecolor=blue,filecolor=blue,urlcolor=magenta}

\newcommand{\sbt}{\,\begin{picture}(-1,1)(1,-3)\circle*{3}\end{picture}\ }

\title[Eccentric Jupiters in disc cavities]{Observational signatures of eccentric Jupiters inside gas cavities in protoplanetary discs}

\author[Baruteau et al.]{Cl{\'e}ment Baruteau,$^{1}$\thanks{E-mail: clement.baruteau@irap.omp.eu}
Gaylor Wafflard-Fernandez,$^{1,2}$
Romane Le Gal,$^{3,1}$
\newauthor
Florian Debras,$^{1}$
Andr{\'e}s Carmona,$^{2}$
Asunci{\'o}n Fuente$^{4}$ 
and Pablo Rivi{\`e}re-Marichalar$^{4}$\\
$^{1}$IRAP, Universit{\'e} de Toulouse, CNRS, UPS, Toulouse, France\\
$^{2}$IPAG, Universit{\'e} Grenoble Alpes, CNRS, Grenoble, France\\
$^{3}$Center for Astrophysics | Harvard \& Smithsonian, Cambridge, MA, USA\\
$^{4}$Observatorio Astron{\'o}mico Nacional, Madrid, Spain\\
}

\date{Accepted 2021 April 8. Received 2021 April 8; in original form 2021 March 2}

\pubyear{2021}
\begin{document}
\label{firstpage}
\pagerange{\pageref{firstpage}--\pageref{lastpage}}
\maketitle

\begin{abstract}
Predicting how a young planet shapes the gas and dust emission of its parent disc is key to constraining the presence of unseen planets in protoplanetary disc observations. We investigate the case of a 2~Jupiter mass planet that becomes eccentric after migrating into a low-density gas cavity in its parent disc. Two-dimensional hydrodynamical simulations are performed and post-processed by three-dimensional radiative transfer calculations. In our disc model, the planet eccentricity reaches $\sim$0.25, which induces strong asymmetries in the gas density inside the cavity. These asymmetries are enhanced by photodissociation and form large-scale asymmetries in $^{12}$CO J=3$\rightarrow$2 integrated intensity maps. They are shown to be detectable for an angular resolution and a noise level similar to those achieved in ALMA observations. Furthermore, the planet eccentricity renders the gas inside the cavity eccentric, which manifests as a narrowing, stretching and twisting of iso-velocity contours in velocity maps of $^{12}$CO J=3$\rightarrow$2. The planet eccentricity does not, however, give rise to detectable signatures in $^{13}$CO and C$^{18}$O J=3$\rightarrow$2 inside the cavity because of low column densities. Outside the cavity, the gas maintains near-circular orbits, and the vertically extended optically thick CO emission displays a four-lobed pattern in integrated intensity maps for disc inclinations $\ga$ 30$\degr$. The lack of large and small dust inside the cavity in our model further implies that synthetic images of the continuum emission in the sub-millimetre, and of polarized scattered light in the near-infrared, do not show significant differences when the planet is eccentric or still circular inside the cavity.
\end{abstract}


\section{Introduction}
About 50\% of exoplanets with orbital periods $\ga$ 100 days and masses between that of Saturn and 5 times that of Jupiter have eccentricities in the range [0.1$-$0.4] \citep{Debras21}. This large fraction of eccentric, so-called warm Jupiters is an important constraint for theories of planet formation and orbital evolution \citep[see review in][]{Baruteau2016SSR}. It tells us that there should be one or several generic mechanisms able to grow substantially the eccentricity ($e$) of Jupiter-mass planets which form on near-circular orbits in their protoplanetary disc. A variety of ways to grow the eccentricity of warm Jupiters have been explored in the literature, including (i) gravitational scattering with one or more planet companions, during the disc phase or after disc dissipation \citep{Chatterjee08,Marzari10,Andersen20}, (ii) disc-planet interactions in a low-density gas cavity \citep{Papaloizou01,DAngelo06,Rice08,Ragusa18,Debras21}, or (iii) secular perturbations from a distant companion \citep{Anderson17}. 

The presence of a dust cavity in a large number of protoplanetary discs (transition discs), with mounting evidence for substantial gas depletion inside these cavities \citep{Carmona14,vDM15,vdM16,Carmona17}, suggests that the formation of a gas cavity could be a common outcome of protoplanetary disc evolution, and that it could therefore constitute a generic mechanism to grow the eccentricity of Jupiter-like planets \citep{Debras21}. As pointed out by \citet{Debras21}, the cavity does not need to be carved by the planet or another planetary companion, but could build up due to photoevaporation, magnetised winds, or a combination thereof (see references in \S~4.2 of \citealp{Debras21}). The hydrodynamical simulations carried out by \citet{Debras21} show that a planet in the Jupiter-mass range could acquire an eccentricity as high as 0.3$-$0.4 after migrating into a low-density gas cavity. One of the key features of the simulations for reaching such large eccentricities was to relax the commonly adopted assumption that the decrease in the gas surface density inside the cavity should be exactly compensated by a proportional increase in the turbulent viscosity, which is generally expected for discs in a steady-state viscous evolution. But, as said above, gas cavities and more generally discs do not have to follow such evolution. In the simulations of \citet{Debras21} and in the present study, the surface density of the gas inside the cavity is about 3 orders of magnitude smaller than outside the cavity.

Motivated by these findings, we investigate what observational signatures to expect in the gas and dust emission from the presence of an eccentric planet inside the gas cavity of a protoplanetary disc. To this aim, we have carried out gas and gas+dust hydrodynamical simulations, which were post-processed by radiative transfer calculations. For the gas emission, we focus on CO since it is the most commonly used gas tracer in discs, and on the J=3$\rightarrow$2 rotational line, which is one of the most observed transitions in the sub-millimetre. In Section~\ref{sec:methods}, we describe the physical model and numerical set-up of the hydrodynamical simulations and the radiative transfer calculations. Their results are analysed and discussed in Section~\ref{sec:results}, which highlights that a 2 Jupiter mass planet around a 2 Solar mass star that reaches an $\approx0.25$ eccentricity inside its disc cavity has clear detectable signatures in both $^{12}$CO J=3$\rightarrow$2 integrated intensity maps and velocity maps. Concluding remarks follow in Section~\ref{sec:conclusion}.

\section{Physical model and numerical methods}
\label{sec:methods}

The hydrodynamical simulations carried out in this work basically adopt the same physical model and numerical setup as the reference simulation of \citet{Debras21}, the main differences being the code used and the value of the disc viscosity. For the sake of convenience, the disc model is recapped in Section~\ref{sec:hydrosetup}. We then describe in Section~\ref{sec:RTsetup} the methodology used to compute synthetic images of the dust and gas emission by post-processing our hydrodynamical simulations with radiative transfer calculations.

\subsection{Hydrodynamical simulations}
\label{sec:hydrosetup}

The disc that we model in our hydrodynamical simulations is two-dimensional (2D) and described by cylindrical polar coordinates $\{R,\varphi\}$ centred on the star. While the simulations performed in \citet{Debras21} used the \href{http://fargo.in2p3.fr}{FARGO3D} code \citep{Benitez2016fargo3d} in 2D, here we have used the code Dusty FARGO-ADSG. It is an extended version of the 2D grid-based code \href{http://fargo.in2p3.fr/-FARGO-ADSG-}{FARGO-ADSG} \citep{Masset2000,BaruteauMasset2008a,BaruteauMasset2008b} with dust modelled as Lagrangian test particles \citep{Baruteau2016,Fuente2017}. As in \citet{Debras21}, our simulations are done in three steps, which are detailed later in this section. A cavity is first built in the disc gas via a preliminary 1D simulation with no planet. The simulation is then restarted in 2D with a giant planet held on a fixed circular orbit slightly outside the cavity. In the third and last step, the planet is allowed to feel the gravitational force exerted by the disc gas, which makes the planet migrate into the cavity and causes eccentricity growth once the planet is sufficiently far from the edge of the cavity.

Although our gas hydrodynamical simulations are scale free, physical units need to be specified for our disc model for two main reasons. First, the radiative transfer calculations performed with RADMC-3D (see Section~\ref{sec:RTsetup}) need input fields, like the gas density and velocity, to have physical units. Second, eccentricity growth by disc-planet interactions in a cavity can be a slow process that takes a few thousand planet orbits at least, depending on the planet mass and the disc's physical properties (like the gas density and turbulent viscosity; see for instance \citealp{Debras21}). We have therefore adopted code units such that a significant planet eccentricity could be reached by a few million years of disc evolution: (i) the code's unit of mass, which is the mass of the central star, is taken to be $M_{\star} = 2M_{\odot}$, and (ii) the code's unit of length, which corresponds to the outer edge of the gas cavity at the beginning of the simulation, is taken to be 30 au. The code's unit of time follows by setting the gravitational constant $G$ equal to unity in the code, and that of temperature by setting the universal gas constant equal to unity and by specifying a mean molecular weight $\umu$ for the gas, which we take equal to 2.3 (corresponding to a solar nebula H$_2$-He gas mixture). The reasons for choosing a 30 au wide gas cavity are three-fold. First, gas cavities of comparable sizes have been reported in several (pre-)transitional discs, like for instance HD 135344B \citep{Carmona14}. Second, our cavity is not too large so that the planet can migrate inside the cavity and have its eccentricity grow substantially in a few million years (Myr). Third, the cavity is large enough for planet-induced features in the cavity to be detectable for an angular resolution similar to that currently achieved in gas line observations with ALMA (50 mas) and for a reasonable disc distance (100 pc, see Section~\ref{sec:RTsetup}). 

With the above set of units, $R$ extends from 9 to 57 au over 500 logarithmically spaced cells, while $\varphi$ covers the full $2\upi$ range with 800 uniformly spaced cells. The grid resolution is enough for our needs, and we have checked that a very similar orbital evolution of the planet in an eccentricity vs. semi-major axis diagram is obtained with a grid resolution reduced to $400\times600$. Wave-killing zones are used for $R \in [9-10.8] \cap [48-57]$ au, where the gas 2D fields are damped toward their initial 1D radial profiles \citep{DVB06}. During the first step of the simulation, a low-density gas cavity is carved by adopting an initial gas surface density profile, $\Sigma_0(R)$, that decreases by three orders of magnitude inside 30 au over two pressure scale heights ($2H$). This decrease in the gas surface density goes with an increase in the disc's turbulent kinematic viscosity, $\nu$, by two orders of magnitude. Apart from the smooth transition region, both $\Sigma_0$ and $\nu$ are chosen uniform on both sides of the cavity: inside the cavity, $\Sigma_0 \approx$ 0.02 g cm$^{-2}$ and $\nu = 7.5\times 10^{-5}$ in code units, while outside the cavity $\Sigma_0 \approx$ 20~g~cm$^{-2}$ and $\nu = 7.5\times 10^{-7}$ (code units). The initial disc-to-star mass ratio is $\sim$ $8.4\times 10^{-3}$. A locally isothermal equation of state is used where the gas temperature varies with $R$ but is kept constant in time. The disc's aspect ratio ($h=H/R$) is taken uniform, equal to 0.05, which makes the temperature $\approx 41\,{\rm K} \times (R/30\,{\rm au})^{-1}$. Our disc's kinematic viscosity can thus be expressed as an equivalent alpha viscosity $\alpha = \nu h^{-2} (G M_{\star}R)^{-1/2}$ $\sim$ 0.03 inside the cavity and $\sim$ $3\times10^{-4}$ outside the cavity (note that $\alpha$ scales as $R^{-1/2}$ in our disc model). The star being fixed in the simulations, the indirect terms that represent the acceleration the star would feel from the disc and the planet are accounted for in the evolution of the disc and of the planet. Gas self-gravity, however, is discarded (the Toomre Q-parameter is $\ga$ 5$-$10 outside the cavity, and is in the range $[10^3-10^5]$ inside the cavity).

The first step of the simulation lasts for 900 orbits at $R=33$~au, which corresponds to $\sim$ 0.12~Myr (one orbit at 33~au is about 134~yr with our set of units). The simulation is then restarted in 2D by inserting a 2~Jupiter mass planet on a fixed circular orbit at 33~au (the planet-to-primary mass ratio is $10^{-3}$). The planet mass is gradually increased over nearly 20 planet orbits, and the planet's gravitational potential is smoothed over a softening length set to 0.5 local pressure scale heights. This second step lasts another 520 planet orbits, after which the planet is allowed to feel the gravitational acceleration of the disc gas. The gas located inside the planet's Hill radius is discarded in the calculation of the disc acceleration on the planet, which is recommended when gas self-gravity is neglected \citep{Crida09}.

\begin{figure}
\includegraphics[width=\hsize]{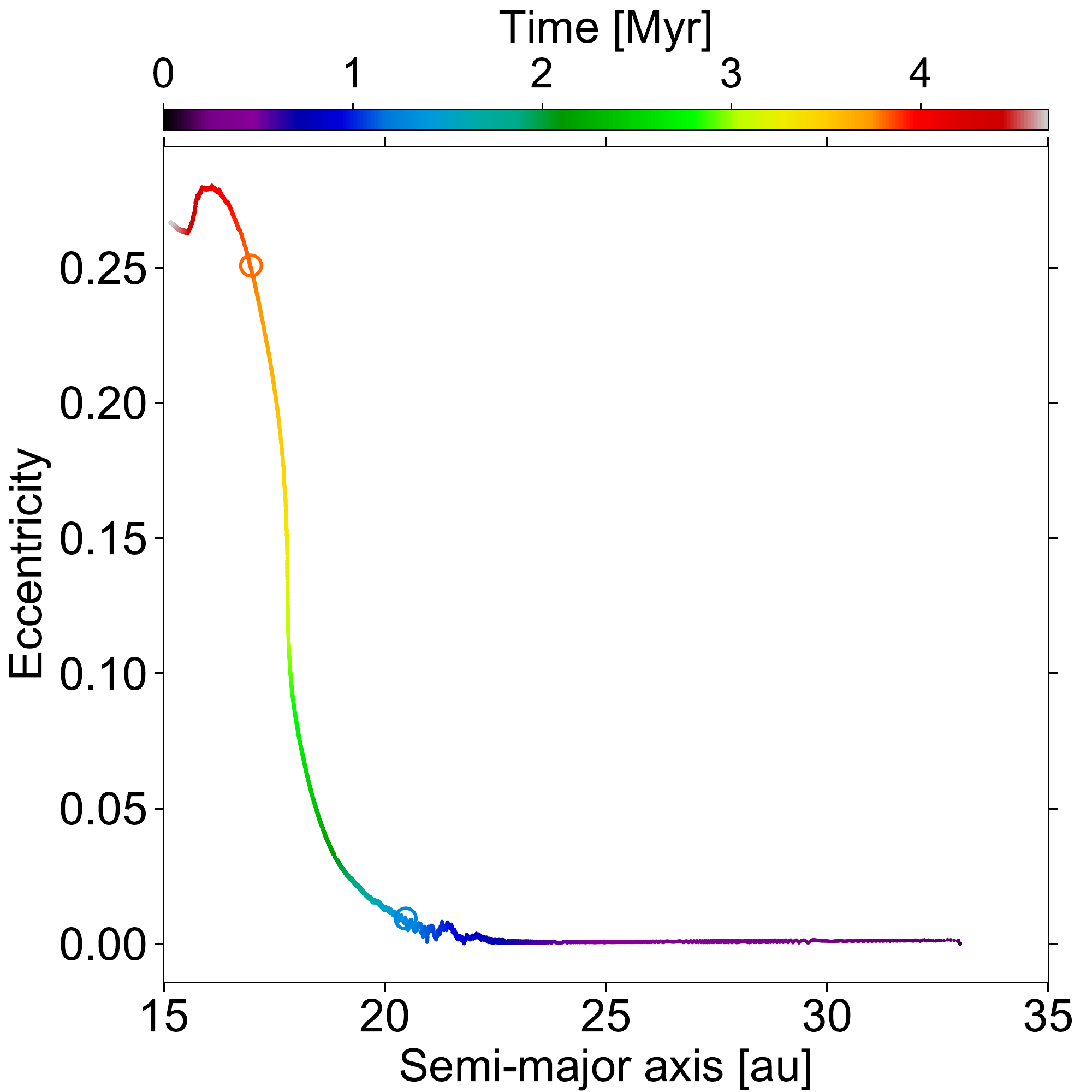}
\caption{Time evolution of the planet's semi-major axis and eccentricity in the hydrodynamical simulation, shown as a scatter plot with time in color. The outer edge of the gas cavity is located at about 30 au. Since the planet migrates inward into the cavity, it moves from right to left in the panel. The open circles mark the two situations for which results are presented in Section~\ref{sec:results}: when the planet has entered the cavity and still has a nearly circular orbit, and when it has acquired a near-maximum eccentricity ($e\approx0.25$).}
\label{fig:fig1}
\end{figure}
The evolution of the planet's eccentricity and semi-major axis is displayed in Fig.~\ref{fig:fig1}, with time shown in color. Once the planet feels the disc gas force (third step of the simulation), it quickly migrates into the cavity, where its orbital migration becomes slower. Once the planet is far enough from the edge of the cavity, its eccentricity grows roughly linearly with time. The evolution shown in Fig.~\ref{fig:fig1} is very similar to that in Figure 3 of \citet{Debras21}, although a gas viscosity about three times smaller has been used here (a preliminary run with the same viscosity as in the reference run of \citet{Debras21} showed no visible differences compared to their Figure 3). The reader is referred to \citet{Debras21} for more details about the planet's orbital evolution in the cavity, and for a more detailed discussion on the disc gas model.

In order to compute synthetic images of the dust's continuum emission at radio wavelengths, the hydrodynamical simulation was restarted by inserting 30000 Lagrangian dust particles at $\approx$ 2.3 Myr, when the planet eccentricity has reached $\approx 0.06$. The particles were inserted between 18 and 27 au, that is between the planet's orbital radius at restart time and approximately the edge of the cavity. They feel the gravity of the star and of the planet, gas drag and turbulent diffusion. However, aerodynamic drag from the dust on the gas is discarded in the simulation. This means that the dust particles are passive spectators of the disc dynamics, and implies that the dust's size distribution $n_{\rm simu}(s)$ can be chosen arbitrarily in the simulation. Following \citet{Baruteau2019}, we take $n_{\rm simu}(s) \propto s^{-1}$ for particle sizes $s$ between 10 $\micron$ and 1 cm, so that there is approximately the same number of dust particles per decade of size. A more realistic (steeper) size distribution will be adopted, however, in the dust radiative transfer calculations. Also, the internal mass volume density of the dust particles is set to 1.3 g cm$^{-3}$, which corresponds to the internal density of the dust mixture in the radiative transfer calculation of the dust continuum emission (see Section~\ref{sec:dustRTsetup}). For more details about the dust modelling by Lagrangian particles in the code Dusty FARGO-ADSG, the reader is referred to \S~2.1.3 of \citet{Baruteau2019}.

\subsection{Radiative transfer calculations}
\label{sec:RTsetup}

We used the 3D radiative transfer code \href{https://www.ita.uni-heidelberg.de/~dullemond/software/radmc-3d/}{RADMC-3D} (version 2.0, \citealp{Dullemond2015}, Dullemond et al. in prep.) to post-process our results of 2D hydrodynamical simulations and generate synthetic emission maps of the gas and dust. More specifically, synthetic images have been produced for several CO isotopologues, for the dust continuum emission in the (sub)millimetre, and for the polarized scattered light in the near-infrared. To this aim, we have extended the public python code \href{https://github.com/charango/fargo2radmc3d}{\texttt{fargo2radmc3d}} \citep{Baruteau2019} to compute synthetic images of gas line emission with RADMC-3D from the results of Dusty FARGO-ADSG simulations. The procedure is described in Section~\ref{sec:gasRTsetup}. The method used to produce synthetic images of polarized scattered light and of continuum emission was detailed in \citet{Baruteau2019}, and Section~\ref{sec:dustRTsetup} briefly mentions the main hypotheses and parameters used to produce the dust images presented in Section~\ref{sec:results}. Note that in all our synthetic images, the disc is assumed to be located at 100~pc. The inclination of the disc mid-plane relative to the sky-plane is set to 30$\degr$, unless stated otherwise. Note finally that our calculations do not include heating from the planet, which could impact the emission of the planet's circumplanetary material.

\subsubsection{Gas line radiative transfer calculations}
\label{sec:gasRTsetup}

For RADMC-3D gas line radiative transfer calculations, we need the number density, temperature and velocity field (in physical units) of a gas species (e.g., $^{12}$CO) on a 3D grid described by spherical coordinates. The grid used in RADMC-3D is a 3D extension in spherical coordinates of the 2D polar grid used in the hydrodynamical simulations. For the gas radiative transfer calculations, the 3D grid spans 5 pressure scale heights on both sides of the disc mid-plane over 80 cells evenly spaced in colatitude.

From the gas surface density $\Sigma_{\rm gas}$ in the simulation, we first infer the gas volume density $\rho_{\rm gas}$ on the 3D grid by assuming for simplicity (i) vertical hydrostatic equilibrium, (ii) that the gas temperature is the same as in the hydrodynamical simulation and (iii) is independent of altitude ($z$). The gas number density is then $n_{\rm gas} = \rho_{\rm gas}/\umu m_{\rm p}$ with $m_{\rm p}$ the proton mass. In this work, we have considered three gas species for the radiative transfer calculations: $^{12}$CO, $^{13}$CO and C$^{18}$O. The number density of each species is first taken to be proportional to the gas number density: $n_{\rm species} = \chi n_{\rm gas}$, with $\chi$ the fractional abundance of the molecular species with respect to hydrogen nuclei. In all the results presented in Section~\ref{sec:results}, $\chi = 10^{-4}$, $2\times10^{-6}$ and $2\times10^{-7}$ for $^{12}$CO, $^{13}$CO and C$^{18}$O, respectively. These values are in line with abundance determinations in the local interstellar medium \citep{WR94_abun}, but note that they may vary from one disc to another and with disc age \citep[see, e.g., ][and references therein]{Zhang20}. We also include the effects of photodissociation by UV irradiation, by simply dropping $n_{\rm species}$ by 5 orders of magnitude wherever the gas column number density above an altitude $z$ falls below $10^{21}$\,cm$^{-2}$ \citep[see, e.g.,][]{Flaherty15}, namely:
\begin{equation}
\int_{z}^{\infty} n_{\rm gas}(R,\varphi,z')dz' < 10^{21}\,{\rm cm}^{-2},
\label{pd_eq1}
\end{equation}
with 
\begin{equation}
n_{\rm gas}(R,\varphi,z) = \frac{\Sigma_{\rm gas}(R,\varphi)}{\umu m_{\rm p} \sqrt{2\upi} H}\exp(-z^2/2H^2).
\label{pd_eq1b}
\end{equation}
Using that
\begin{equation}
\int_{z}^{\infty} \frac{1}{\sqrt{2\upi}H} \exp(-z'^2/2H^2) dz' = \frac{1}{2} {\rm erfc}(z/\sqrt{2}H),
\label{pd_eq2}
\end{equation}
where erfc denotes the complementary error function \citep[see, e.g., page 348 of][]{Krumholz15}, Eq.~(\ref{pd_eq1}) can be conveniently recast as
\begin{equation}
\frac{1}{2} {\rm erfc}(z/\sqrt{2}H) \times \frac{\Sigma_{\rm gas}(R,\varphi)}{\umu m_{\rm p}} < 10^{21}\,{\rm cm}^{-2}.
\label{pd_eq3}
\end{equation}
A similar strategy to account for photodissociation was used, for instance, in \citet{Simon15pd}. Note that CO isotope-selective photodissociation \citep[see, e.g.,][]{vdM16} is discarded: $n_{\rm species}$ is decreased by the same amount for $^{12}$CO, $^{13}$CO and C$^{18}$O wherever Eq.~(\ref{pd_eq3}) is satisfied. Moreover, CO freezeout onto dust grains is not taken into account since the gas temperature does not fall below 20 K in our disc model. Finally, to expand in 3D the gas velocity field of our simulations, we simply assume that the radial and azimuthal components of the gas velocity are independent of altitude, and that the vertical velocity component is zero.

Our gas radiative transfer calculations focus on the J=3$\rightarrow$2 line for $^{12}$CO, $^{13}$CO and C$^{18}$O, whose rest wavelength is approximately 0.866 mm, 0.906 mm and 0.909 mm, respectively. The reason for choosing this rotational transition is that it usually provides a good compromise in terms of signal-to-noise ratio in sub-millimetre observations: the higher J, the higher the excitation temperature, the larger the integrated intensity for optically thick emission, but also the higher the emission frequency and the higher the resulting noise level for a given observing time.

The calculations further assume a local thermodynamic equilibrium (LTE) and RADMC-3D computes the partition function for the level populations from the molecular data files provided by the Leiden \href{https://home.strw.leidenuniv.nl/~moldata/}{LAMBDA} database \citep{Leiden2005}. For the LTE assumption to be valid, the number density of a given gas species should exceed a critical value for energy levels to be populated primarily by collisions (and not by radiation). For instance, for $^{12}$CO J=3$\rightarrow$2, the CO number density at $\sim$ 40 K needs to be $\ga 10^4$ cm$^{-3}$ for the LTE hypothesis to be valid \citep[see \S~2.1 of][]{Weaver18}. This is indeed the case outside the cavity at most altitudes about the midplane, but is marginally verified inside the cavity. We have checked that non-LTE calculations carried out with the Large Velocity Gradient approximation (Sobolev method) or with the optically thin non-LTE population method in RADMC-3D yield undistinguishable results from our LTE calculations, whether photodissociation is included or not.

The specific intensity of the gas line emission is computed in 101 channel maps covering $\pm$ 9 km s$^{-1}$ around the systemic velocity, which encompasses the range of line-of-sight velocities in our disc model for disc inclinations up to 40$\degr$. Channel maps are thus spaced by $\Delta v \approx 0.18$ km s$^{-1}$, which is comparable to the spectral resolution currently achieved in ALMA disc gas observations. It is also comparable to the thermal broadening of the CO lines in our disc model, $\delta v_{\rm th} = (2\umu m_{\rm p} / m_{\rm CO})^{1/2} \times c_{\rm s}$, with $m_{\rm CO} = 28 m_{\rm p}$ the mass of CO molecules and $c_{\rm s}$ the sound speed ($\delta v_{\rm th}$ varies between 0.11 and 0.28 km s$^{-1}$ depending on the gas temperature). A turbulent broadening of the lines should also be induced by the gas turbulent viscosity, with a turbulent line width $\sim$ $\sqrt\alpha c_{\rm s}$. Given the values of $\alpha$ inside and outside the cavity (see Section~\ref{sec:hydrosetup}), the thermal line width is $\sim$ 2.4 times larger than the turbulent line width inside the cavity, and $\sim$ 24 times larger outside. The effective line width, which is the quadratic sum of the thermal and turbulent line widths, is therefore weakly affected by the gas turbulence in our disc model, despite $\alpha$ reaching a quite substantial value inside the cavity. For this reason, calculations including turbulent broadening of the lines yield integrated intensity maps that differ from those without turbulent broadening by only a few percent.

The specific intensity in each channel is then convolved with a circular beam of full-width at half-maximum (FWHM) set to 50 milli-arcseconds (mas), which corresponds to the solid angle subtended by about 40 pixels in our synthetic maps. This angular resolution has been achieved, for instance, in the $^{12}$CO DSHARP observations of the discs around HT Lup \citep{Kurtovic_dsharp} and HD 143006 \citep{Perez_dsharp}. Moment maps of the gas emission are finally generated with the code \href{https://bettermoments.readthedocs.io/}{\texttt{bettermoments}} \citep{Teague18_bm}. The use of \texttt{bettermoments} was motivated by the inclusion of a simple noise model in the channel intensities for some of the results of radiative transfer calculations presented in Section~\ref{sec:results}. When noise is included, random numbers with a Gaussian probability distribution of zero mean and standard deviation $\sigma$ are added in each channel map prior to beam convolution. Each convolved channel map then includes white noise with a spatial scale that is similar to the beam size. We take $\sigma$ to be 1~mJy/beam per channel map, which is similar to the rms noise level attained in the $^{12}$CO DSHARP observations of HT Lup \citep{Kurtovic_dsharp}. Note that this noise model is very similar to that used in our synthetic maps of dust continuum emission \citep{Baruteau2019}.

\subsubsection{Dust radiative transfer calculations}
\label{sec:dustRTsetup}

The procedure to compute synthetic images of the dust continuum emission and of polarized scattered light with RADMC-3D from the results of Dusty FARGO-ADSG simulations was detailed in \S~2.2.1 and \S~2.2.2 of \citet{Baruteau2019}. We will therefore not repeat it here, but, for the sake of completeness, we list below the main parameters of our dust radiative transfer calculations:

\begin{figure*}
\centering
\resizebox{\hsize}{!}
{
\includegraphics{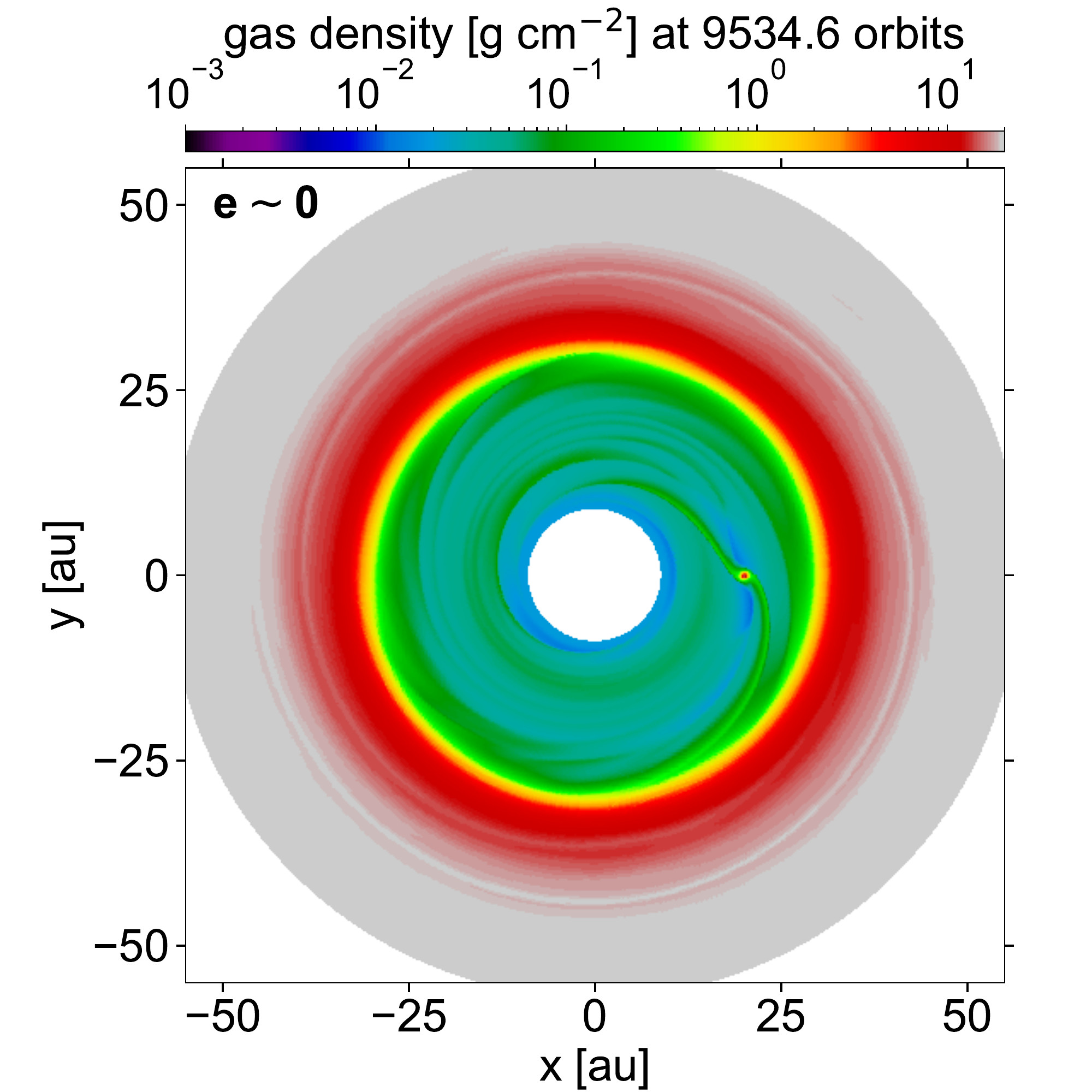}
\includegraphics{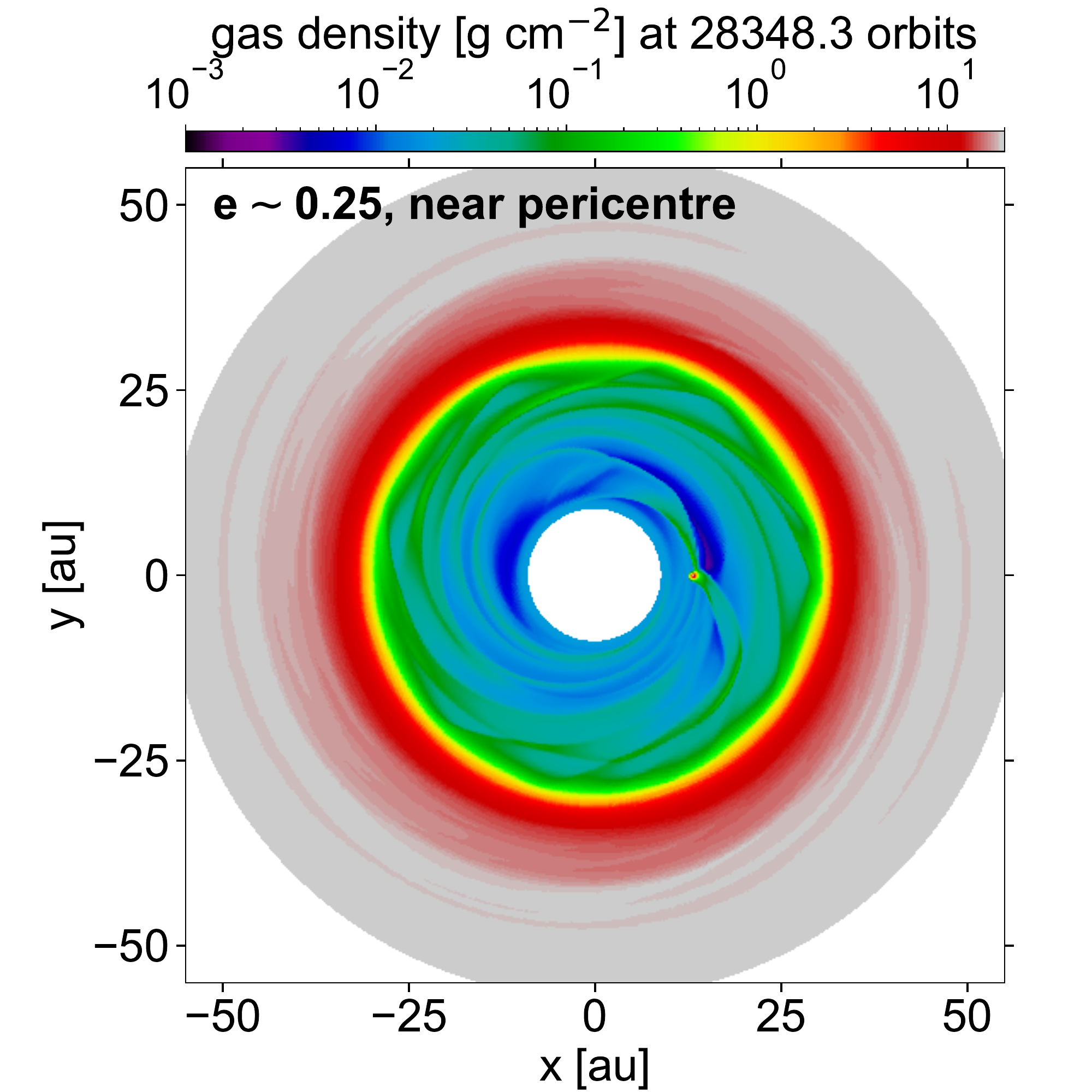}
\includegraphics{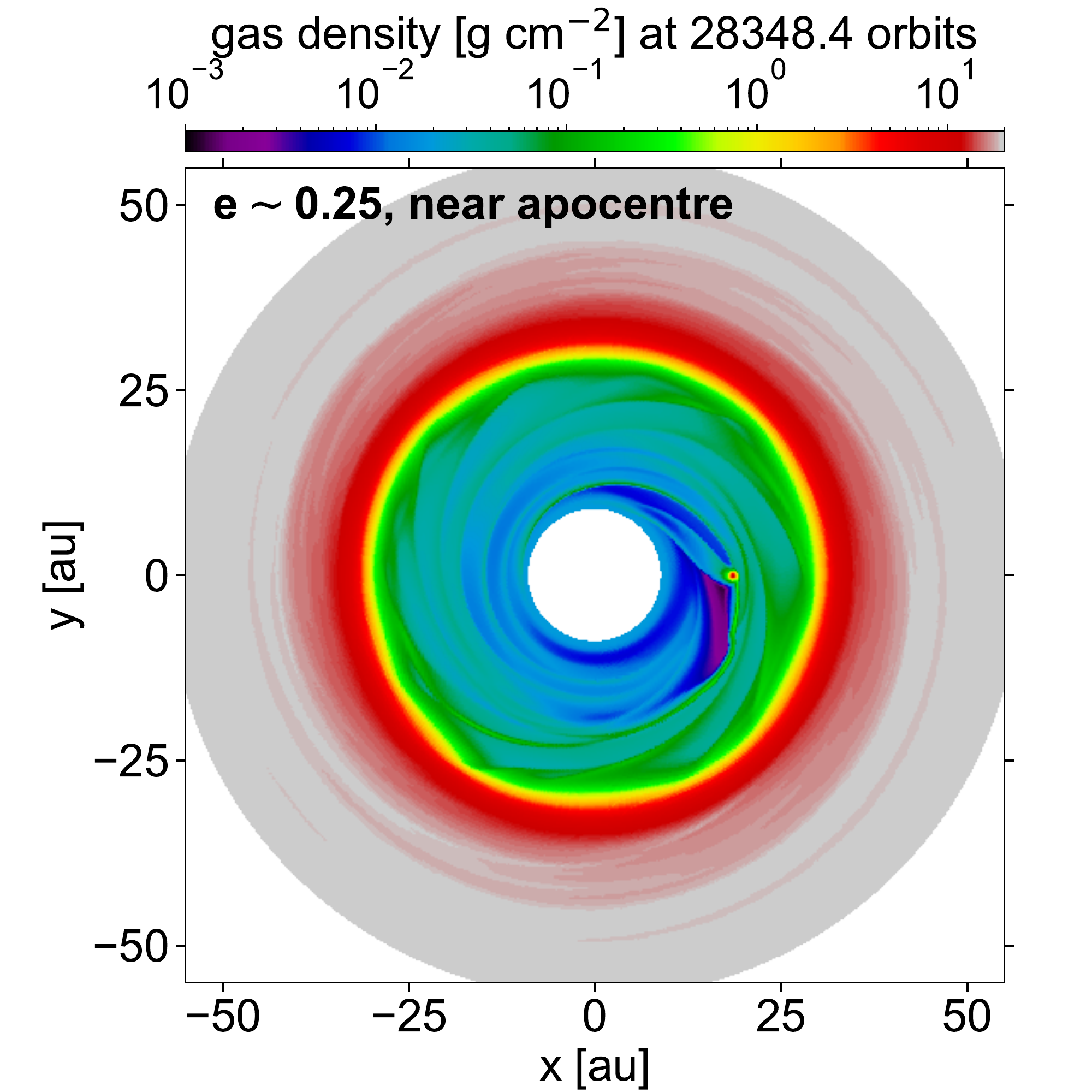}
}
\caption{Gas surface density in the hydrodynamical simulation when the planet still has a nearly circular orbit in the cavity (left panel, at about 1.3 Myr) and when it has reached a nearly maximum eccentricity ($e \approx 0.25$, at about 3.8 Myr), close to pericentre (middle panel) and apocentre (right panel).}
\label{fig:fig2}
\end{figure*}

\begin{enumerate}
\item Dust continuum images are generated at 0.9 mm from our restart gas+dust simulation (see Section~\ref{sec:hydrosetup}). The same size range for the dust as in the simulation, namely [10 $\micron$ $-$ 1 cm], is taken in the radiative transfer calculation, but the size distribution $n(s)$ is taken $\propto s^{-3.5}$ \citep{Mathis77}. We further assume that the dust-to-gas mass ratio is 0.01 for this range of dust sizes, and that the dust is a mixture of 70 per cent silicates and 30 per cent water ices (corresponding to a mean internal density $\approx$ 1.3 g cm$^{-3}$, consistent with the particles' internal density adopted in the simulations). The 3D grid used in RADMC-3D spans 2 pressure scale heights on both sides of the disc mid-plane with 40 cells logarithmically spaced in colatitude (see Equation~4 of \citealp{Baruteau2019} for the dust's scale height used in the vertical density profile). Unlike in \citet{Baruteau2019}, the dust temperature is taken to be the gas temperature of the hydrodynamical simulation (with no vertical dependence). No preliminary thermal Monte-Carlo was therefore carried out to compute the dust temperature. Henyey-Greenstein anisotropic scattering is included in the dust radiative transfer calculations, and the specific intensity of the continuum emission is finally convolved with a circular beam of FWHM 50 mas (the same beam as in the gas emission maps).
\smallskip
\item Polarized intensity images are computed at 1.04 $\micron$ (Y-band). The scattered light is assumed to arise from small dust grains in the size range [0.01$-$0.3] $\micron$, with a size distribution $\propto s^{-3.5}$ and a corresponding dust-to-gas mass ratio of $2\times10^{-3}$. As in \citet{Baruteau2019}, these small grains are assumed to form a mixture of 60 per cent silicates and 40 per cent amorphous carbons, thus having a mean internal density $\rho_{\rm int} \approx 2.7$ g cm$^{-3}$. These small grains are further assumed to be perfectly coupled to the gas as long as their Stokes number $\sim$ $s\rho_{\rm int} / \Sigma_{\rm gas}$ remains smaller than a threshold value which we set to $10^{-4}$ (i.e., a small fraction of the alpha turbulent viscosity inside the cavity). When their Stokes number is larger than $10^{-4}$, which actually occurs only inside the cavity, the mass volume density of the small grains is set to 0, as will be justified and discussed in Section~\ref{sec:res_pol}. The 3D grid used in RADMC-3D is the same as that in the gas radiative transfer calculations. Again, the dust temperature is directly obtained from the hydrodynamical simulation. The Stokes maps computed by RADMC-3D are convolved with a circular beam of FWHM 25 mas, which is similar to the angular resolution in the Y-band SPHERE observations of MWC 758 \citep{Benisty2015}. The final maps of polarized intensity are those of the (convolved) $Q_{\varphi}$ Stokes parameter scaled with the square of the deprojected distance from the central star, and further normalized to its maximum value.
\end{enumerate}

\section{Results and discussion}
\label{sec:results}
We present in this section the results of our hydrodynamical simulations and radiative transfer calculations. As we have seen in Fig.~\ref{fig:fig1}, the eccentricity of the 2 Jupiter mass planet rises smoothly after it has migrated into the gas cavity, reaching a maximum value of about 0.25 in a few Myr. The planet eccentricity drives strong asymmetries in the surface density of the gas inside the cavity, as described in Section~\ref{sec:res_hydro}. We show in Section~\ref{sec:res_12CO} that these asymmetries in the gas density manifest themselves as large-scale asymmetries in integrated intensity maps of the $^{12}$CO J=3$\rightarrow$2 line emission, which are detectable with the angular resolution and sensitivity currently achievable in ALMA disc gas observations. Section~\ref{sec:res_12CO} also highlights (i) the key role of photodissociation in enhancing the asymmetries in the $^{12}$CO integrated intensity maps, and (ii) detectable signatures of eccentric gas inside the cavity in the velocity map of $^{12}$CO J=3$\rightarrow$2. Outside the cavity, the gas emission is not affected by the planet eccentricity, and integrated intensity maps display a four-lobed pattern due to optical depth effects when the disc is inclined relative to the sky-plane. Next, sections~\ref{sec:res_13COC18O} to~\ref{sec:res_pol} serve to illustrate that in contrast to $^{12}$CO, the $^{13}$CO and C$^{18}$O line emissions, the polarized intensity in the near-infrared as well as the dust continuum emission in the sub-millimetre show no detectable signatures of the planet eccentricity in the disc cavity.

\subsection{Gas surface density}
\label{sec:res_hydro}
Fig.~\ref{fig:fig2} displays the gas surface density of our hydrodynamical simulation when the planet still has a near circular orbit in the cavity (left panel), and when its eccentricity is close to its maximum value of $\approx 0.25$ (middle and right panels). The left panel is obtained at $\sim$ 9500 orbital periods at the planet's initial orbital radius, which corresponds to about 1.3 Myr. The middle and right panels are obtained at $\sim$ 28300 orbital periods ($\approx$ 3.8 Myr). The gas surface density changes appreciably over a planet orbit when the planet is eccentric, and the two rightmost panels in Fig.~\ref{fig:fig2} highlight two situations: when the planet is near pericentre (middle panel) or near apocentre (right). The planet eccentricity alters the gas surface density in the cavity in mainly two ways: (i) the gap carved by the planet around its orbit is no longer annular, and (ii) the planet wakes are non-steady shock waves in a frame rotating with either the planet or the guiding centre of its orbit. That the gas density varies strongly with the orbital phase of an eccentric planet has been already documented in the literature, and we refer the reader to, for instance, \citet{DAngelo06} or \citet{Calcino20}. 

For the purpose of this work, we stress the very different pitch angles of the inner and outer primary wakes of the planet, which arise from the large velocity difference between the planet and the background gas. Near pericentre, the inner wake is much more tightly wound than the outer wake, and vice versa near apocentre. This large velocity difference also creates substantial gas depressions ahead or behind the planet in the azimuthal direction depending on the planet's orbital phase. For instance, when the planet is near apocentre in the right panel of Fig.~\ref{fig:fig2}, the gas surface density just behind the planet in azimuth is about 1$-$1.5 orders of magnitude smaller than the surface density just ahead of the planet. This density contrast can also be appreciated in the top-left panel of Fig.~\ref{fig:fig12}. A gas depression of similar amplitude occurs ahead of the planet when the latter approaches its pericentre (middle panel in Fig.~\ref{fig:fig2}). As we will see in the next section, these density contrasts can be observable signatures in the $^{12}$CO line emission of eccentric Jupiters inside a disc cavity.

\subsection{$^{12}$CO J=3$\rightarrow$2 line}
\label{sec:res_12CO}

We focus in this section on the $^{12}$CO J=3$\rightarrow$2 rotational line at $\sim$ 0.87 mm, with the aim to highlight how moment maps of the line emission are changed by the planet eccentricity inside the disc cavity. 

\begin{figure*}
\centering
\resizebox{0.96\hsize}{!}
{
\includegraphics{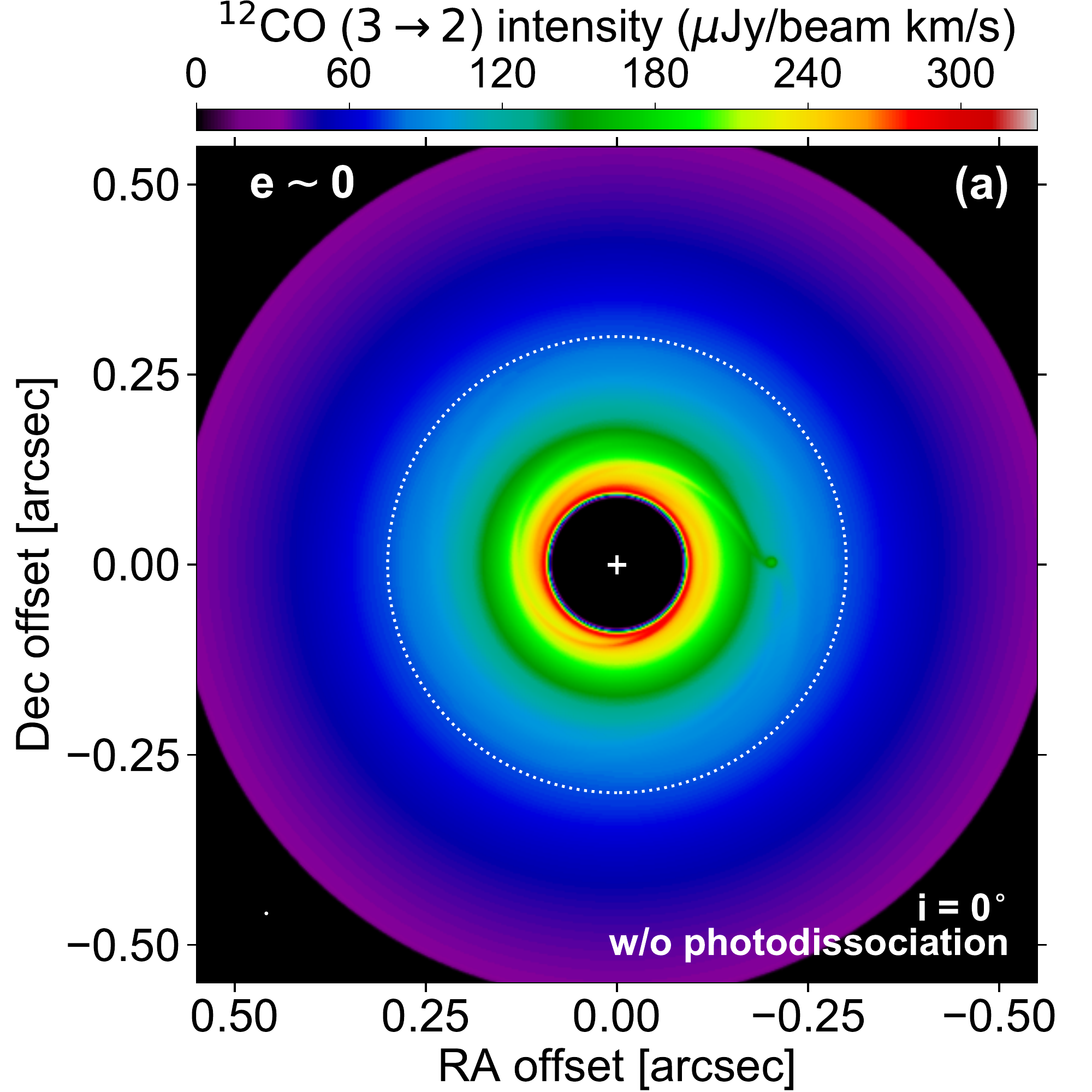}
\includegraphics{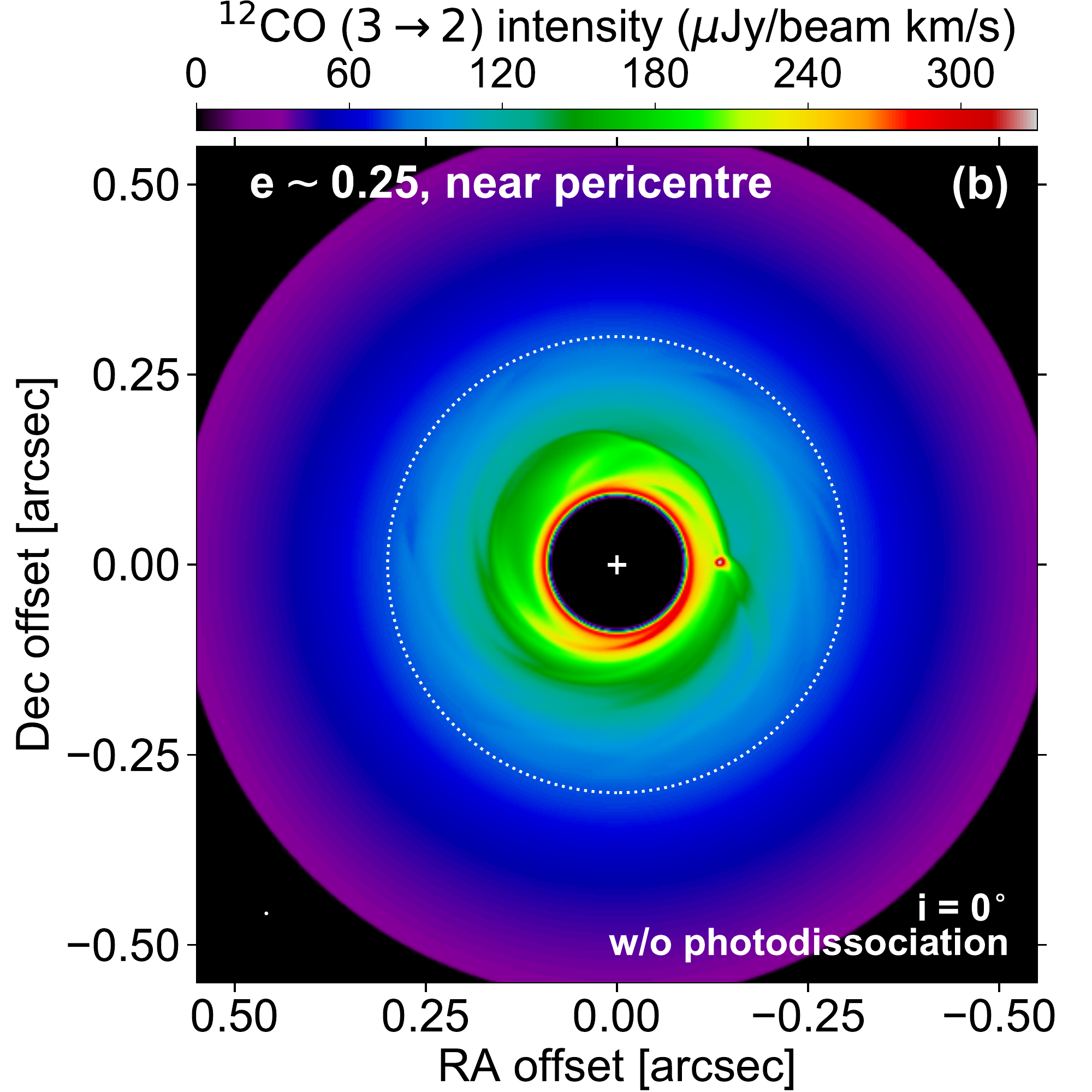}
\includegraphics{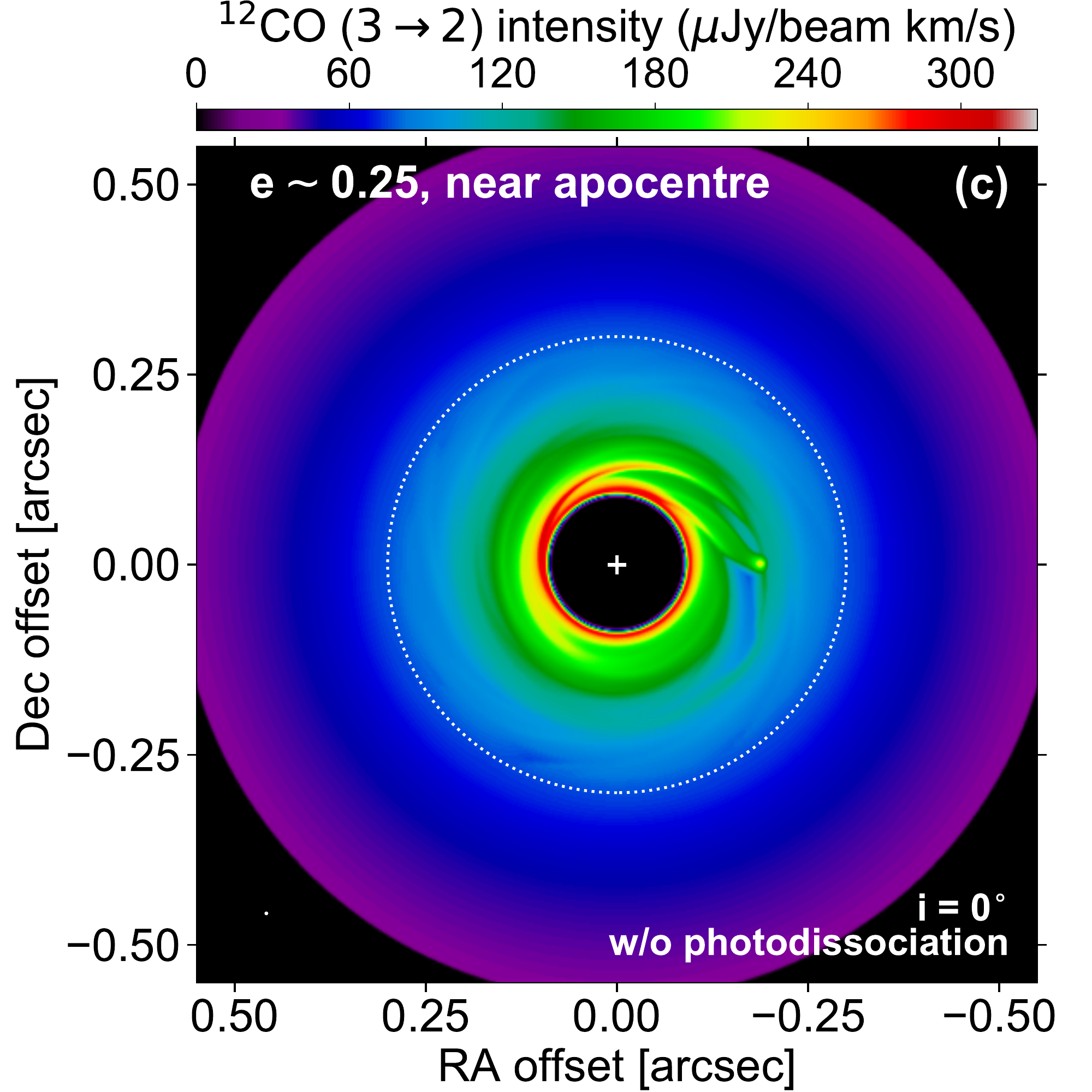}
}
\resizebox{0.96\hsize}{!}
{
\includegraphics{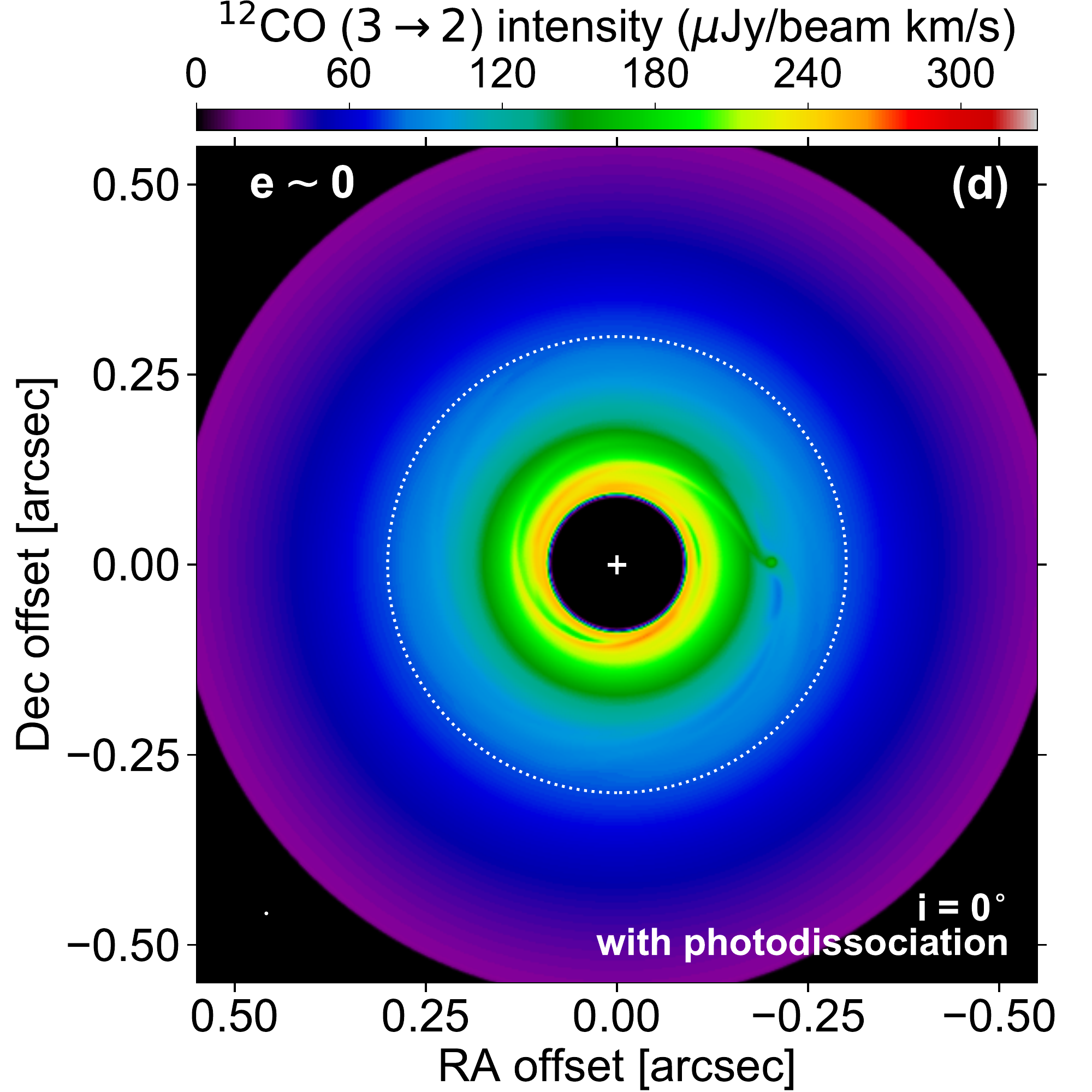}
\includegraphics{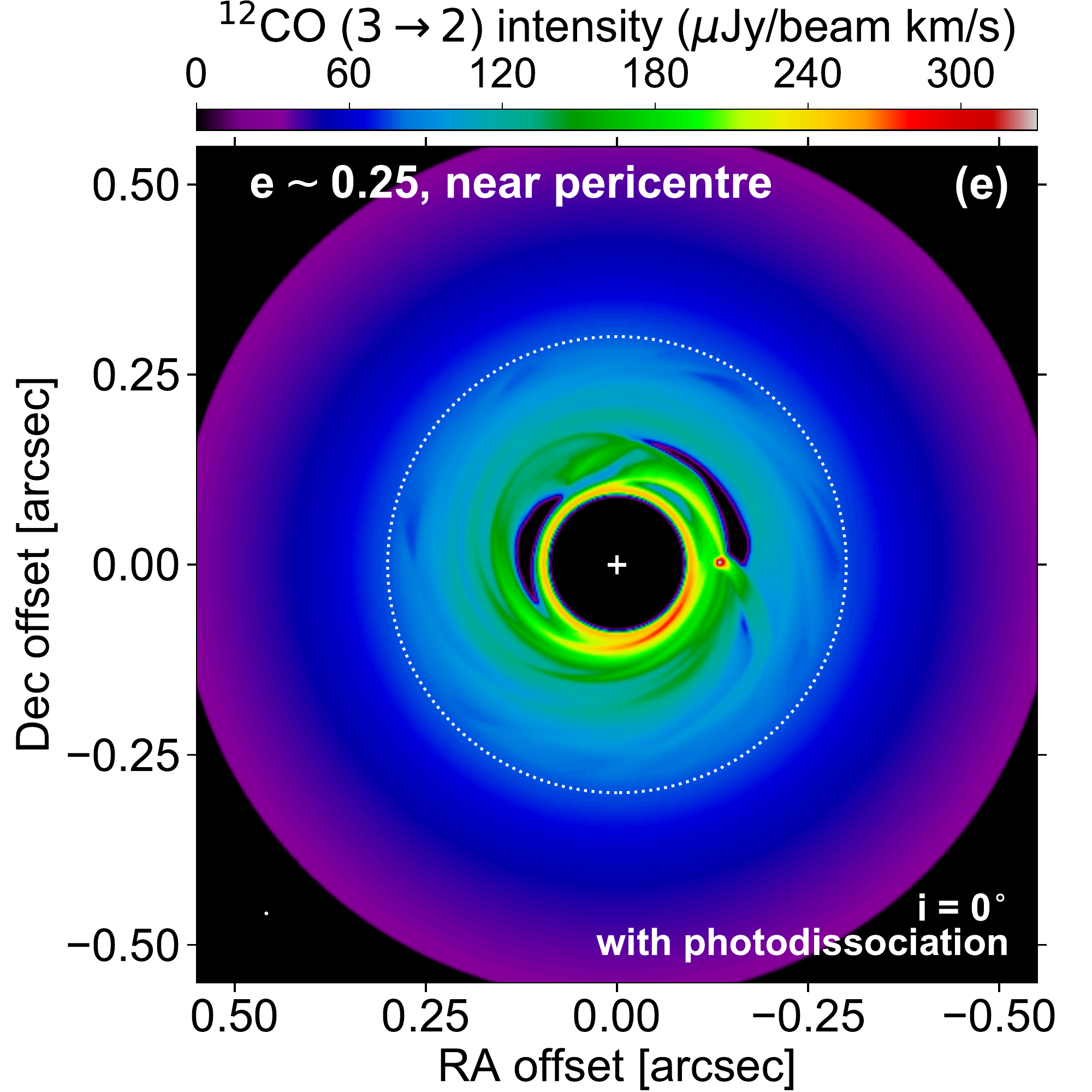}
\includegraphics{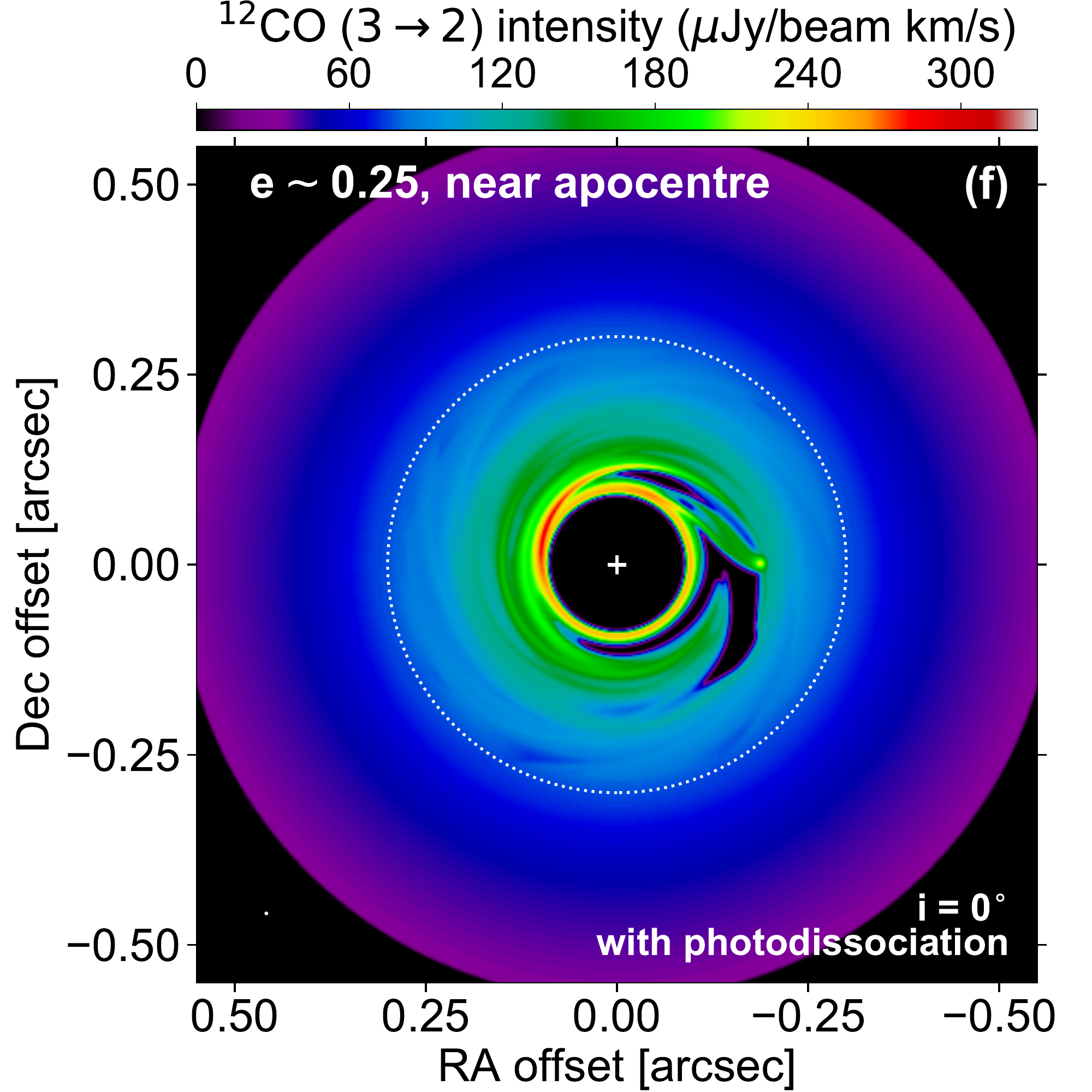}
}
\resizebox{0.96\hsize}{!}
{
\includegraphics{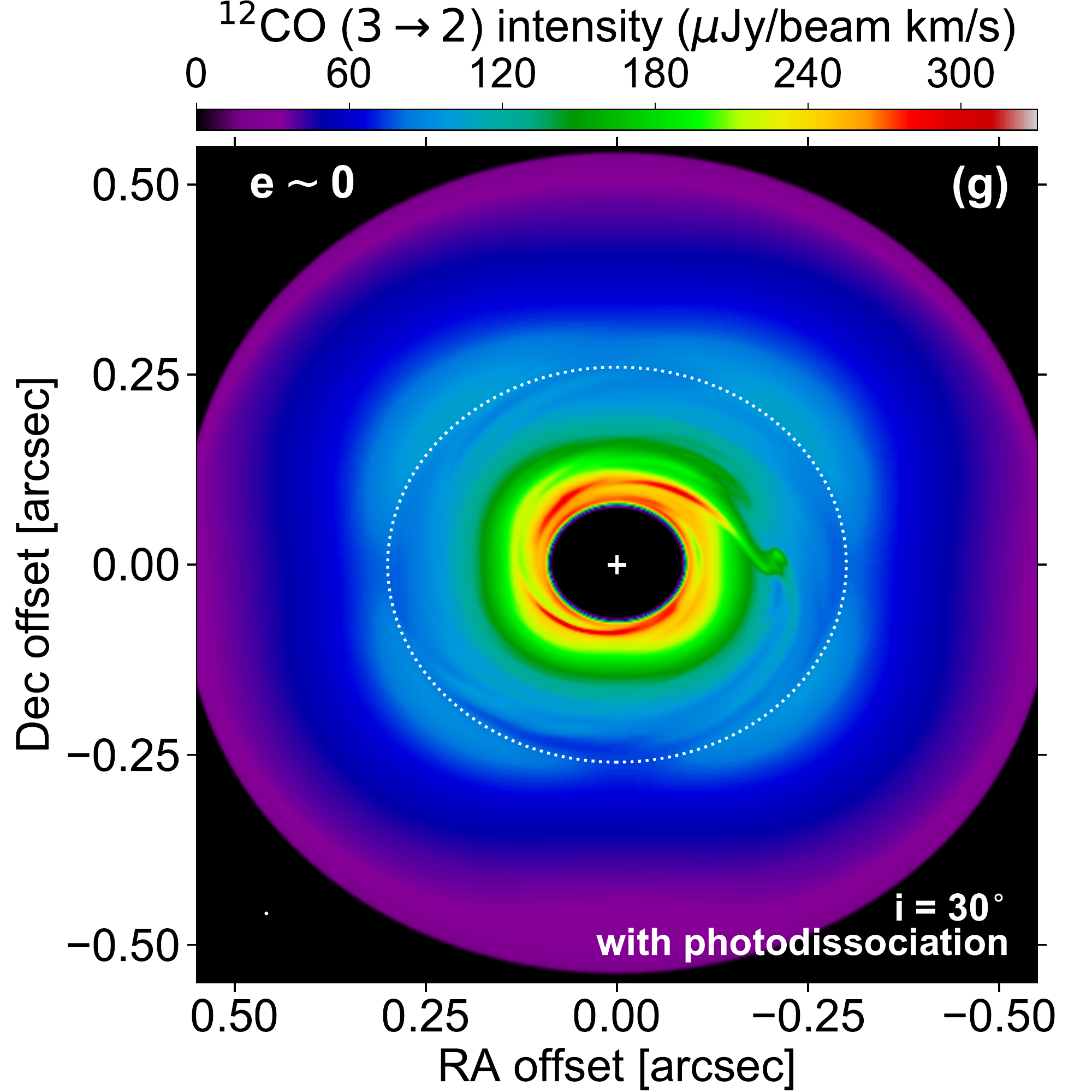}
\includegraphics{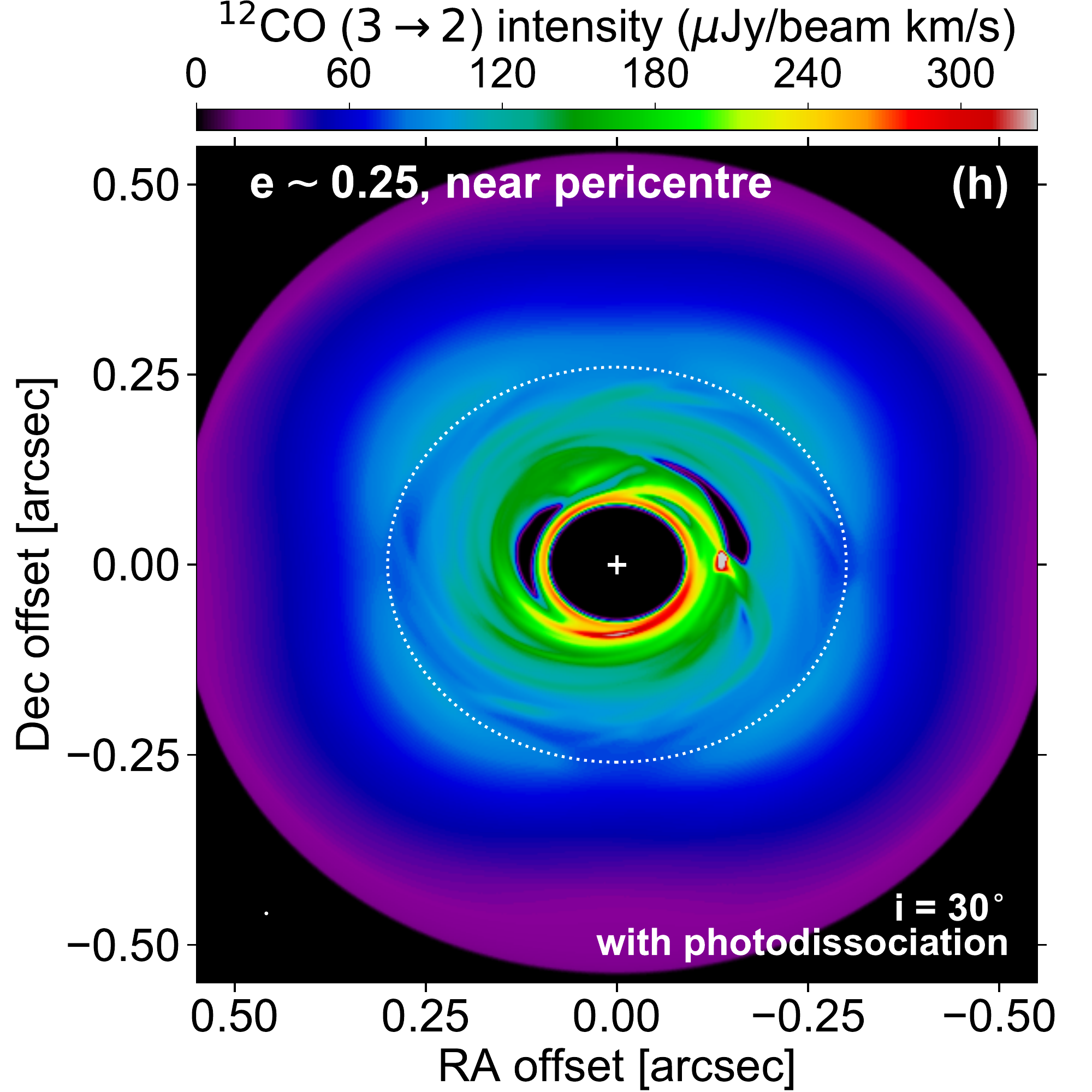}
\includegraphics{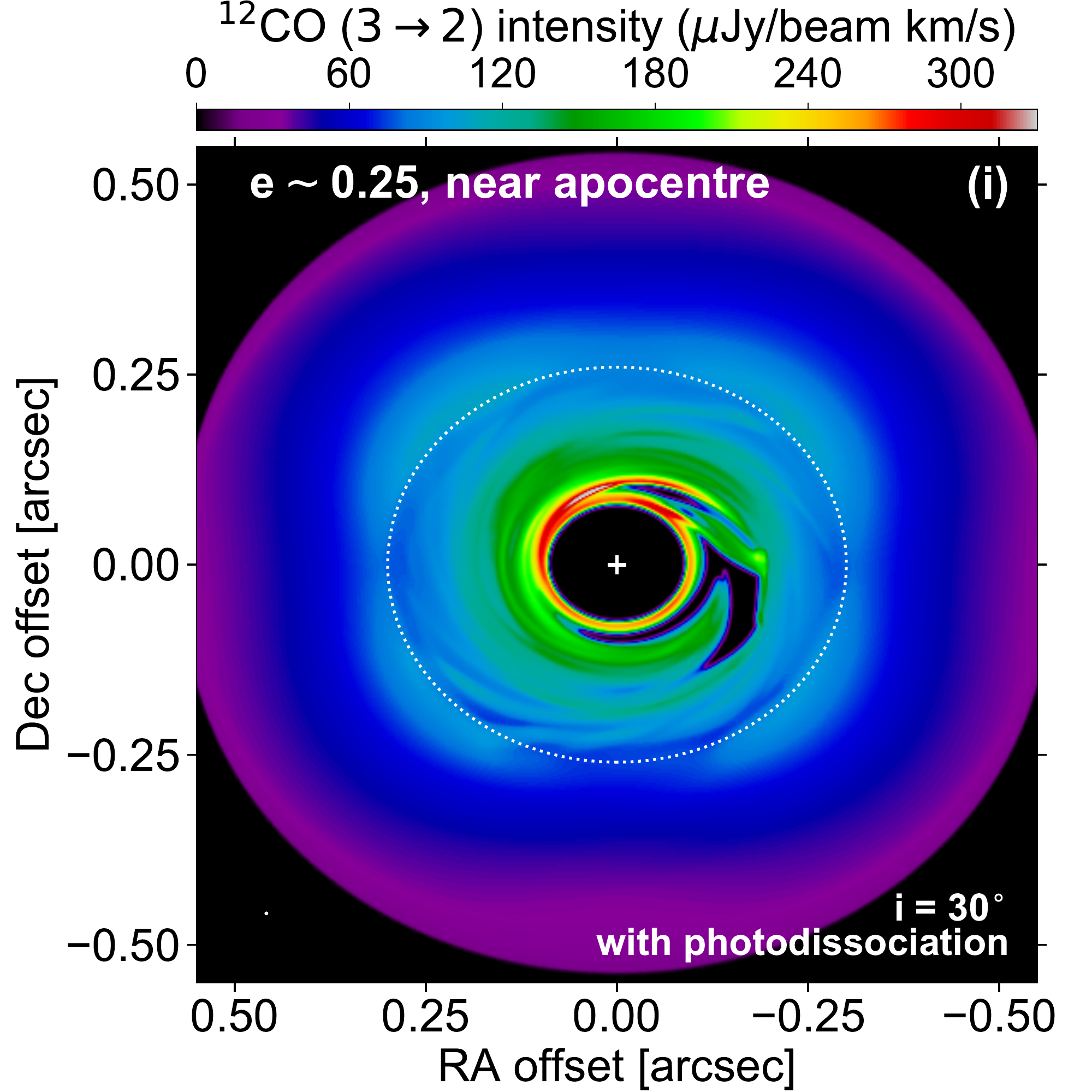}
}
\resizebox{0.96\hsize}{!}
{
\includegraphics{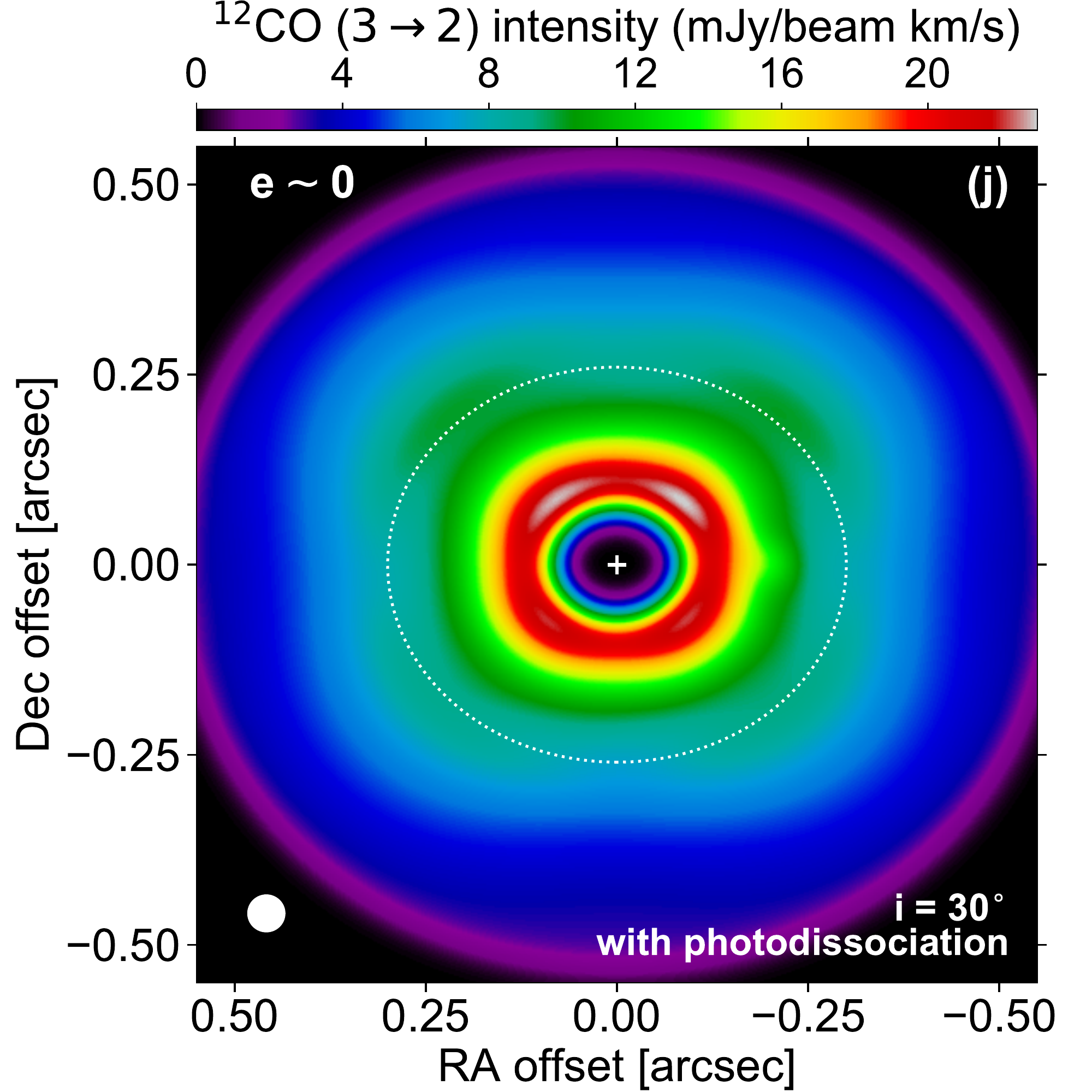}
\includegraphics{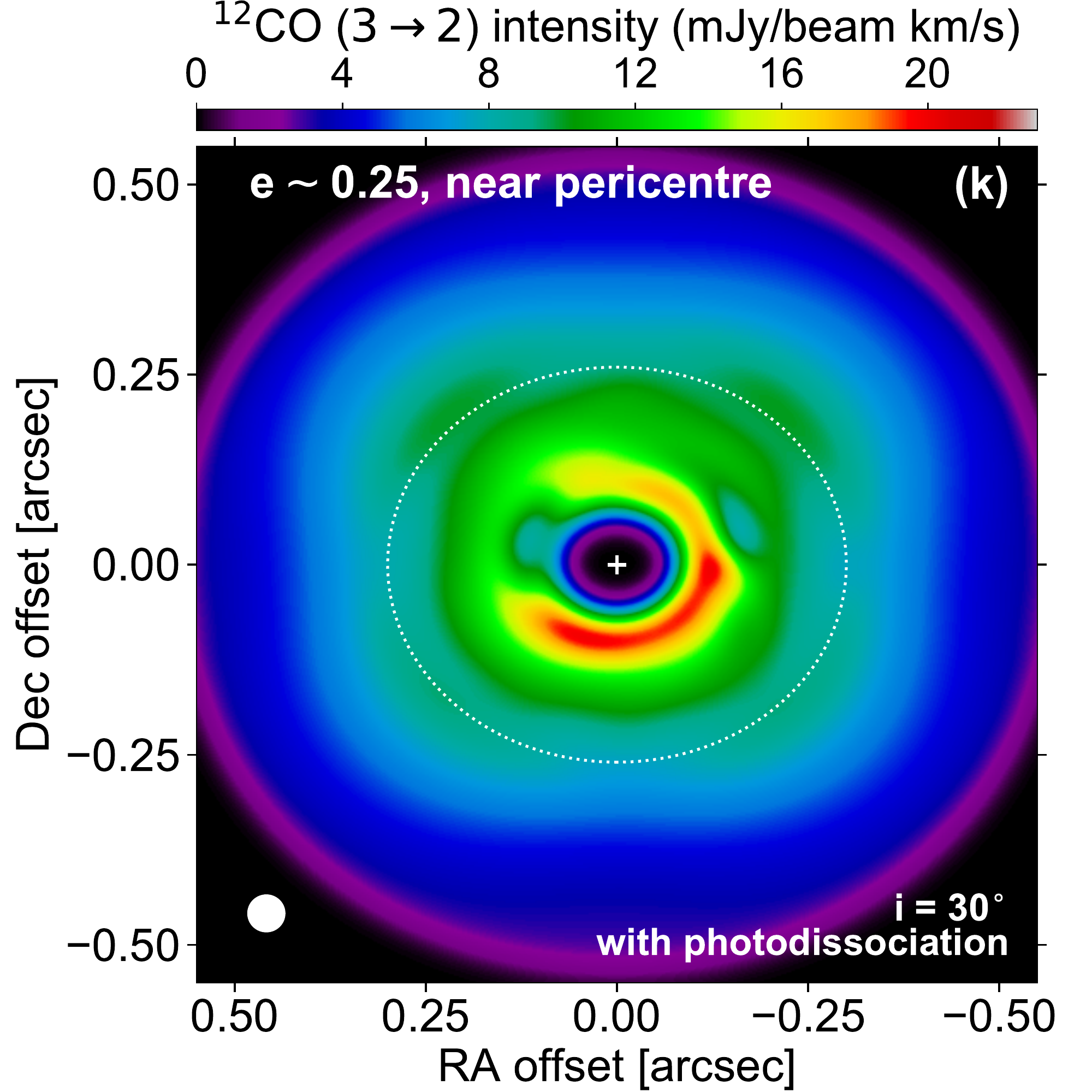}
\includegraphics{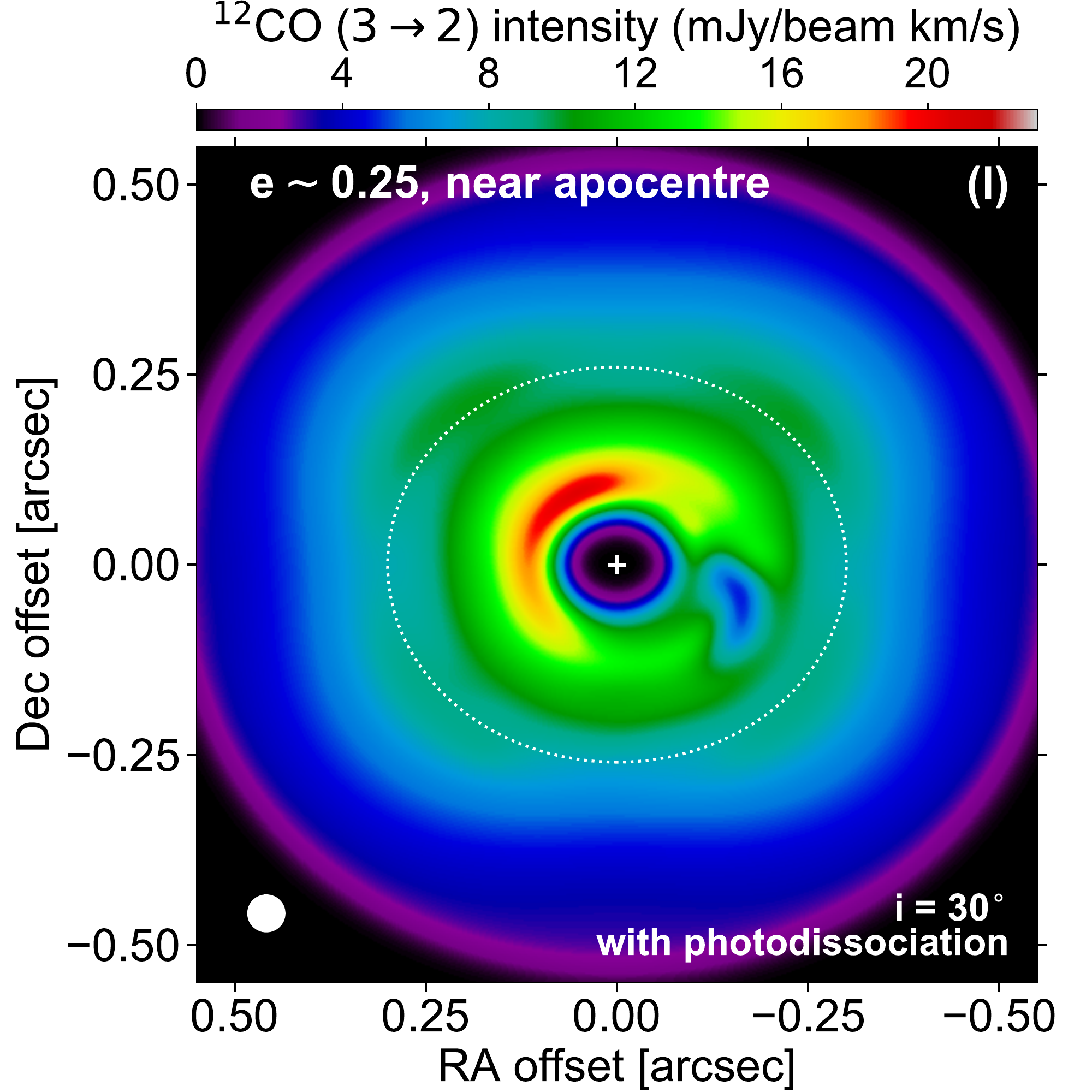}
}
\caption{Integrated intensity maps of the $^{12}$CO J=3$\rightarrow$2 line emission, at the same three times as the gas surface density panels of Fig.~\ref{fig:fig2}. The rows of panels show separately the effects of photodissociation, disc inclination ($i$) and beam convolution (see the lower right corners). For instance, rows 2 and 3 are both with photodissociation, but $i$ is $0\degr$ in row 2 and $30\degr$ in row 3. The white circle in the lower left corners shows the beam size, which is 5 mas in panels (a) to (i) and 50 mas in panels (j) to (l). In each panel, the dotted curve shows the approximate location of the gas cavity, and the white cross marks the star position.}
\label{fig:fig3}
\end{figure*}

\subsubsection{Integrated intensity maps}
\label{sec:res_12CO_mom0}

We display in Fig.~\ref{fig:fig3} a series of integrated intensity maps (i.e., moment 0 maps) at the same times as in Fig.~\ref{fig:fig2}: when $e\sim0$ (left panels), $e\sim0.25$ near pericentre (middle panels) and $e\sim0.25$ near apocentre (right panels). The different rows help to illustrate separately the effects of photodissociation, disc inclination and beam convolution. 

The first row of panels shows the moment 0 maps obtained without photodissociation, for a disc with zero inclination and for a small 5 mas circular beam (equivalent to 4 pixels of our synthetic maps). This idealised resolution is intended to highlight the structures induced by the planet eccentricity but might not be achievable with observations today. We see that the integrated intensity smoothly decreases with increasing distance from the star. This is due to the line emission being optically thick even inside the $\sim0\farcs3$ cavity, so that the radial profile of the integrated intensity reflects mostly radial variations of the gas temperature. Still, the planet and its wakes are visible inside the cavity, more clearly when the planet is eccentric. This tells us that the optical depth of the $^{12}$CO J=3$\rightarrow$2 line emission cannot be too large inside the cavity for the asymmetries in the gas surface density to translate into asymmetries in the moment 0 maps (recall that the gas temperature in the simulation and in the radiative transfer calculations only depends on the radial distance from the star). This is confirmed by inspecting maps of optical depth ($\tau$) computed by RADMC-3D: while $\tau \in [3-7]\times 10^4$ outside the cavity, $\tau$ is typically less than a few tens inside the cavity when the planet is circular. It even approaches unity where the gas density is smallest when the planet is eccentric. This range of variation of $\tau$ is consistent with that of $\Sigma_{\rm gas}$ in Fig.~\ref{fig:fig2} despite the opacity dependence on the radial temperature profile.

The second row of panels in Fig.~\ref{fig:fig3} displays the same maps with photodissociation. We see that photodissociation has little impact when the planet is still on a near circular orbit (panel d). It mostly makes the planet gap a little more discernible in the vicinity of the planet. The effect of photodissociation is, however, much clearer in panels (e) and (f) when the planet is eccentric. In that case, the depressions in the gas surface density brought about by the planet wakes are sufficiently strong for the photodissociation criterion Eq.~(\ref{pd_eq1}) to be met down to the midplane. At these locations, the CO number density is greatly reduced (by 5 orders of magnitude, see Section~\ref{sec:gasRTsetup}), and so are the optical depth and the integrated intensity.

\begin{figure*}
\centering
\resizebox{\hsize}{!}
{
\includegraphics{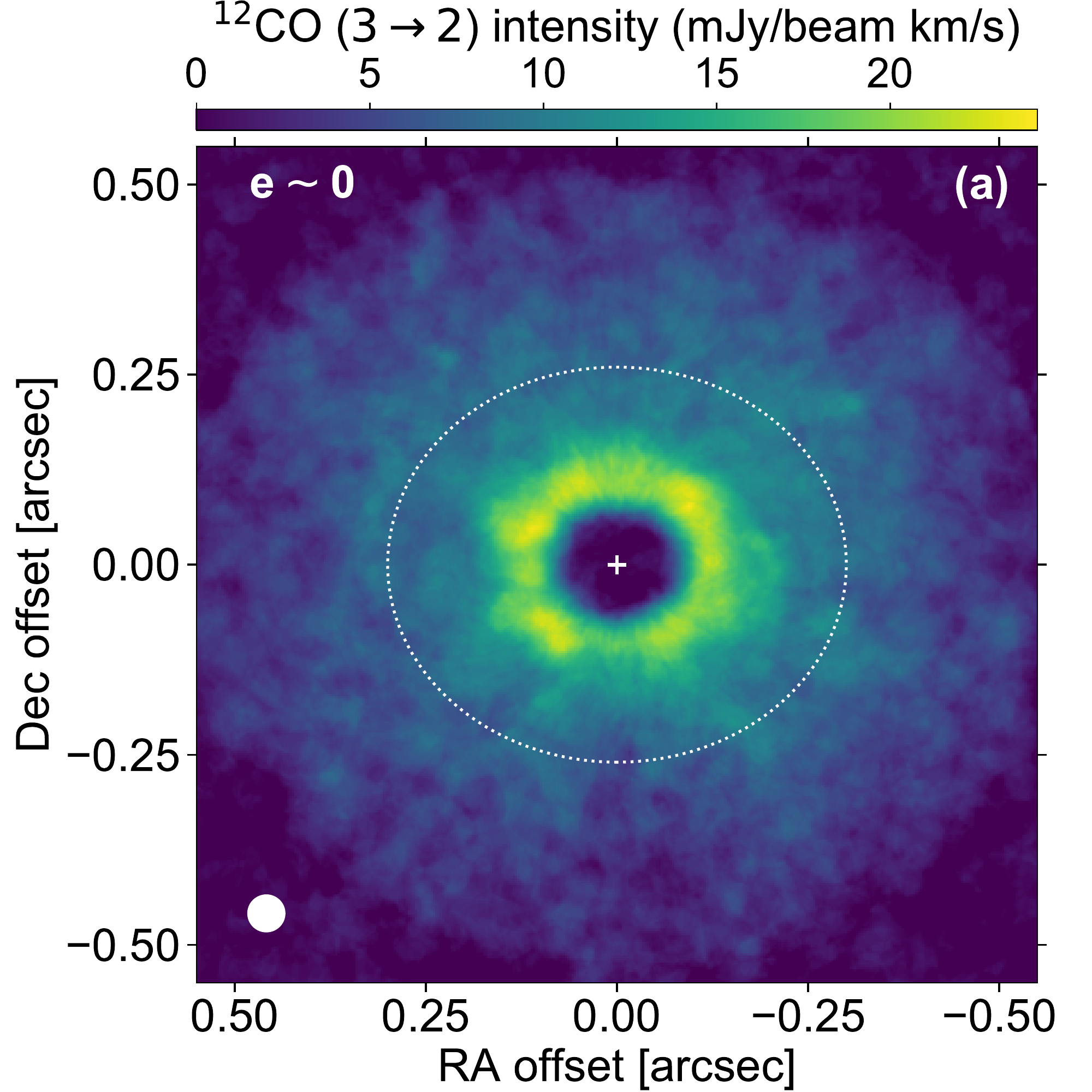}
\includegraphics{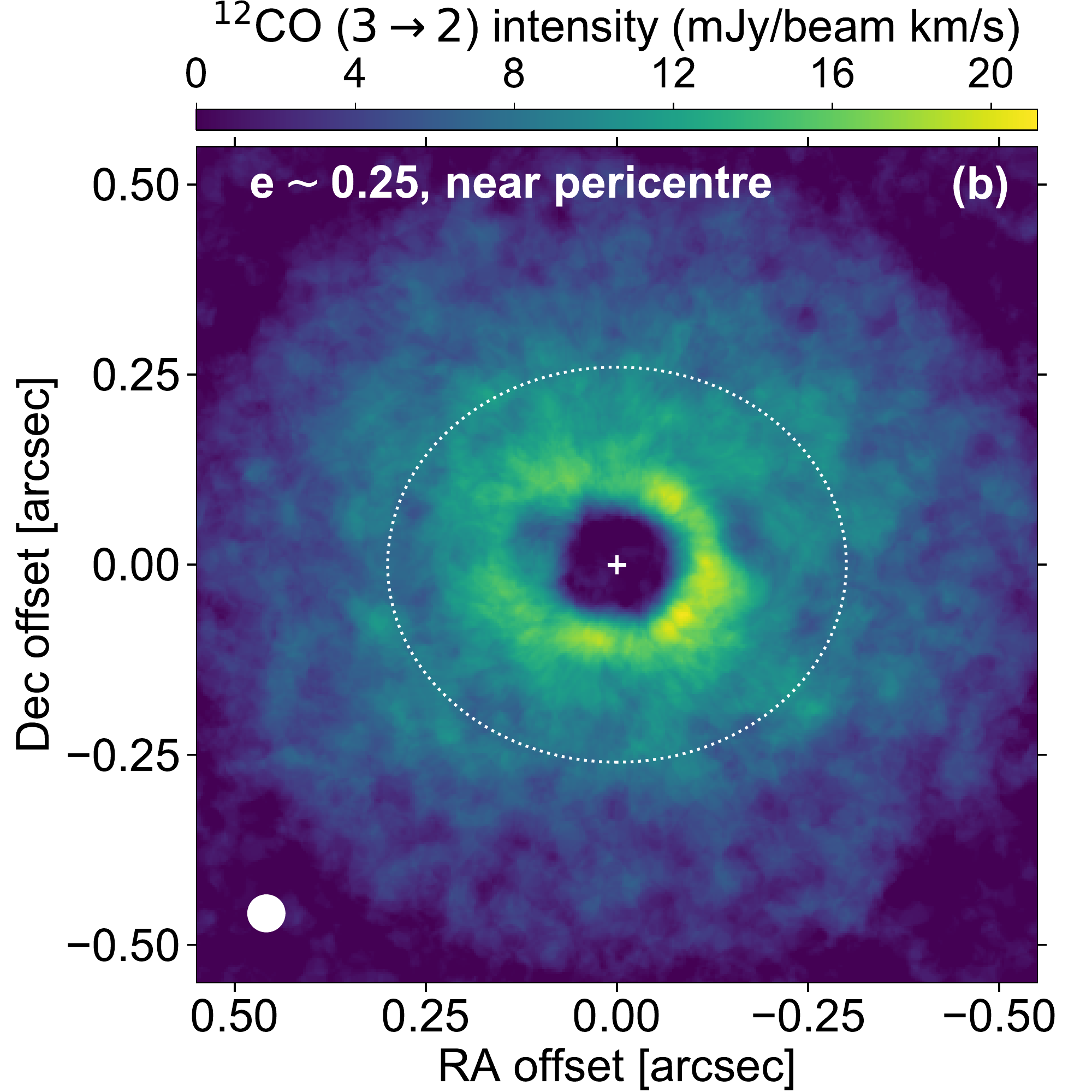}
\includegraphics{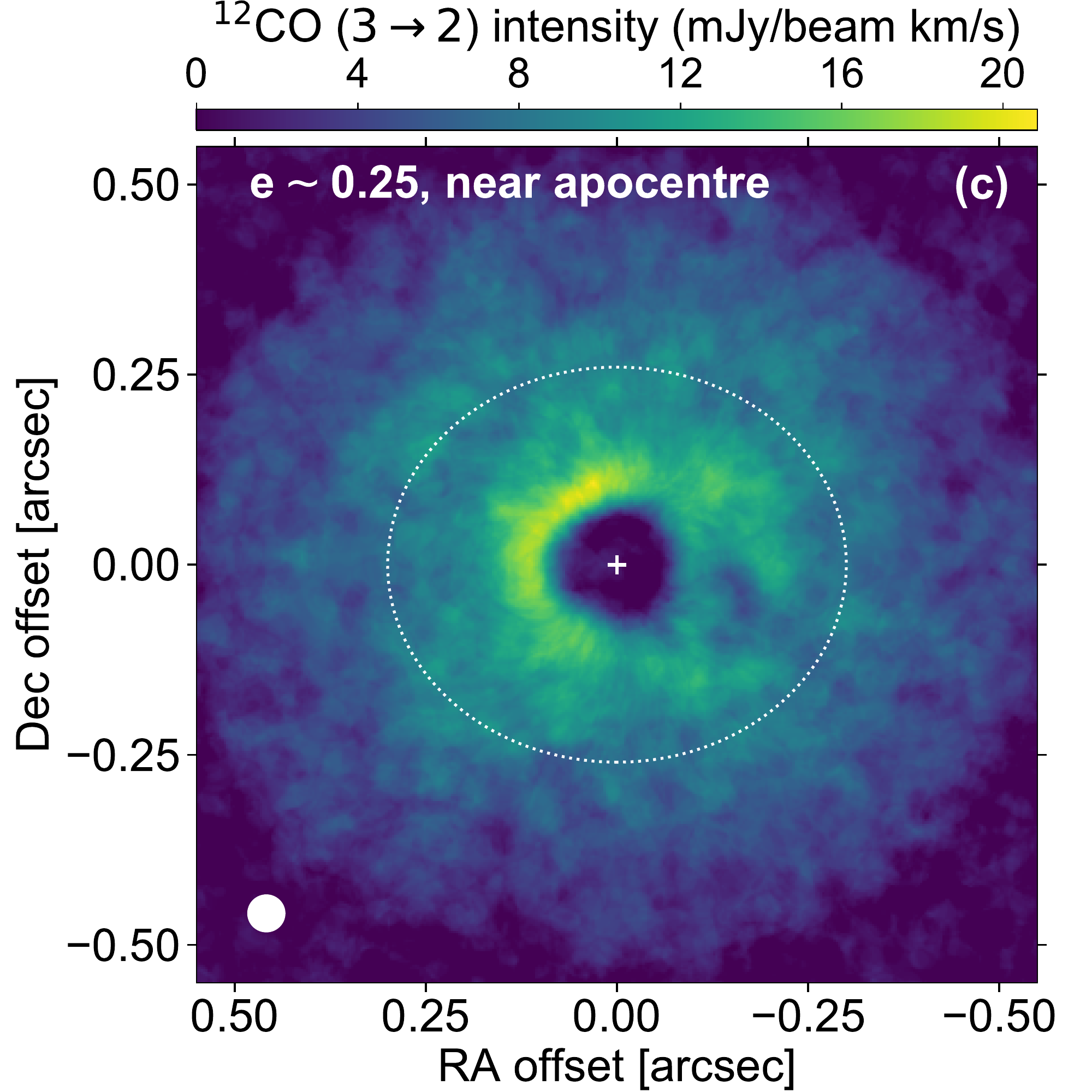}
}
\caption{$^{12}$CO J=3$\rightarrow$2 integrated intensity maps with photodissociation, a disc inclination of 30$\degr$, a 50 mas beam, and inclusion of white noise with 1~mJy/beam rms in each channel map. The same maps without noise are shown in panels (j) to (l) in Fig.~\ref{fig:fig3}. In each panel, the dotted curve shows the approximate location of the gas cavity, the white circle shows the beam and the white cross marks the star position.}
\label{fig:fig4}
\end{figure*}

In the third row of panels, a disc inclination of 30$\degr$ is further taken into account. Since the beam remains very small in these three maps, the number of spectral channels has been increased to 401 so that the contributions of each channel to the moment 0 maps remain indiscernible. This effect is of course absent when the disc inclination is zero, since only the channel with zero line-of-sight velocity then contributes to the integrated intensity map. The effect of the disc inclination is essentially two-fold: (i) a projection effect mainly inside the cavity (near-circular contours now appear elliptical) and (ii) a four-lobed pattern of emission outside the cavity. This four-lobed pattern, which is due to optical depths effects, will be discussed in Section~\ref{sec:res_12CO_4lobe}.

The last row of panels finally shows the moment 0 maps obtained after convolution with our fiducial 50 mas beam. For the circular planet case (panel j), the beam convolution makes the planet and its wakes barely visible. However, for the eccentric planet case (panels k and l), the asymmetries in the gas density due to the planet eccentricity clearly manifest themselves as large-scale asymmetries in the moment 0 maps. In all three panels, the four-lobed emission pattern outside the cavity is now more subtle and causes iso-intensity contours to have an approximately square shape.

We end up this section by showing in Fig.~\ref{fig:fig4} the moment 0 maps obtained after including white noise with standard deviation $\sigma =$ 1~mJy/beam in all channels (see last paragraph of Section~\ref{sec:gasRTsetup}). 2$\sigma$ clipping has been applied to the beam-convolved channel maps when computing the resulting moment 0 map with \href{https://bettermoments.readthedocs.io/}{\texttt{bettermoments}} \citep{Teague18_bm}. The comparison between the three panels makes it clear that, when the planet is eccentric, the asymmetries in the integrated intensity inside the cavity have a contrast well above the noise level. These asymmetries therefore constitute a clear detectable signature of the presence of an eccentric Jupiter in the gas cavity of its protoplanetary disc. While photodissociation helps enhance the asymmetries, they are still visible, albeit to a lesser extent, in moment 0 maps where photodissociation is discarded, even with a similar level of noise in the channel maps. This is shown in Fig.~\ref{fig:figa1} of Appendix~\ref{sec:res_12CO_mom0_nopd}. We finally point out that the four-lobed emission pattern outside the cavity is not discernible in the panels of Fig.~\ref{fig:fig4}, but becomes more prominent with our noise model upon increasing disc inclination, as will be shown in Fig.~\ref{fig:figb1} of Appendix~\ref{sec:res_12CO_mom0_incl}.

\subsubsection{A four-lobed pattern at high optical depths for inclined discs}
\label{sec:res_12CO_4lobe}
The $^{12}$CO J=3$\rightarrow$2 integrated intensity maps described in the previous section display a four-lobed pattern outside the cavity when the disc is inclined by 30$\degr$ relative to the sky-plane (panels g to l in Fig.~\ref{fig:fig3}). The integrated intensity outside the cavity has indeed local minima along the disc major and minor axes, i.e. along the horizontal and vertical axes in the images (the disc's position angle being taken equal to zero). A four-lobed pattern in synthetic integrated intensity maps has been obtained in several radiative transfer calculations with disc inclinations $\ga40\degr$, like for instance in \citet{Liu18}, \citet{Keppler19_pds70} and \citet{Zhu19}. This pattern is also clearly discernible in observed integrated intensity maps of the PDS 70 disc, for the $^{12}$CO J=3$\rightarrow$2 line \citep{Keppler19_pds70,Facchini21} and for other molecular lines like HCO$^{+}$ J=4$\rightarrow$3 \citep[see Figure 5 in][]{Facchini21}. It is also highly suggestive in the $^{12}$CO, $^{13}$CO and C$^{18}$O observations of HD 163296 \citep[e.g.,][]{Rosenfeld13,Isella16}. Note that the HD 163296 and PDS 70 discs have inclinations of about 42$\degr$ \citep{Isella16} and 52$\degr$ \citep{Keppler19_pds70}, respectively.

\begin{figure*}
\centering
\resizebox{\hsize}{!}
{
\includegraphics{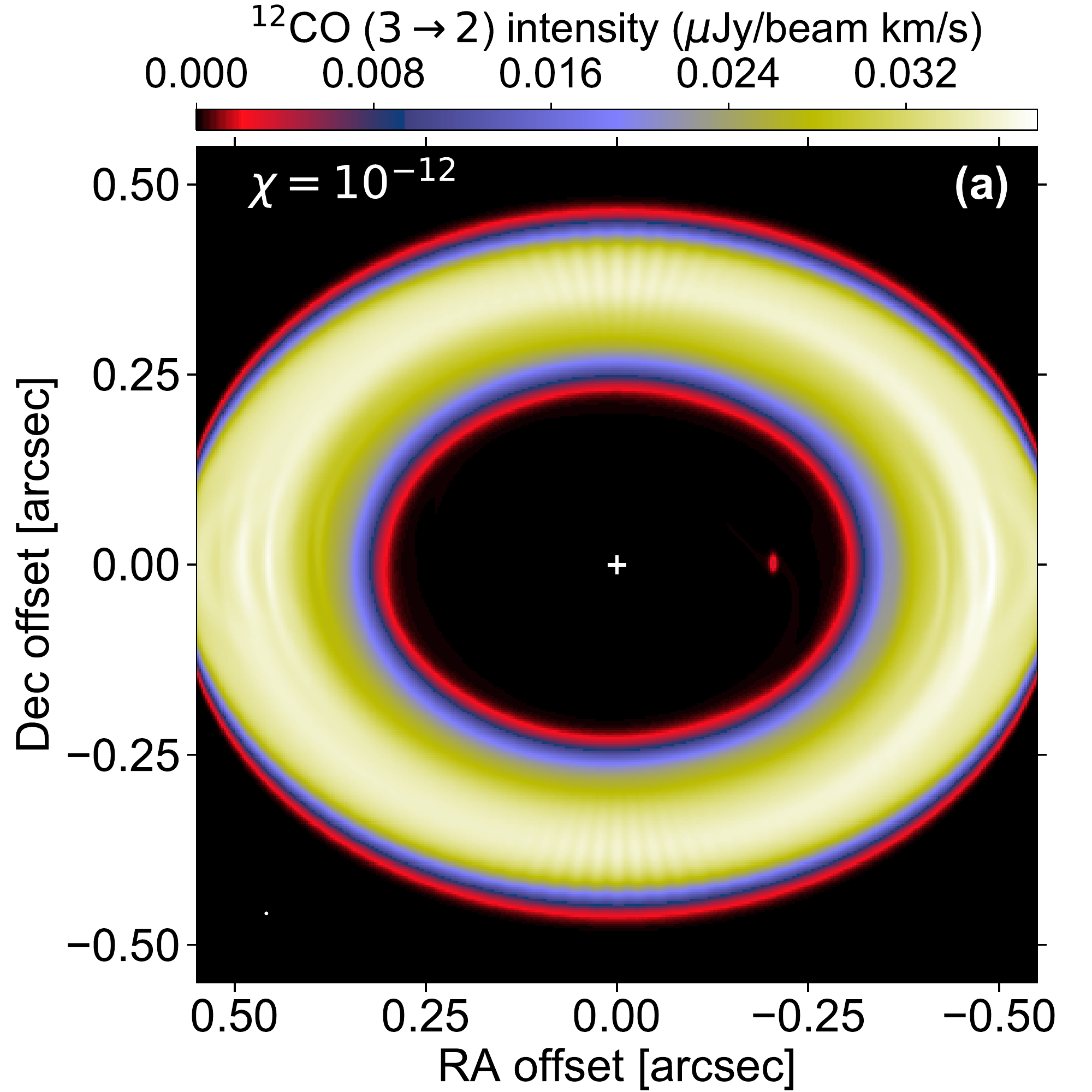}
\includegraphics{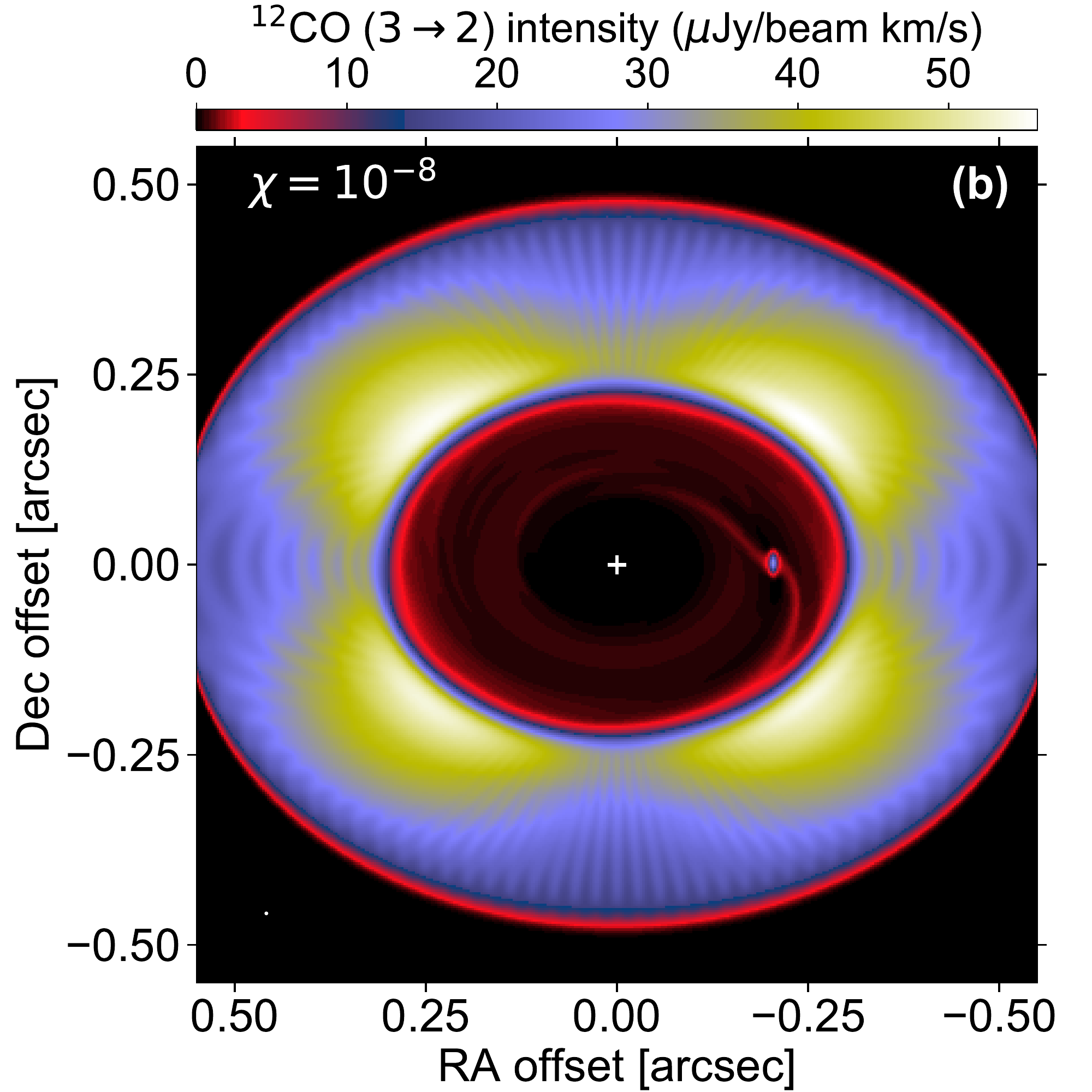}
\includegraphics{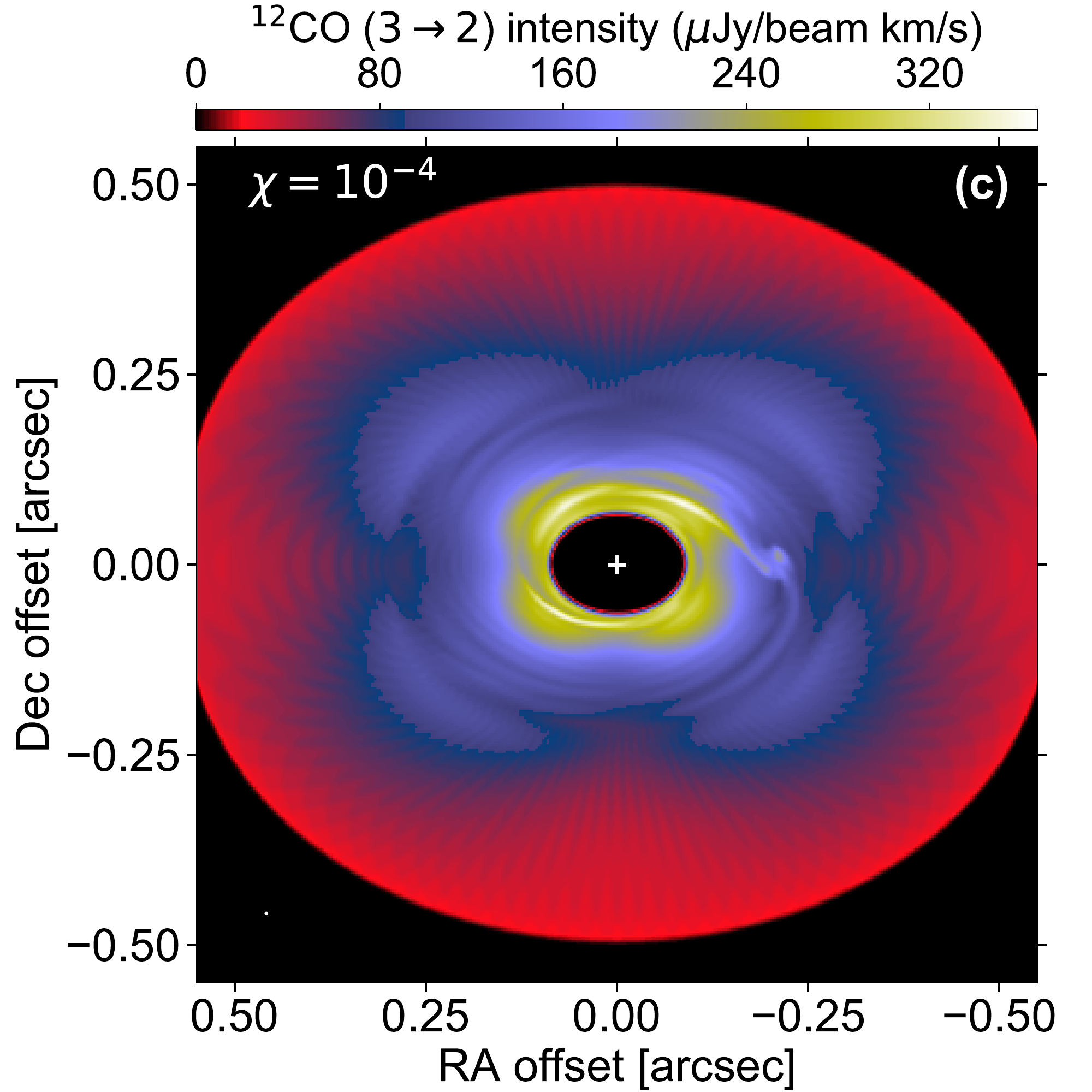}
}
\resizebox{\hsize}{!}
{
\includegraphics{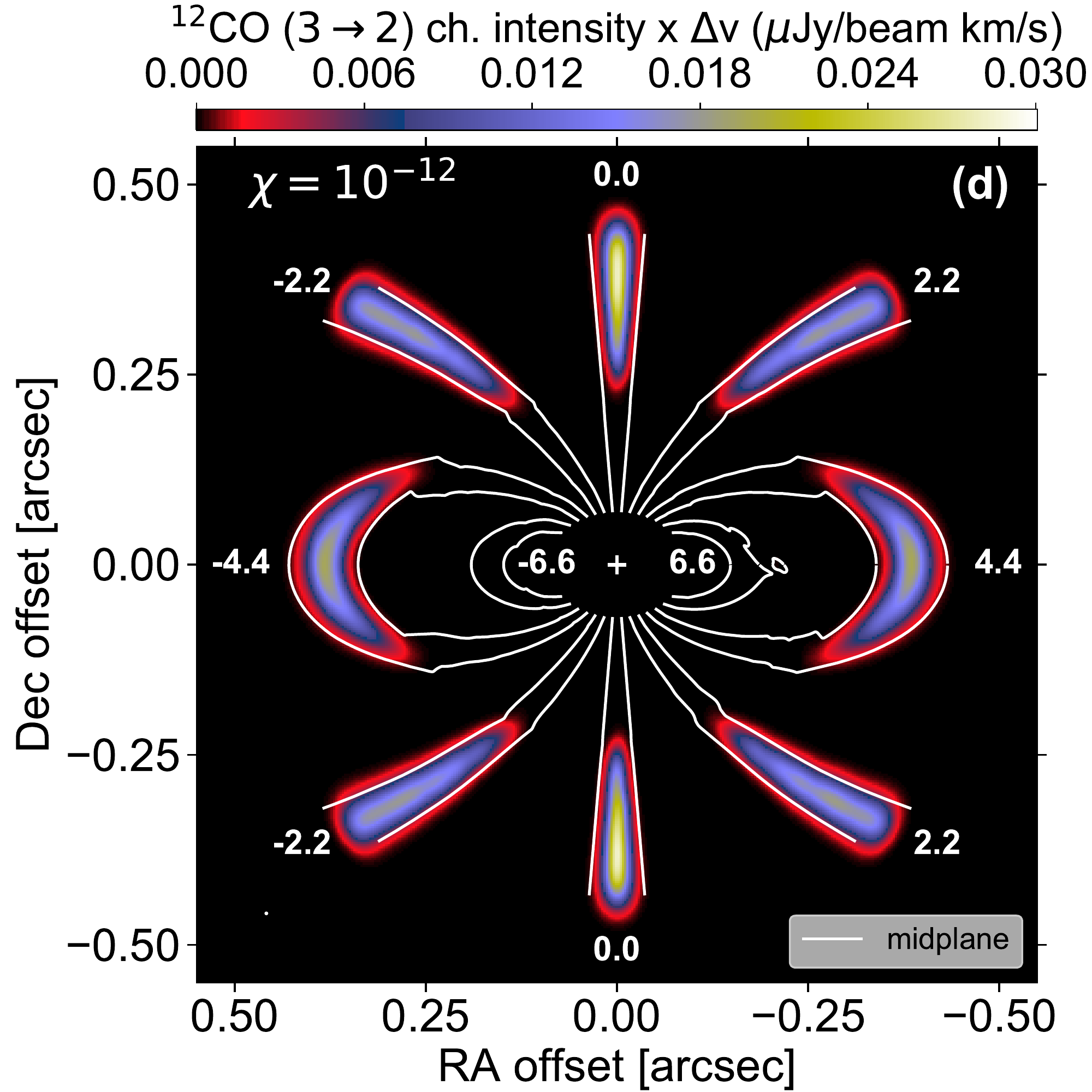}
\includegraphics{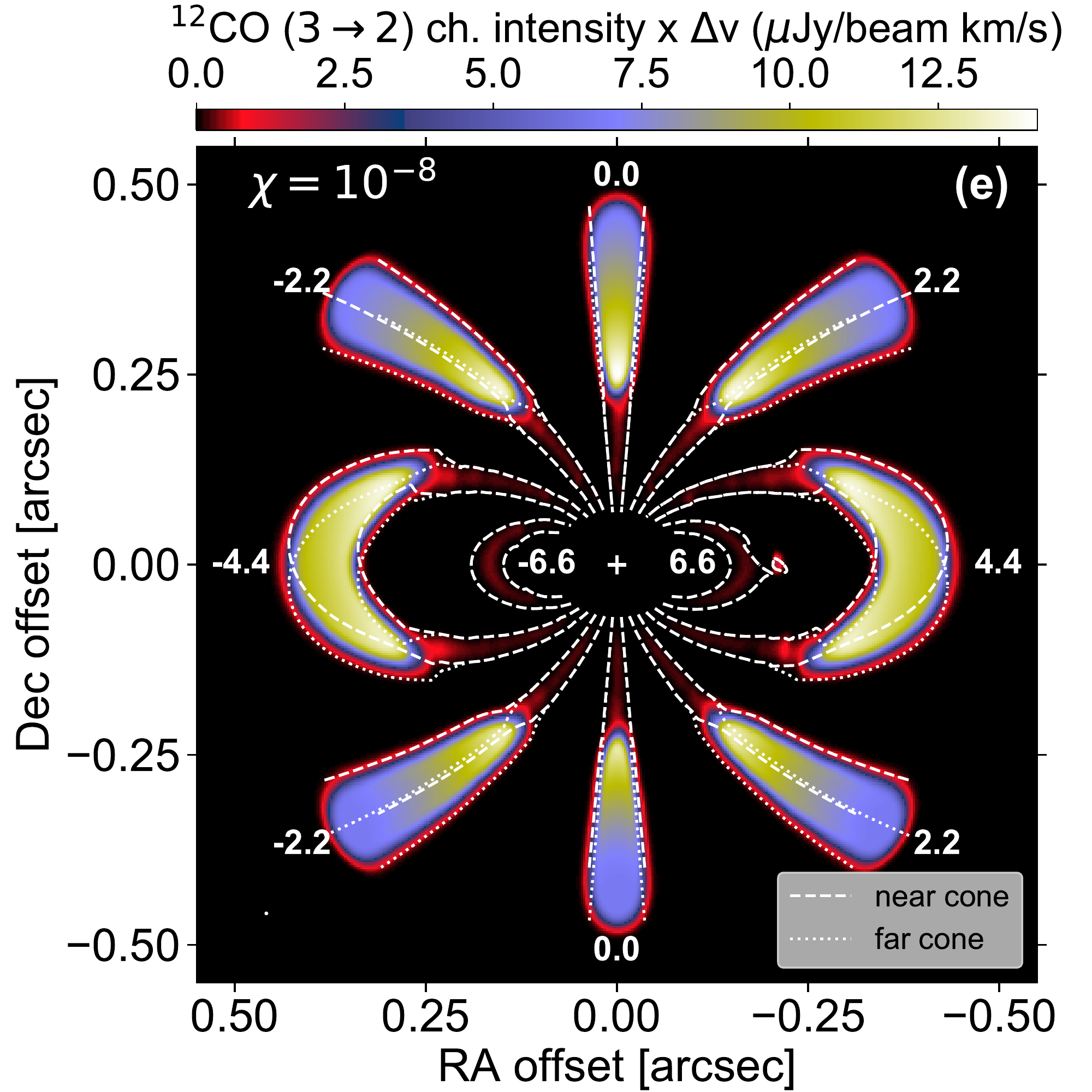}
\includegraphics{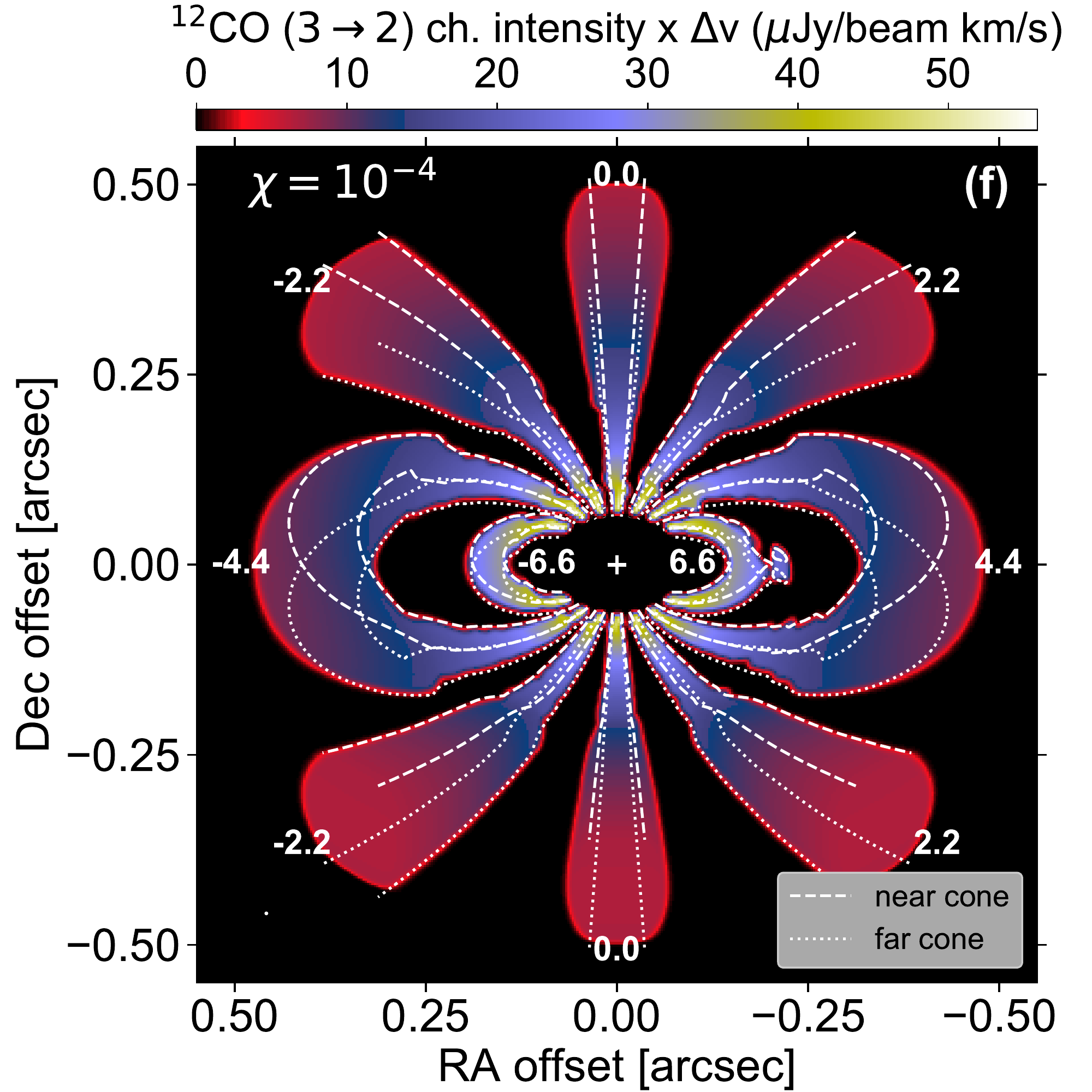}
}
\caption{{\it Top}: $^{12}$CO J=3$\rightarrow$2 integrated intensity maps for different values of the $^{12}$CO-to-H$_2$ number density ratio ($\chi$, see top-left corner in each panel). The planet is on a near-circular orbit inside the cavity, the disc inclination is 40$\degr$. {\it Bottom}: superposition of 7 channel maps of the $^{12}$CO J=3$\rightarrow$2 line emission, where the beam-convolved intensity in each channel is multiplied by the channel width $\Delta v$ in km s$^{-1}$. The numbers in the panels indicate the line centroids in km s$^{-1}$ (relative to systemic velocity). Iso-velocity contours are overplotted by solid, dashed or dotted curves to highlight where the line emission comes from (see legend in the bottom-right corner). While the line emission is confined to the disc midplane for $\chi=10^{-12}$ (left panels), it extends to the surface of a double cone for $\chi=10^{-8}$ and $10^{-4}$ (middle and right panels; see text for the vertical extent of the cone). As in previous figures, the white circle in the bottom-left corner of each panel shows the beam size (5 mas) and the white cross marks the star position.}
\label{fig:fig5}
\end{figure*}

The four-lobed pattern is due to the fact that the gas emission originates from a disc region that has a finite vertical thickness on both sides of the midplane \citep{Rosenfeld13,Keppler19_pds70}. To illustrate this in our disc model, we display in Fig.~\ref{fig:fig5} results of $^{12}$CO J=3$\rightarrow$2 radiative transfer calculations that differ by the $^{12}$CO-to-H$_2$ number density ratio ($\chi$). For these calculations, the disc inclination has been increased to 40$\degr$ so that the four-lobed pattern manifests more clearly. Results are shown at the same time as in the left row of panels in Figs.~\ref{fig:fig2} to~\ref{fig:fig4}, i.e. when the planet is still on a near circular orbit in the cavity. 

The upper panels in Fig.~\ref{fig:fig5} show integrated intensity maps after convolution with a small 5 mas beam. The lower panels display a superposition of 7 channel maps with line centroids $v_{\circ}$ spaced by 2.2 km s$^{-1}$. More specifically, it is the beam-convolved intensity in each channel multiplied by the channel width that is shown in each panel. The white curves show contours of constant line-of-sight velocities which delimit the disc region that most contributes to the emission in each channel map. When the gas is on nearly circular orbits, which is the case in our disc model so long as the planet retains a small eccentricity, the line-of-sight velocity is $v_{\varphi}\cos\varphi \sin i$, with $v_{\varphi}$ the azimuthal component of the gas velocity. Since our disc model features a gas cavity with a rather strong density gradient across the cavity, and because the planet also perturbs the gas velocity, instead of adopting the Keplerian velocity for $v_{\varphi}$ we set it equal to the (2D) gas azimuthal velocity in our hydrodynamical simulation. Like in \citet{Rosenfeld13}, we further assume that the gas emitting region is located within a double cone that forms an angle $\psi$ with respect to the disc midplane. A parcel of gas that is located at cylindrical radius $R$, azimuthal angle $\varphi$ and altitude $z = \pm R\tan\psi$ with respect to the disc midplane is thus located in the image plane at a right ascension offset\footnote{We adopt the convention that right ascension increases leftward in our synthetic images; i.e., north is up and east is to the left.} $-(R/d)\cos\varphi$ and at a declination offset $(R/d)\sin\varphi\cos i \pm (R/d)\tan\psi\sin i$, with $d$ the disc distance. This implies that the extent of the channel maps in the image plane depends not only on the thermal broadening of the line ($\delta v_{\rm th}$) but also on $\psi$.

When $\chi$ is (artificially) as small as $10^{-12}$, the integrated intensity takes the form of an axisymmetric annulus of emission outside the cavity (panel a). This is due to the optical depth being very small even outside the cavity ($\tau \in [5-10]\times10^{-4}$) so that everywhere in the disc the emission remains confined to the midplane. We thus have $\psi = 0$, and the spatial extent of the channel maps in the image plane depends solely on the thermal line width. The solid curves overplotted in panel (d), which show where in the image plane $v_{\varphi}\cos\varphi \sin i = v_{\circ} \pm 2\delta v_{\rm th}$, match very well the spatial extent of the channels maps. Note that the slight discontinuities in the iso-velocity contours are due to the radial variation of $v_{\varphi}$ across the cavity. The twists in the velocity contours for $v_{\circ}= 6.6$ km s$^{-1}$ are due to the planet and its inner wake.

By increasing $\chi$ to $10^{-8}$, the four-lobed pattern becomes visible in the integrated intensity map (panel b) and the channel maps are now more extended in the image plane (panel e). Since $\tau$ is increased by four orders of magnitude compared to $\chi = 10^{-12}$, the line emission is now moderately optically thick outside the cavity ($\tau \in [5-10]$) so that the emission is no longer confined to the disc midplane. Emission inside the cavity is visible in panels (b) and (e) but remains small due to small values of the optical depth there ($\tau \la 10^{-2}$, except in the planet vicinity). By letting $\tan\psi$ equal to 0 inside the cavity, but increasing it to $2h$ outside the cavity, we find that the iso-velocity contours $v_{\varphi}\cos\varphi \sin i = v_{\circ} \pm 2\delta v_{\rm th}$ reproduce again very well the extent of the channel maps in the image plane. More precisely, $\tan\psi$ is increased smoothly from 0 to $2h$ for $R \in [25-35]$ au. This means that the gas emission inside the cavity is still confined to the disc midplane, whereas outside the cavity it arises from a double cone of vertical extent $2H$ on both sides of the disc midplane. The dashed curves in panel (e) are for $z = R\tan\psi$ and thus represent the near side of the disc in the image plane (which we refer to as the "near cone"). The dotted curves are for $z = -R\tan\psi$ and thus represent the far side of the disc ("far cone"). Note that the dashed and dotted curves overlap inside the cavity since $\psi = 0$ there. 

When $\chi = 10^{-4}$ (our fiducial value for $^{12}$CO), not only is the four-lobed pattern clearly visible outside the cavity, it is also discernible inside of it, although to a lesser extent (panel c). This is due to the emission being optically thick everywhere in the disc, including inside the cavity. The spatial extent of the channel maps in the image plane is increased again due to a larger $\psi$, and is well reproduced by the iso-velocity contours $v_{\varphi}\cos\varphi \sin i = v_{\circ} \pm 2\delta v_{\rm th}$ with $\tan\psi$ equal to $2h$ inside the cavity and $4h$ outside (and a smooth transition in between like for $\chi=10^{-8}$). The gas emission is thus contained in a double cone whose vertical extent increases from $2H$ to $4H$ across the cavity.

\begin{figure}
\centering
\includegraphics[width=0.9\hsize]{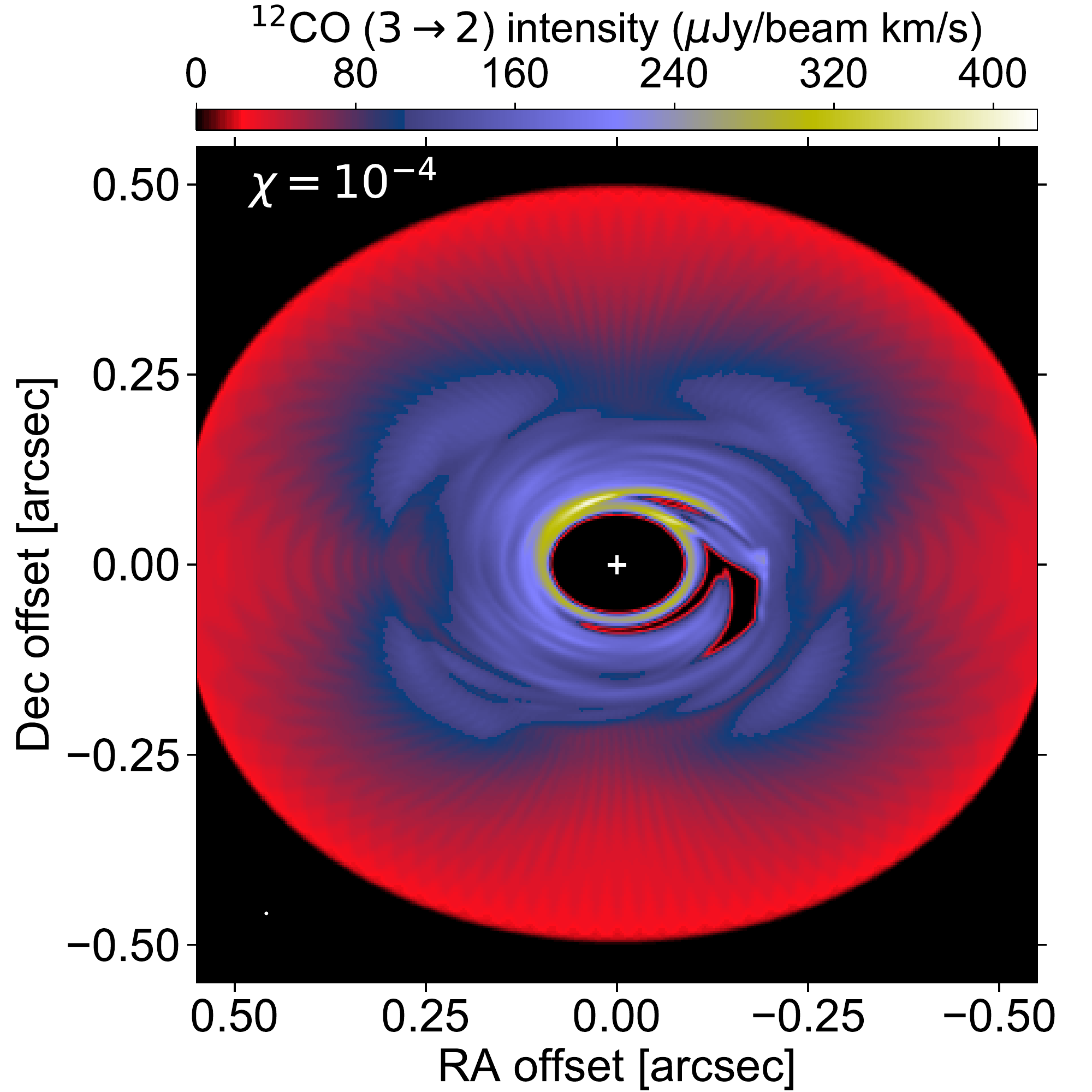}
\includegraphics[width=0.9\hsize]{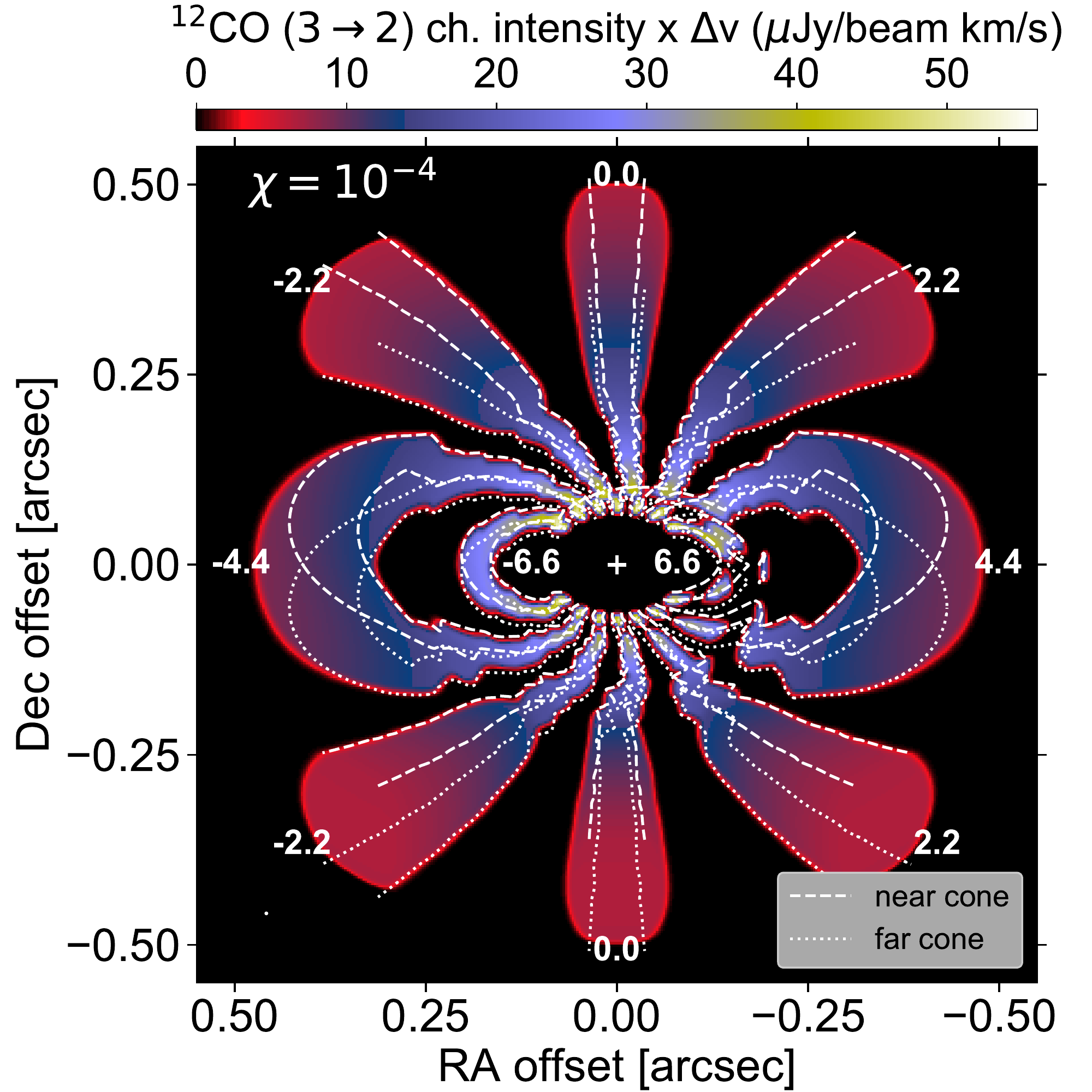}
\caption{Same as Fig.~\ref{fig:fig5}, but for $\chi=10^{-4}$ and when the planet is eccentric, near the apocentre of its orbit inside the cavity.}
\label{fig:fig6}
\end{figure}
\begin{figure*}
\centering
\resizebox{\hsize}{!}
{
\includegraphics{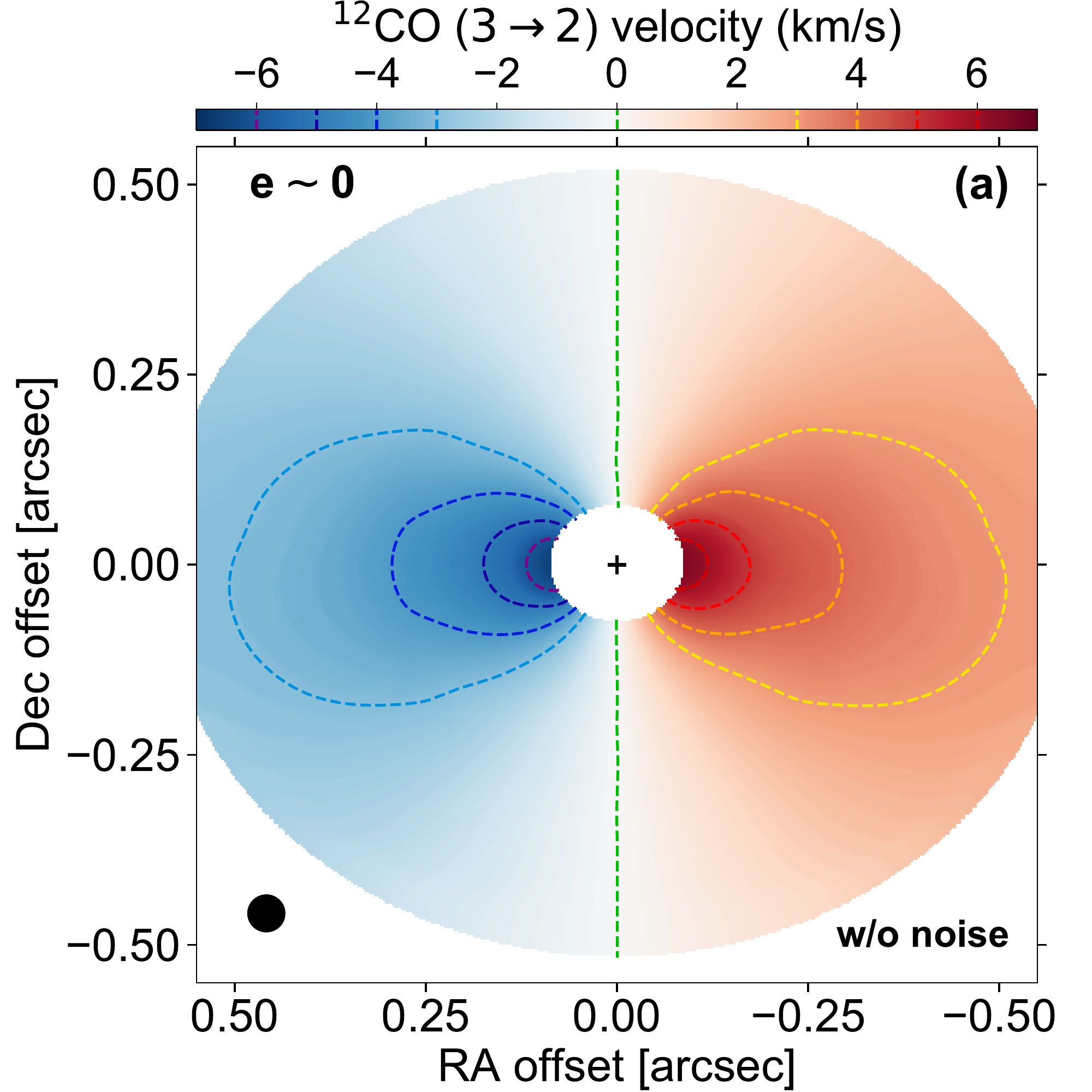}
\includegraphics{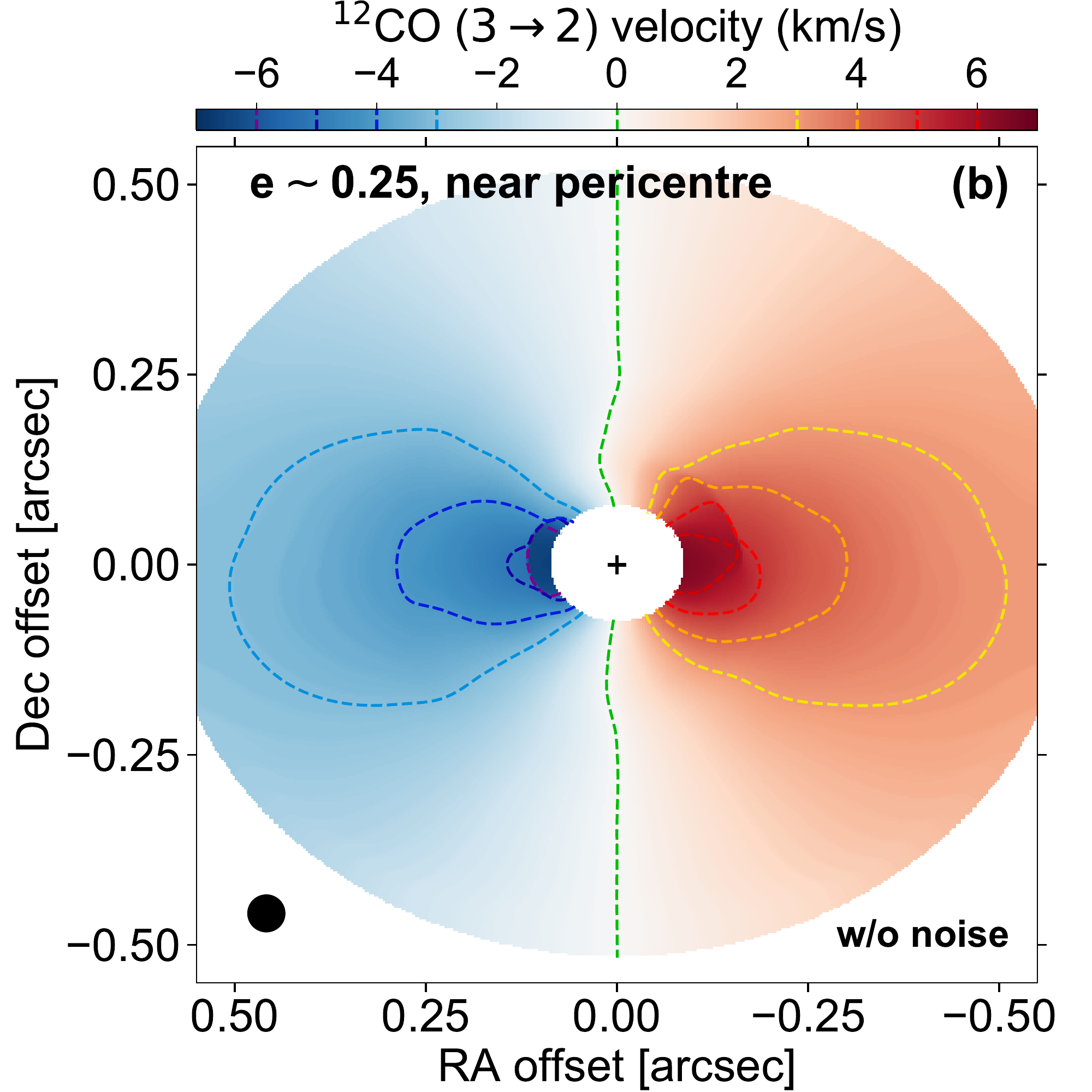}
\includegraphics{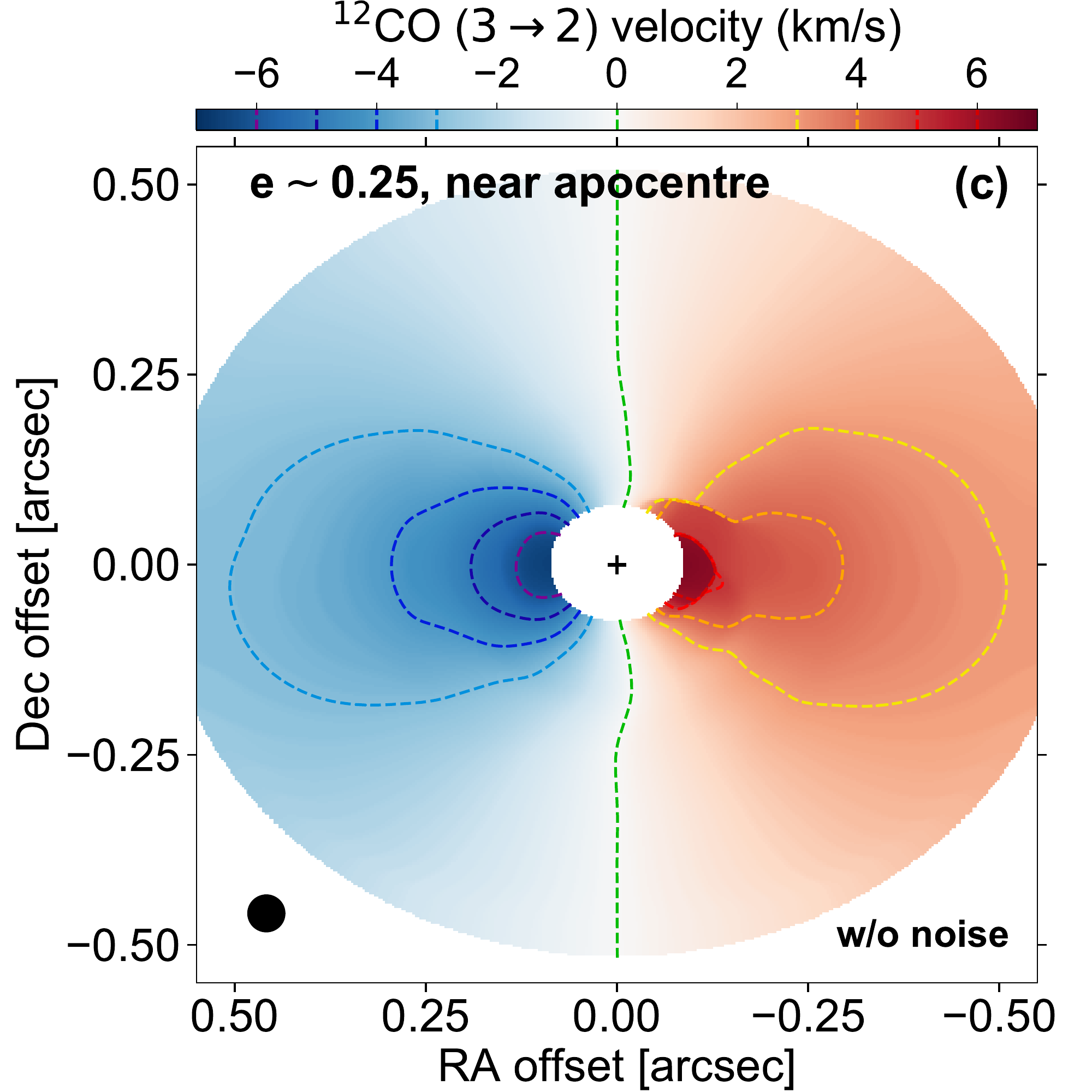}
}
\resizebox{\hsize}{!}
{
\includegraphics{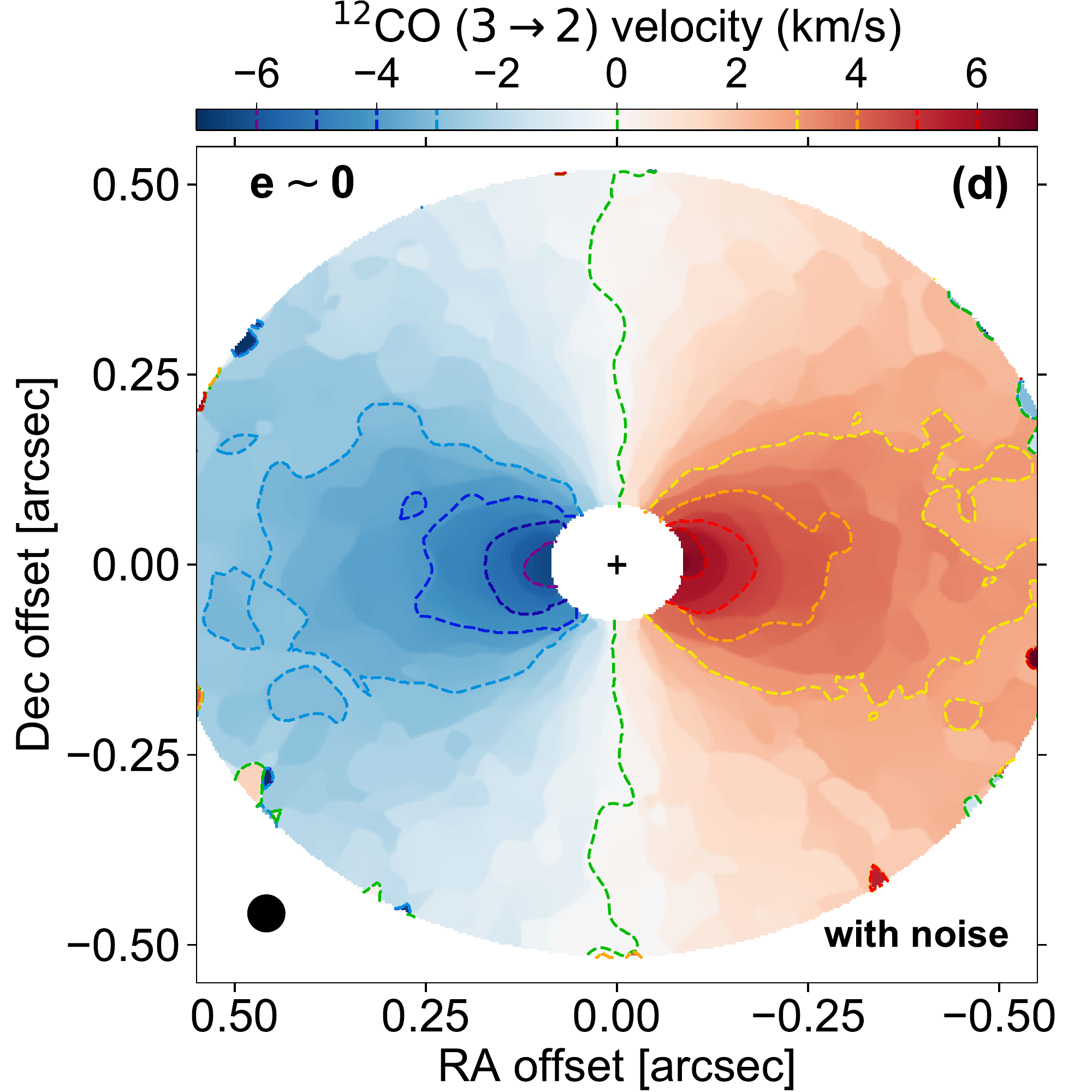}
\includegraphics{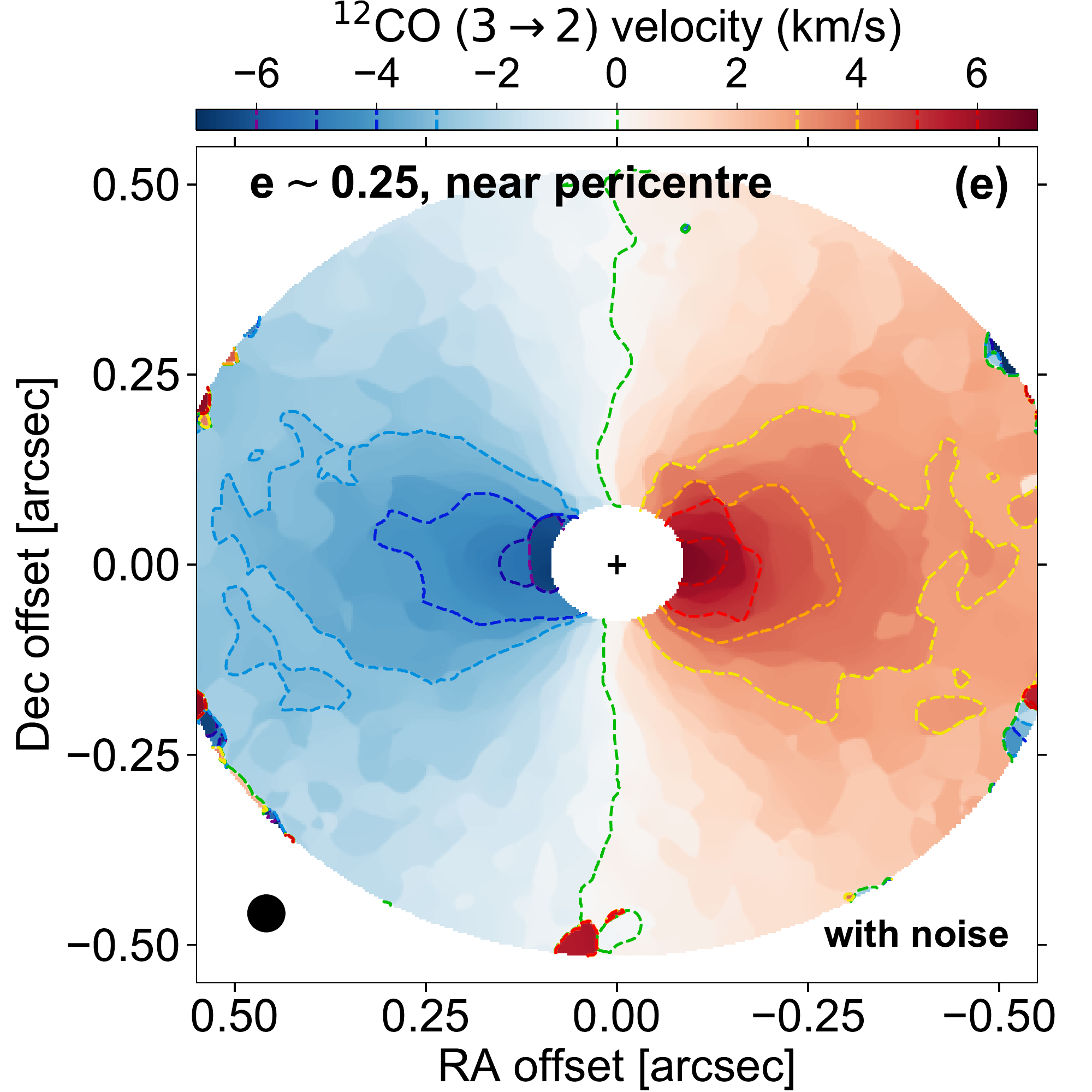}
\includegraphics{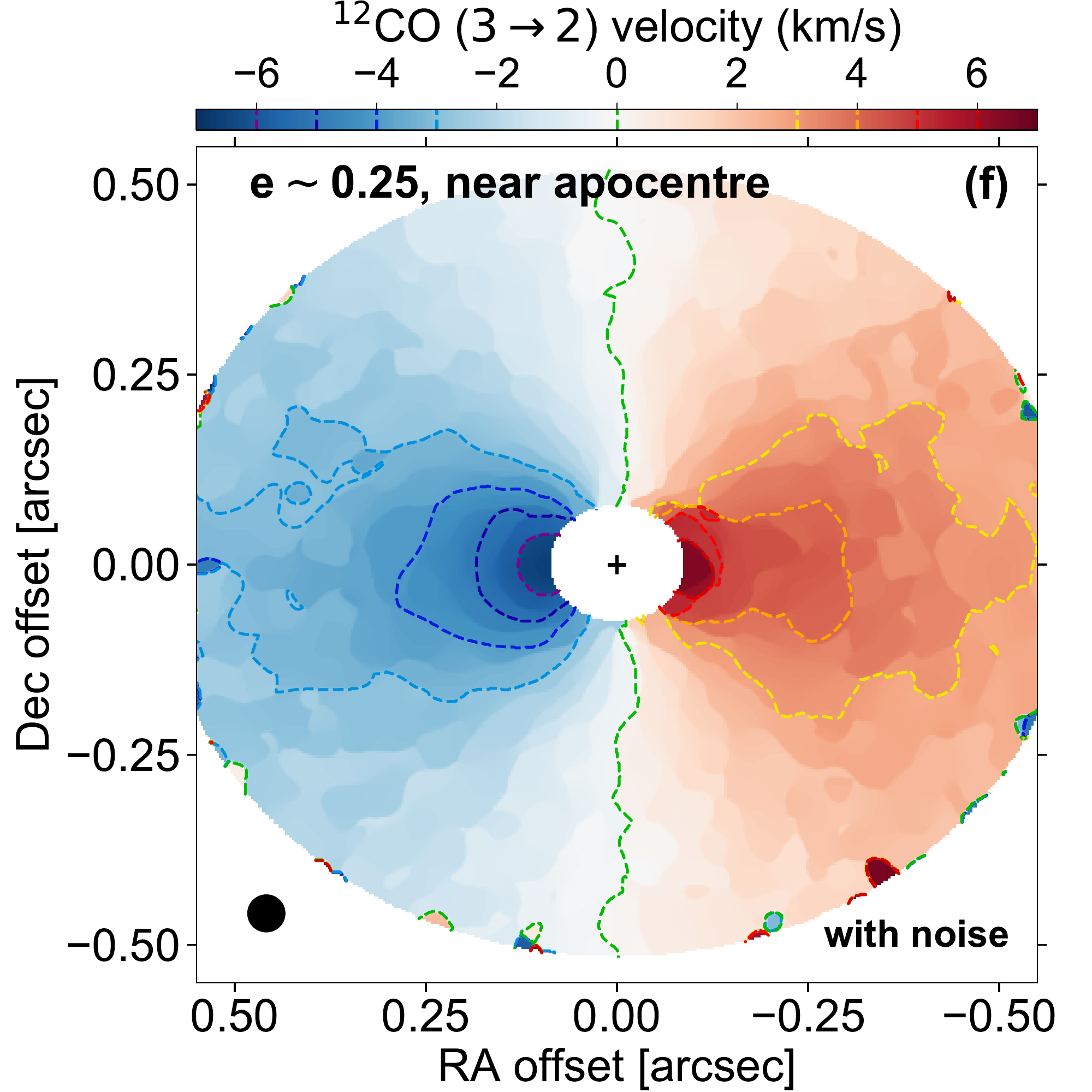}
}
\caption{$^{12}$CO J=3$\rightarrow$2 velocity maps obtained with the quadratic method of \citet{Teague18_bm}. Several iso-velocity contours are overplotted as dashed curves, their level is shown in the color bar on top of the panels. White noise in the channels is included in panels (d) to (f). The black circle in the bottom-left corner of each panel shows the beam size and the black cross marks the star position.}
\label{fig:fig7}
\end{figure*}

From the lower panels in Fig.~\ref{fig:fig5} we make the following comments:
\begin{itemize}
\item[\sbt] The vertical extent of the double cone agrees well with the expected location of the $\tau \sim 1$ surface. More specifically, we have checked that the cone's vertical extent agrees well with the altitude above the disc midplane where the $^{12}$CO column density, integrated toward the midplane, reaches about $10^{15}$ cm$^{-2}$ which, according to \href{http://var.sron.nl/radex/radex.php}{\texttt{RADEX}} radiative transfer calculations in an isothermal homogeneous medium \citep{VanderTak07}, is the typical column density from which the optical depth at the centre of the $^{12}$CO J=3$\rightarrow$2 line becomes $\ga 1$ for line widths and gas temperatures representative of our disc model.
\smallskip
\item[\sbt] For $\chi=10^{-12}$, the intensity distribution varies from one channel to another; for instance, the peak intensity is $\approx 1.5$ larger for $v_{\circ} = 0$ than for $v_{\circ} = \pm 2.2$ km s$^{-1}$. This reflects how the optical depth itself varies from one channel to another. In contrast, for $\chi = 10^{-8}$ and $10^{-4}$ the intensity distribution outside the cavity is very similar across the channels and depends mainly on the deprojected distance from the star. We attribute this behaviour to the line emission being very optical thick outside the cavity, since in this case the intensity depends essentially on the gas temperature. When the line emission is optically thick, the smaller spatial extent of the channel maps along the disc minor and major axes, which is due to a larger overlapping of the iso-velocity curves along both axes, leads to a smaller integrated intensity and thus naturally accounts for the four-lobed pattern (see also \citealp{Keppler19_pds70}, their \S~3.2). The smaller spatial extent of the channel maps is particularly visible when comparing the extent of the iso-velocity contours for $v_{\circ} =$ 0, 2.2 and 4.4 km s$^{-1}$ between panels (d) and (e) in Fig.~\ref{fig:fig5}.
\smallskip
\item[\sbt] For a given channel map, we see, perhaps more clearly for $\chi=10^{-4}$, that the intensity varies nearly uniformly across the iso-velocity contours of the near and far cones. This shows that the line emission actually comes from the entire cone, and is not localized near its surface (where $\tau \sim 1$). The existence of a vertical temperature gradient, or of an absorbing dust layer around the midplane, could change this picture.
\smallskip
\item[\sbt] The above analysis can be extended to the case where the planet is eccentric. Since the gas inside the cavity also gets eccentric, the line-of-sight velocity is $(v_{\varphi}\cos\varphi + v_{\rm r}\sin\varphi) \sin i$, with $v_{\rm r}$ the radial component of the gas velocity. Fig.~\ref{fig:fig6} complements Fig.~\ref{fig:fig5} by showing the case where $e \sim 0.25$ and the planet is near apocentre for $\chi = 10^{-4}$. The upper panel of Fig.~\ref{fig:fig6} is very similar to panel (i) of Fig.~\ref{fig:fig3}, the only difference being that the disc inclination is increased from 30$\degr$ to 40$\degr$. In the lower panel of Fig.~\ref{fig:fig6}, iso-velocity contours now show where, in the image plane, $(v_{\varphi}\cos\varphi + v_{\rm r}\sin\varphi) \sin i = v_{\circ} \pm 2\delta v_{\rm th}$. These contours reproduce well the twisted shape of the channel maps inside the cavity. We will come back to this twisted pattern when analysing the corresponding velocity map in Section~\ref{sec:res_12CO_mom1}. We point out that for $v_{\circ} = 4.4$ and 6.6 km s$^{-1}$ the channel maps show no to very little emission to the south-east inside the cavity, which is due to photodissociation, as already seen in Section~\ref{sec:res_12CO_mom0}.
\end{itemize}

Some additional material on the appearance of the four-lobed pattern with varying disc inclination is presented in Appendix~\ref{sec:res_12CO_mom0_incl}.

\subsubsection{Velocity maps}
\label{sec:res_12CO_mom1}
In this section, we examine how the planet eccentricity impacts the velocity map of the  $^{12}$CO J=3$\rightarrow$2 line emission. Like in Section~\ref{sec:res_12CO_mom0}, the disc inclination is 30$\degr$ and the beam FWHM 50 mas. We display in Fig.~\ref{fig:fig7} velocity maps obtained with \href{https://bettermoments.readthedocs.io/}{\texttt{bettermoments}} by applying the quadratic method of \citet{Teague18_bm} to the beam-convolved channel intensities. Results are shown from left to right at the same times as in Figs.~\ref{fig:fig2} to~\ref{fig:fig4}. The upper panels show the velocity maps without including noise in the channels. In the lower panels, white noise with standard deviation $\sigma = 1$~mJy/beam was added to the channel intensities prior to beam convolution (as in the integrated intensity maps of Fig.~\ref{fig:fig4}). The quadratic method used to build the velocity maps did not need clipping of the data. However, a mask has been applied to the velocity maps for deprojected distances $\la 0\farcs09$ and $\ga 0\farcs60$ to remove the noise emission beyond the radial edges of our disc model. We actually see, in the lower panels of Fig.~\ref{fig:fig7}, a few patches of residual noise emission near the disc's outer edge. For a better comparison, the same masking procedure has been applied to the velocity maps without noise.

A few iso-velocity contours are superimposed as dashed curves in the panels. In panel (a), which is when the planet still has a near-circular orbit in the cavity, the contours are symmetric about the disc's minor axis, and the iso-velocity contour with zero velocity (green dashed curve) basically coincides with that axis. The contours are also approximately symmetric about the disc's major axis. We notice that the contours at $\pm 3$ km s$^{-1}$ are slightly bent towards the lower half of the image (i.e., towards negative declination offsets), and that this effect seems more prominent in the disc's outer parts (i.e., outside the cavity). With the help of panel (f) in Fig.~\ref{fig:fig5}, we see that this bending is directed towards the far cone of emission, that is towards the disc parts further from the observer, which is expected for inclined discs at high angular resolution \citep[see, e.g.,][]{DDC20}. Bending away from the disc's major axis is hardly visible for contours of higher absolute velocity, as they mostly trace emission inside the cavity which is located closer to the midplane. If angular resolution is high enough, a more pronounced bending of the velocity lobes in the outer parts of a disc could be a way to hint at the presence of a gas cavity in the disc's inner parts.

\begin{figure}
\centering
\includegraphics[width=0.9\hsize]{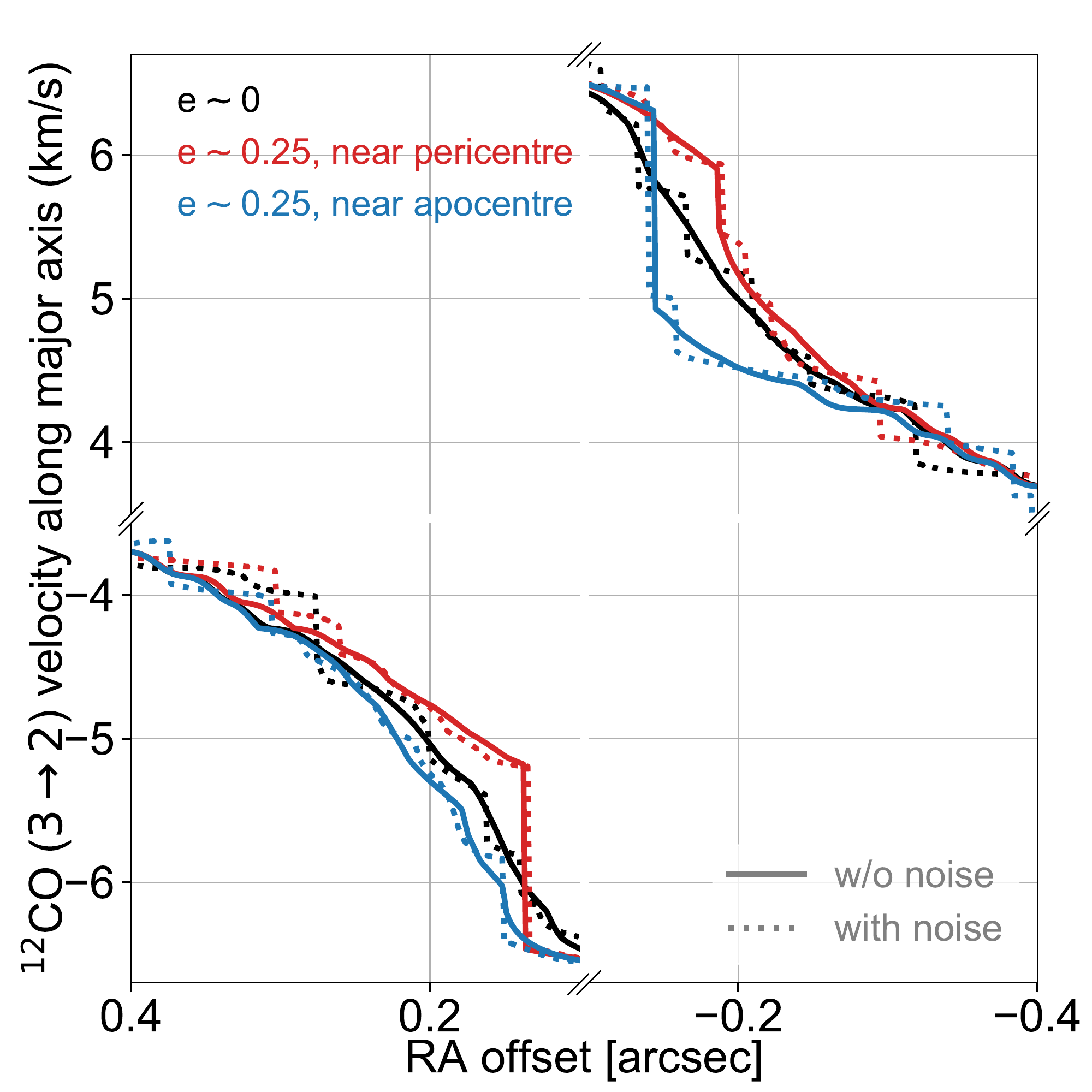}
\caption{Cut along the disc's major axis of the $^{12}$CO J=3$\rightarrow$2 velocity maps shown in Fig.~\ref{fig:fig7}. Both the x- and y-axes are broken to highlight the velocity differences inside the $\sim$ $0\farcs3$ cavity.}
\label{fig:fig8}
\end{figure}

By comparing panels (a) to (c) in Fig.~\ref{fig:fig7}, it is clear that the (near-)symmetry about the disc's minor and major axes is broken with an eccentric planet inside the cavity. When the planet becomes eccentric, the gas inside the cavity also becomes eccentric and precesses. The radial and azimuthal components of the gas velocity inside the cavity thus vary with the orbital phase of the planet: 
\begin{itemize}
\item[\sbt] When the planet is near pericentre, $v_{\varphi}$ and $v_{\rm r}$ are such that the line-of-sight velocity $(v_{\varphi}\cos\varphi + v_{\rm r}\sin\varphi) \sin i$ takes more positive values everywhere inside the cavity compared to the case where the planet is still on a near-circular orbit. This is illustrated by cuts of the velocity maps along the disc's major axis in Fig.~\ref{fig:fig8}. The red and black solid curves in that figure show velocity differences inside the cavity up to $\sim$ 0.5 km s$^{-1}$ (i.e., a $\sim$ 10\% relative difference). These differences explain why, in panel (b) of Fig.~\ref{fig:fig7}, contours of high negative velocity seem to narrow down while those of high positive velocity seem to stretch up. This effect can also be seen by the bending of the contour of zero line-of-sight velocity toward the blue-shifted (east) part of the velocity map within the cavity. Additional twisting of the contours, which is more pronounced in the red-shifted (west) part of the map, arises because of the proximity of the planet and its wakes. 
\smallskip
\item[\sbt] In contrast, when the planet is near apocentre, our simulation results show that $v_{\varphi}$ and $v_{\rm r}$ make the line-of-sight velocity more negative everywhere inside the cavity compared to the case where the planet is on a near-circular orbit. This can also be seen by comparing the black and blue solid curves in Fig.~\ref{fig:fig8}. This now explains why, in panel (c) of Fig.~\ref{fig:fig7}, contours of high negative velocity are more extended and those of high positive velocity shrunk. It also explains the twist of the channel maps in the lower panel of Fig.~\ref{fig:fig6}. 
\end{itemize}

One should bear in mind that the difference of line-of-sight velocities between the eccentric and the near-circular planet cases depends on the planet's orbital phase, and we have checked that its sign generally varies with azimuthal angle. The two orbital phases that we have adopted (near pericentre and near apocentre) can be seen as particular cases where this difference is either positive or negative in the whole cavity. 

Quite remarkably, the lower panels in Fig.~\ref{fig:fig7} show that the features seen above (narrowing, stretching, twisting of the iso-velocity contours inside the cavity, bending of the disc's minor axis) are still visible when noise is included in the channels. This is further illustrated by the dotted curves in Fig.~\ref{fig:fig8}. Velocity maps of the $^{12}$CO J=3$\rightarrow$2 line therefore constitute another useful means of tracking the presence of eccentric Jupiters in the cavities of protoplanetary discs. Interestingly, the aforementioned features have all been observed in high spatial resolution interferometric gas observations of protoplanetary discs; see, for instance, \citet{DDC20} for a recent review.

Before leaving this section, we briefly note that, without noise, moment 1 maps of the $^{12}$CO J=3$\rightarrow$2 line emission computed from beam-convolved channel intensities are overall very similar to the velocity maps obtained with the quadratic method of \citet{Teague18_bm}. The main difference is that the velocity variations in the cavity when the planet is eccentric are not as sharp in the moment 1 maps as in the upper panels of Fig.~\ref{fig:fig7} or in Fig.~\ref{fig:fig8}.

\subsection{$^{13}$CO and C$^{18}$O J=3$\rightarrow$2 lines}
\label{sec:res_13COC18O}

\begin{figure*}
\centering
\resizebox{0.96\hsize}{!}
{
\includegraphics{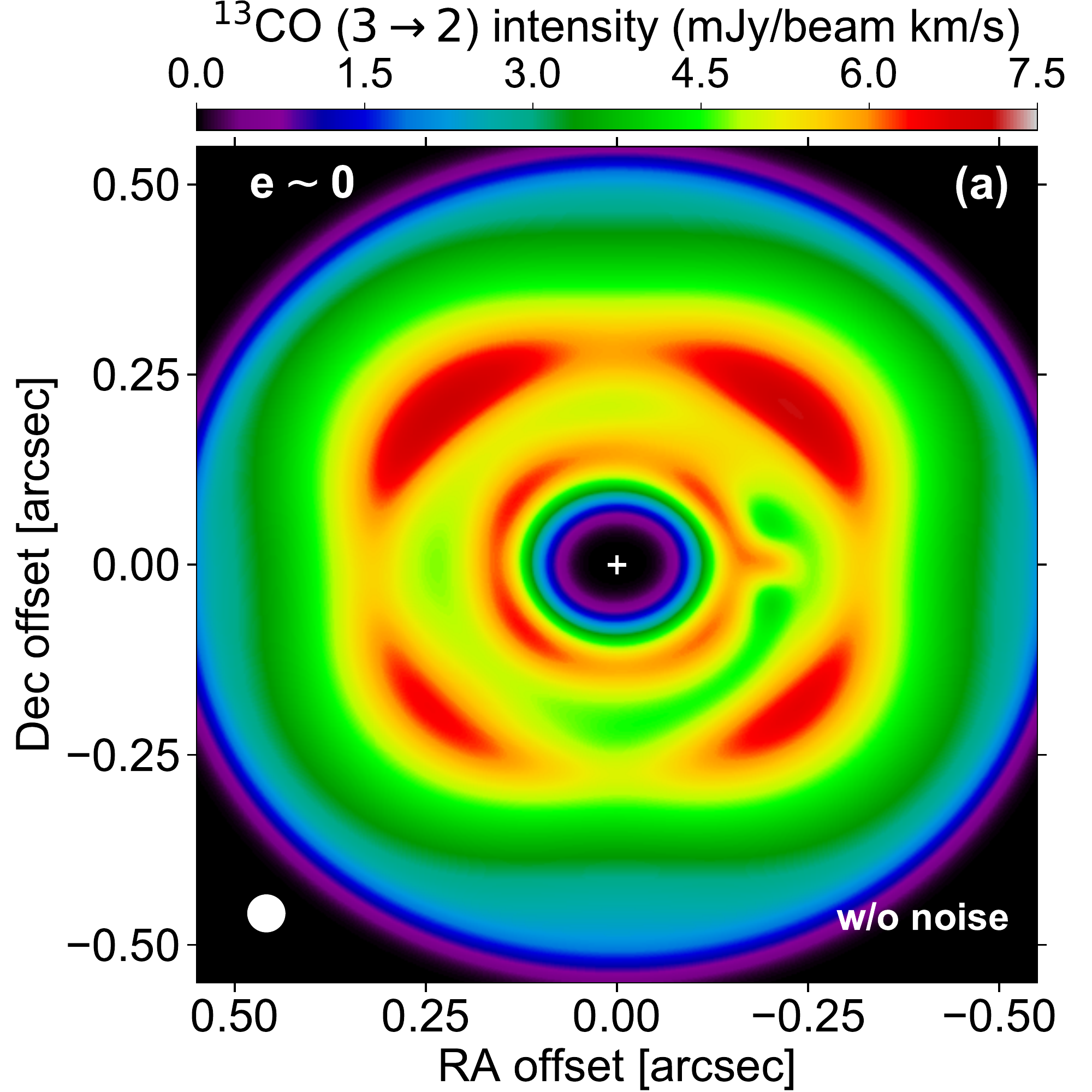}
\includegraphics{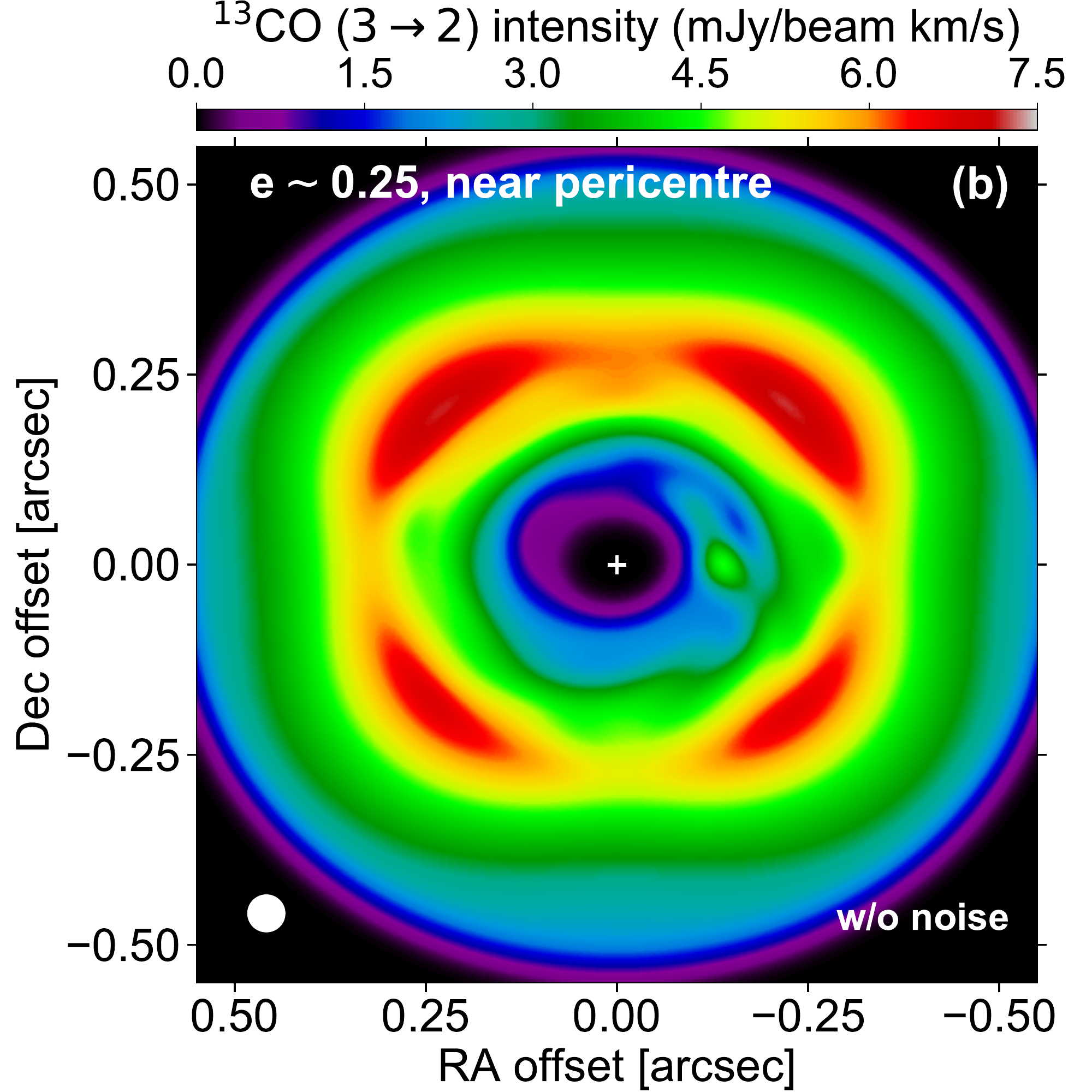}
\includegraphics{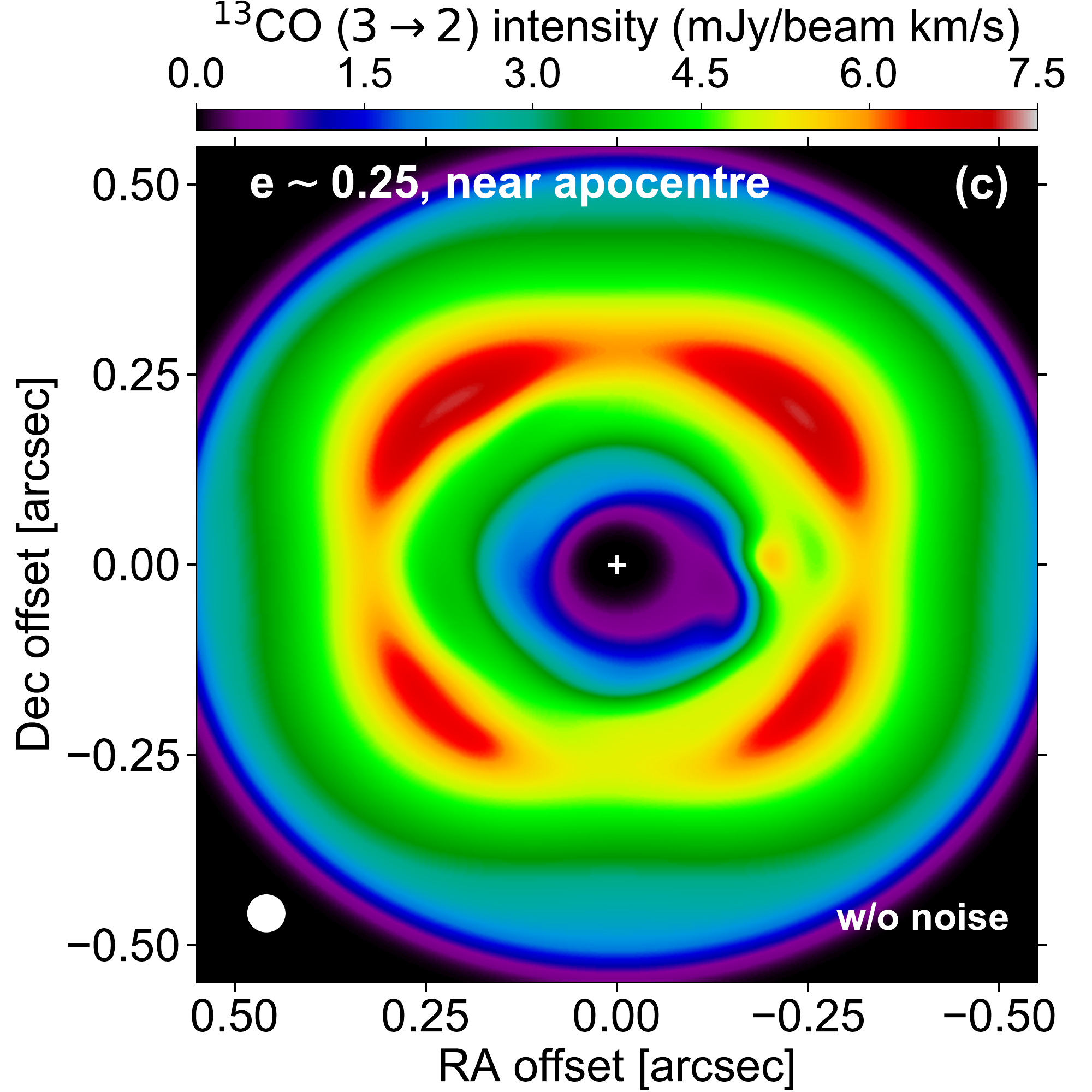}
}
\resizebox{0.96\hsize}{!}
{
\includegraphics{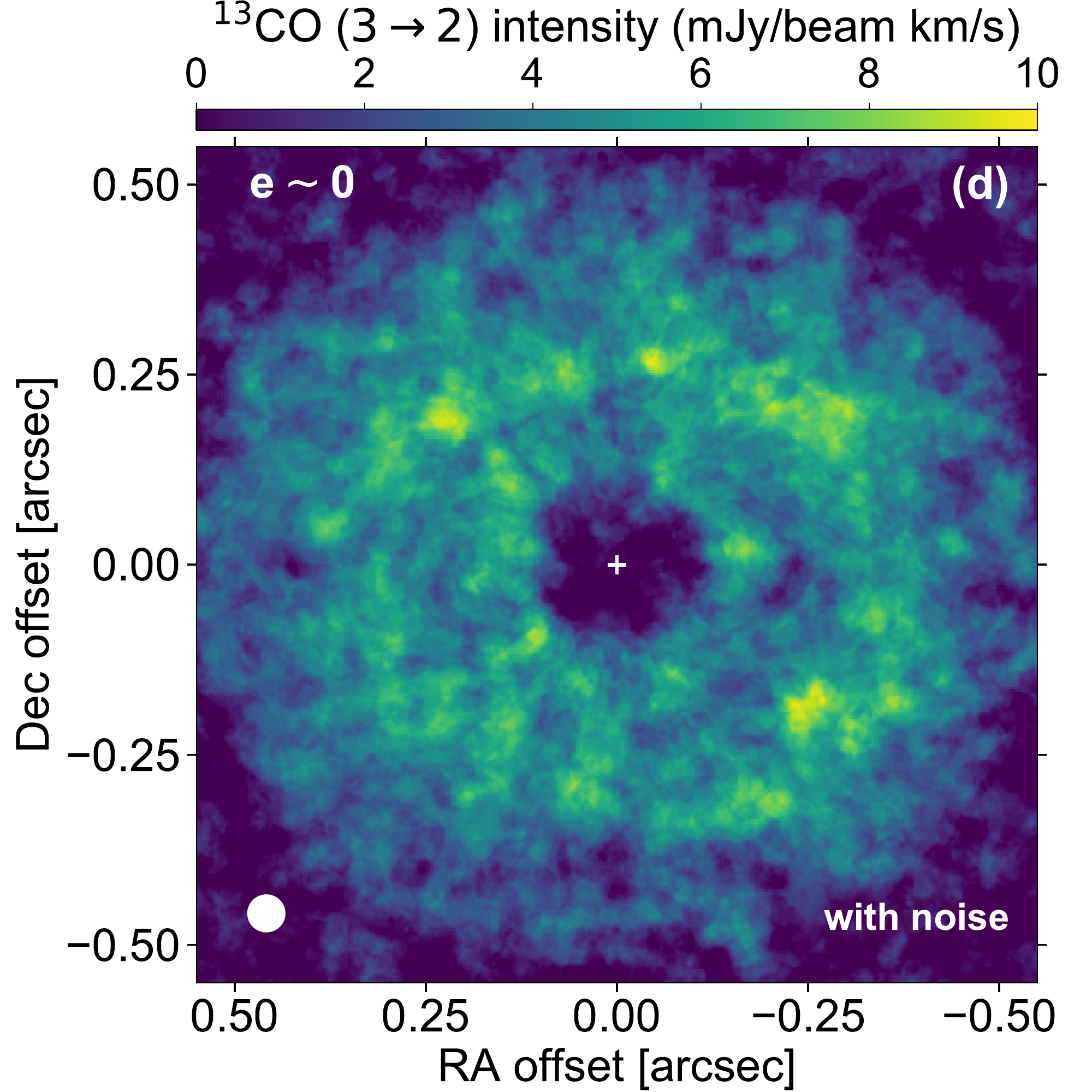}
\includegraphics{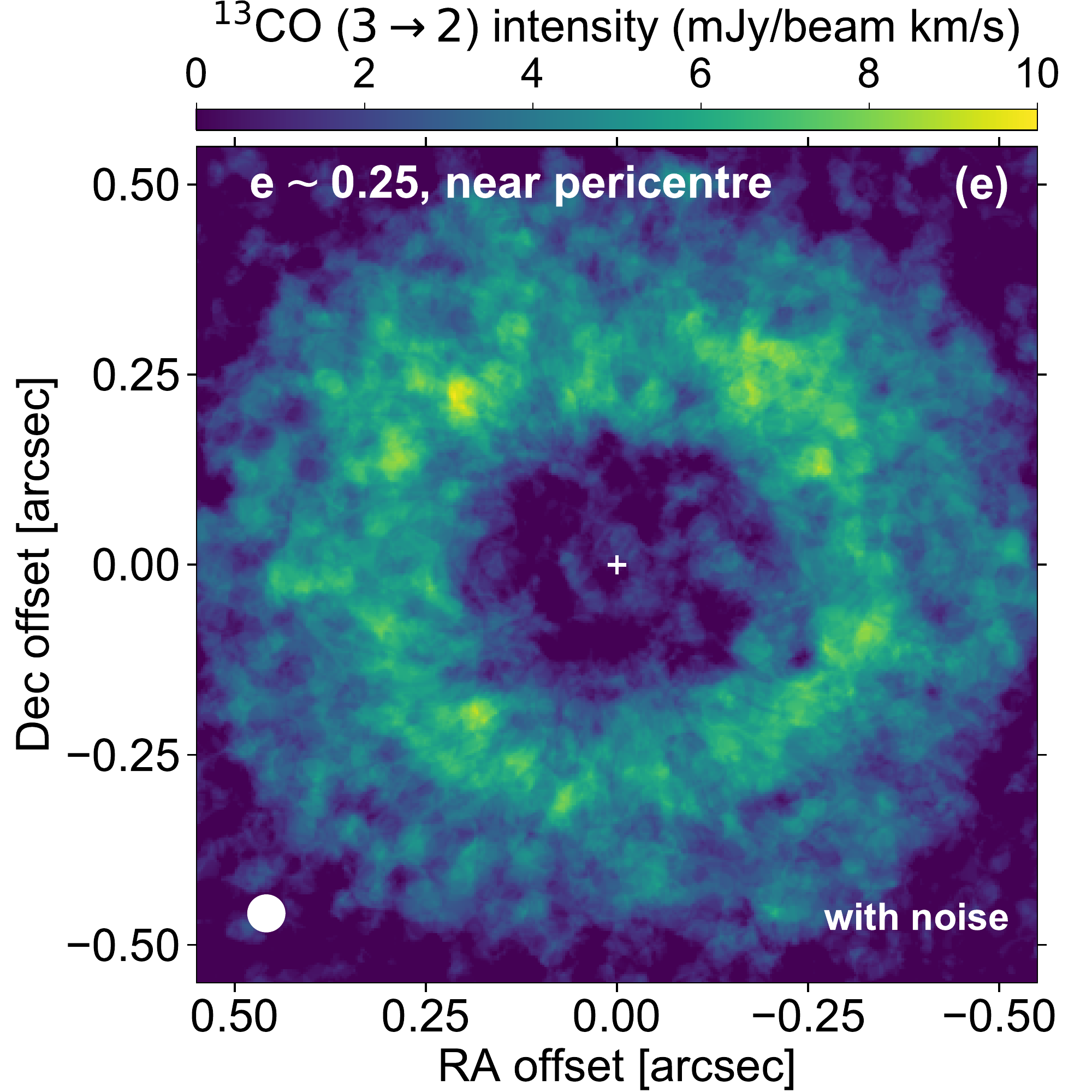}
\includegraphics{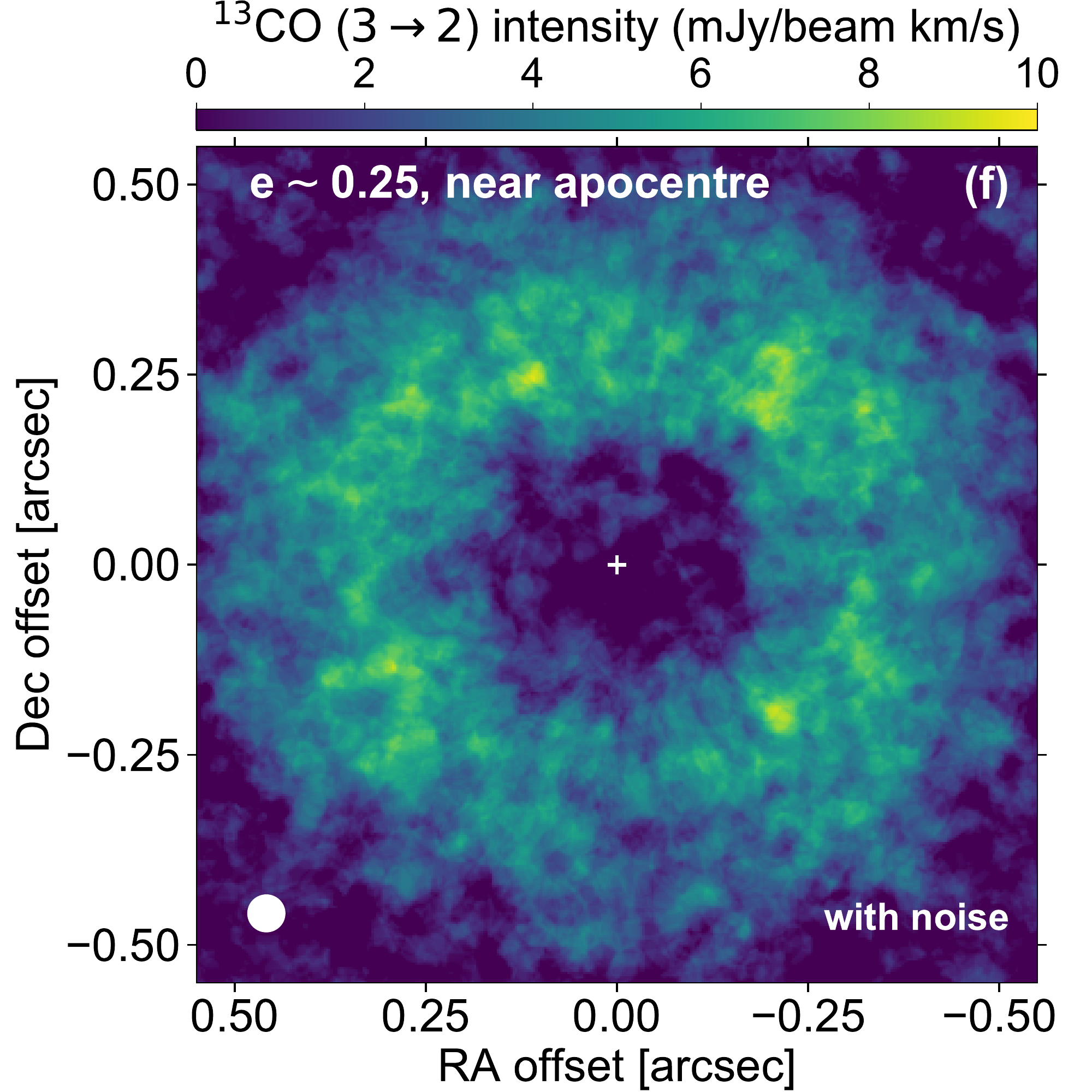}
}
\resizebox{0.96\hsize}{!}
{
\includegraphics{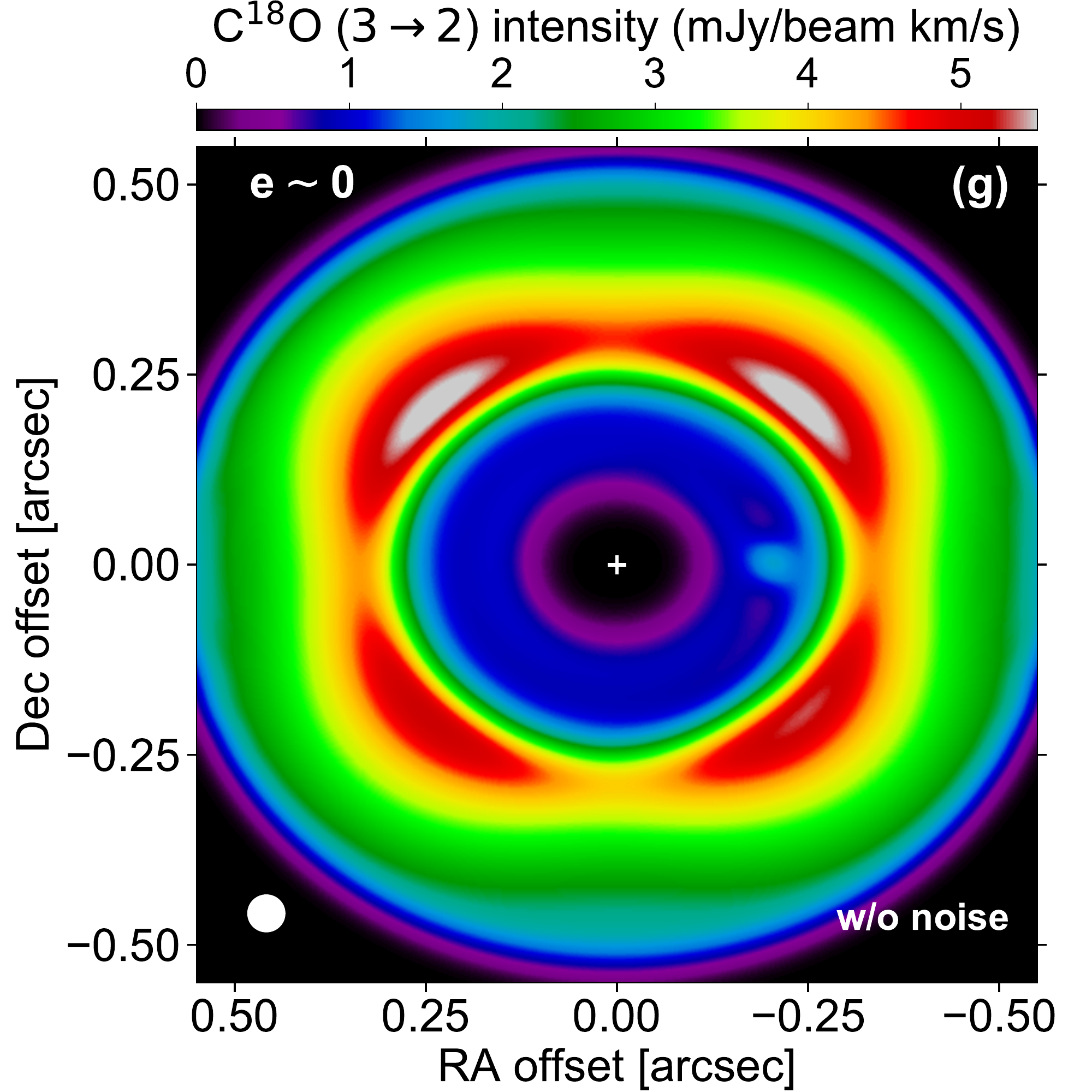}
\includegraphics{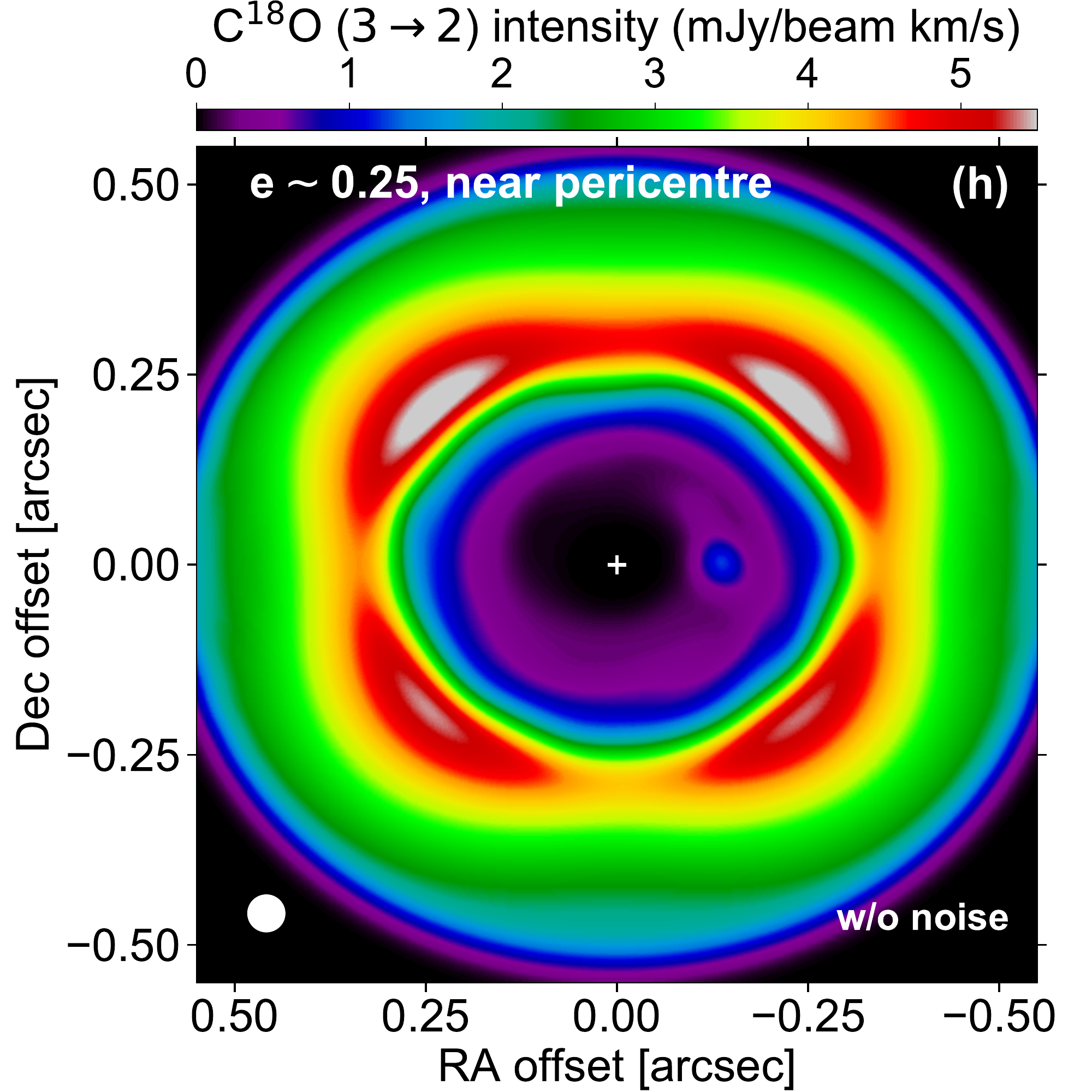}
\includegraphics{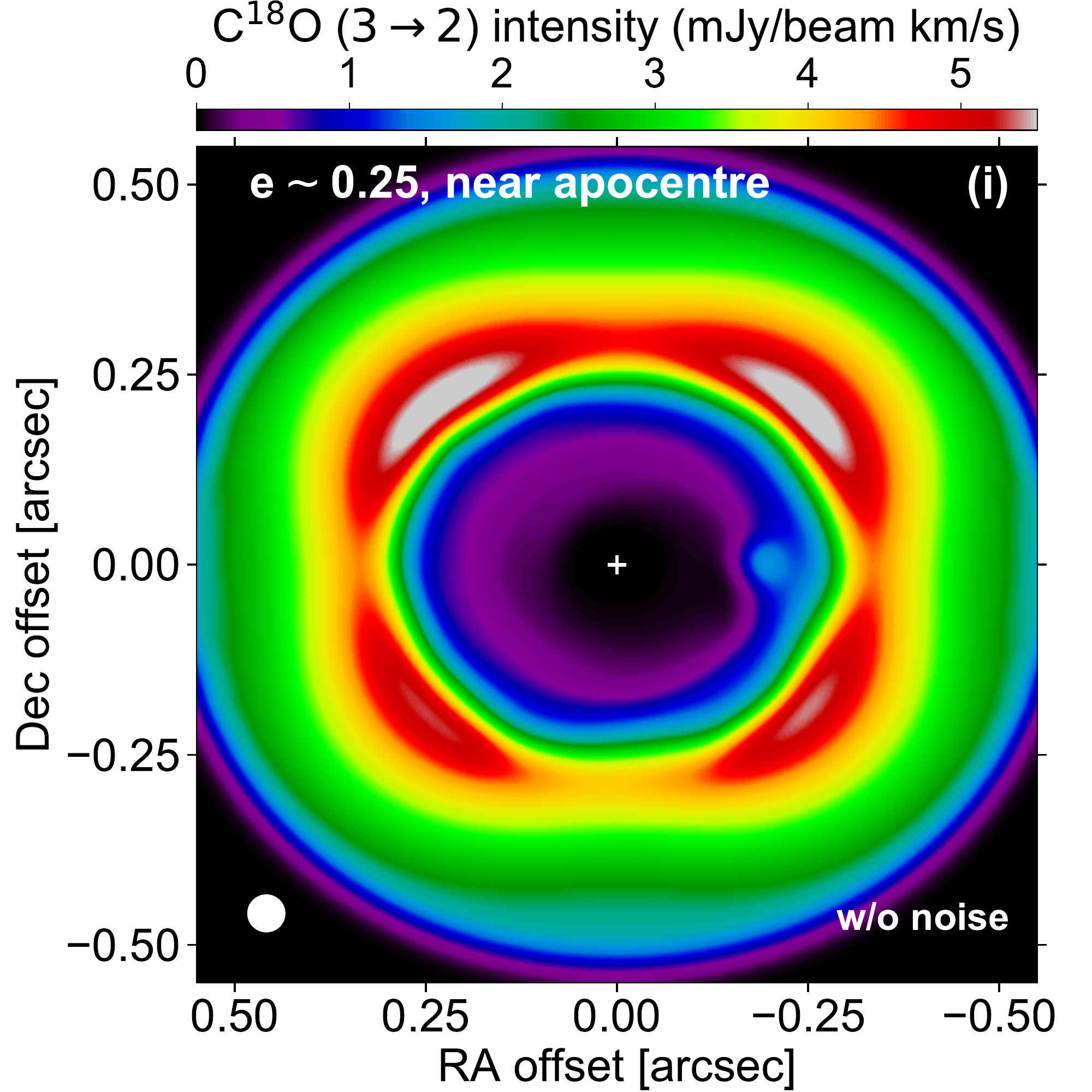}
}
\resizebox{0.96\hsize}{!}
{
\includegraphics{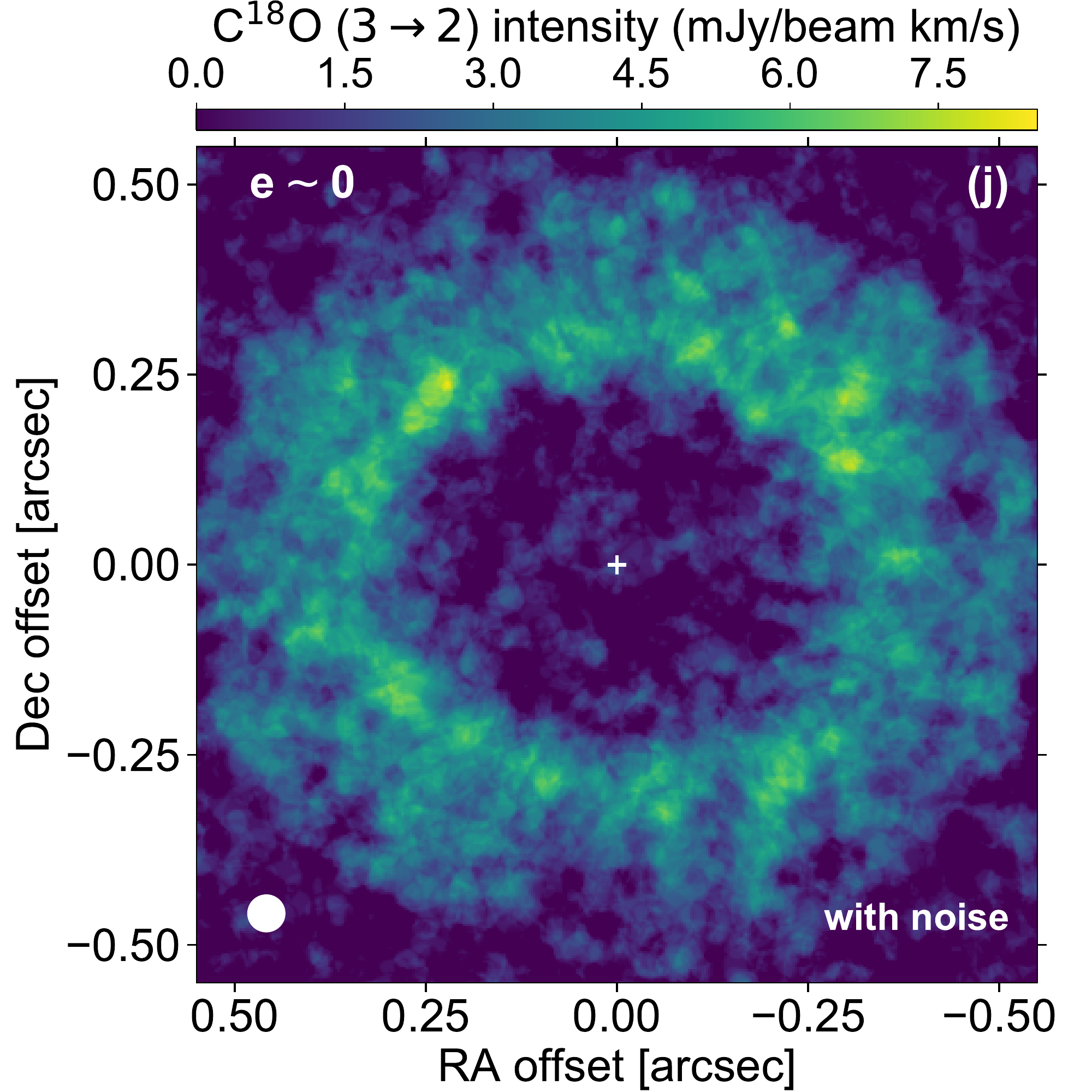}
\includegraphics{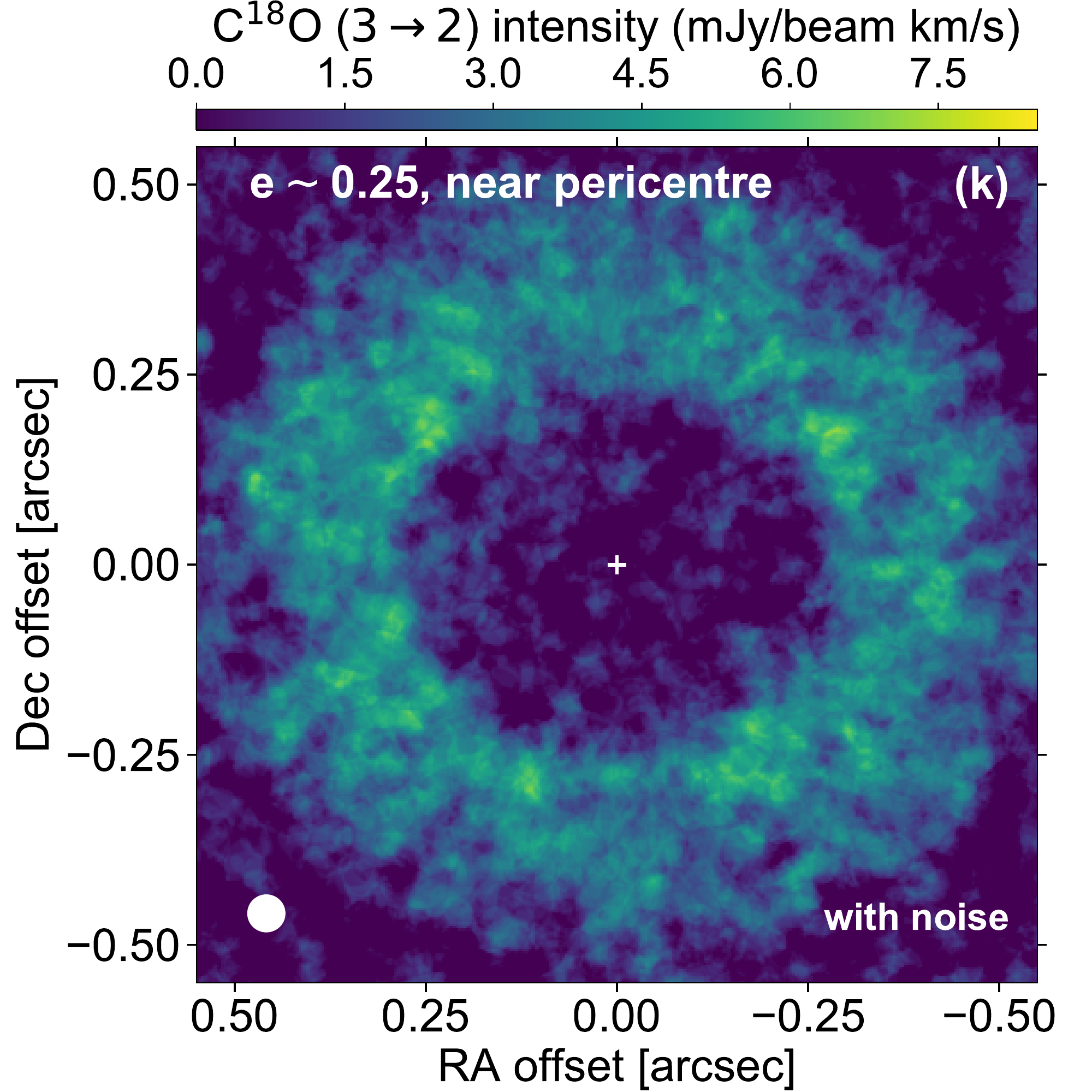}
\includegraphics{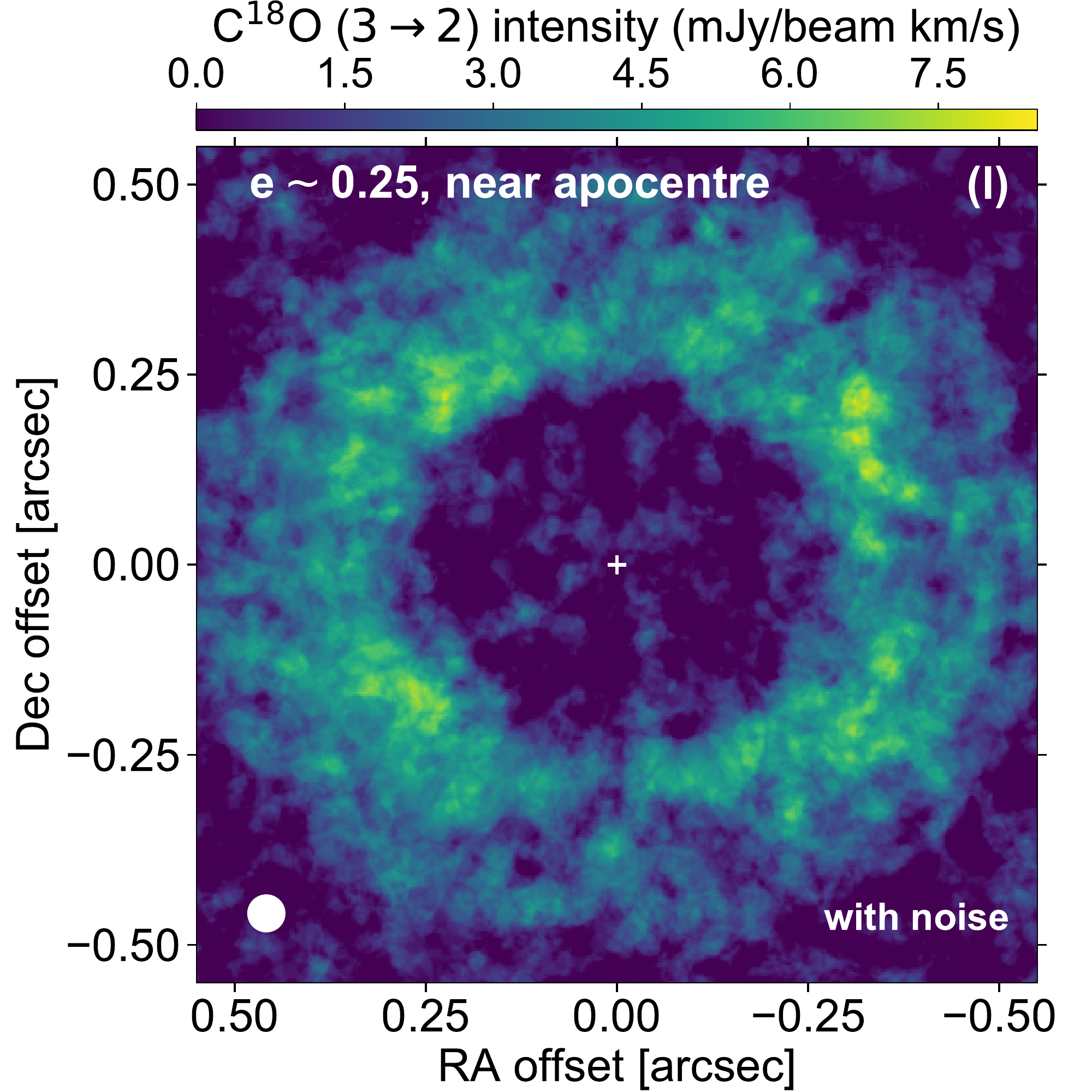}
}
\caption{$^{13}$CO and C$^{18}$O J=3$\rightarrow$2 integrated intensity maps. Panels (d) to (f) and (j) to (l) include noise in the channels.}
\label{fig:fig9}
\end{figure*}
\begin{figure*}
\centering
\resizebox{0.9\hsize}{!}
{
\includegraphics[width=0.53\hsize]{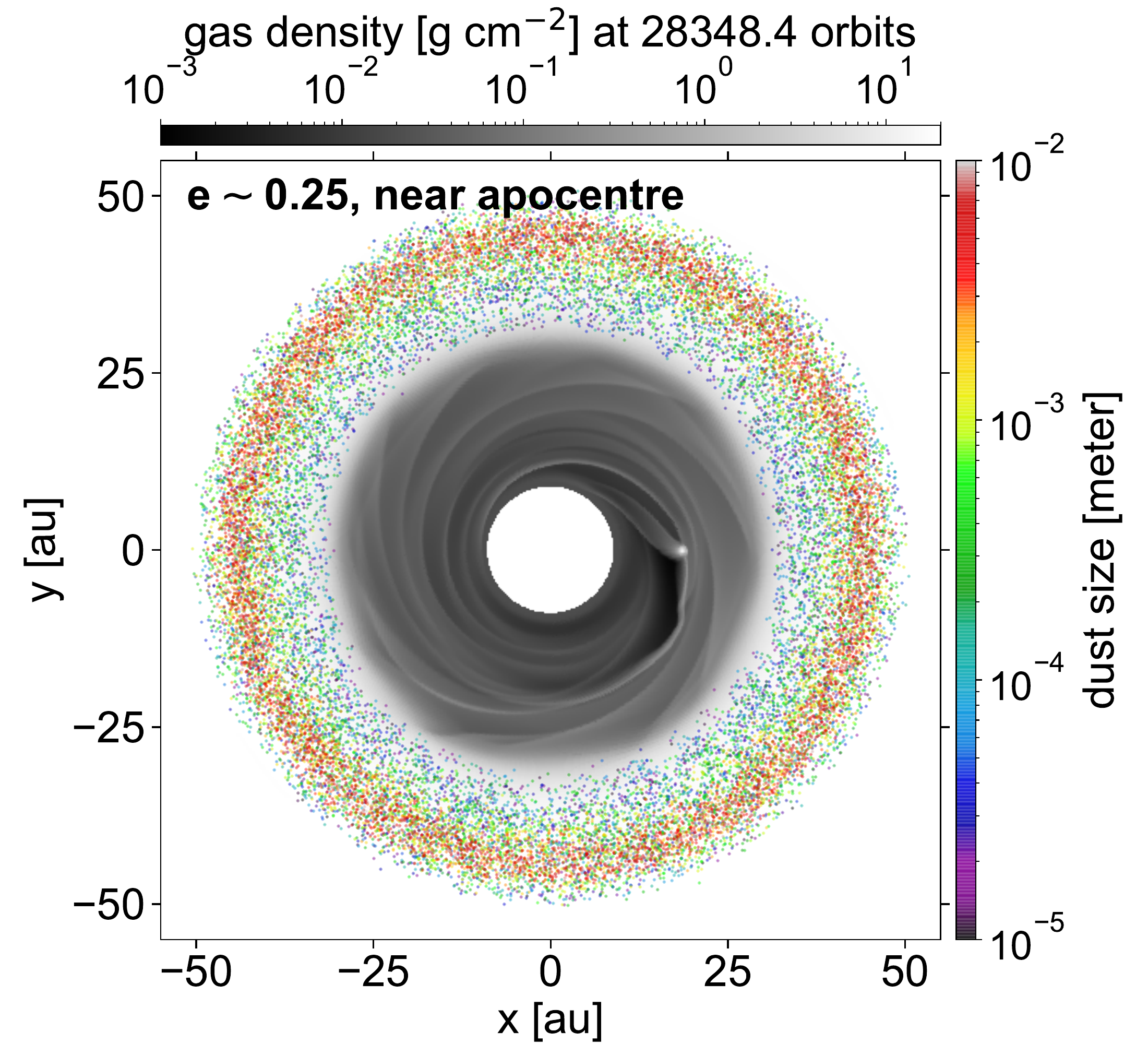}
\includegraphics[width=0.47\hsize]{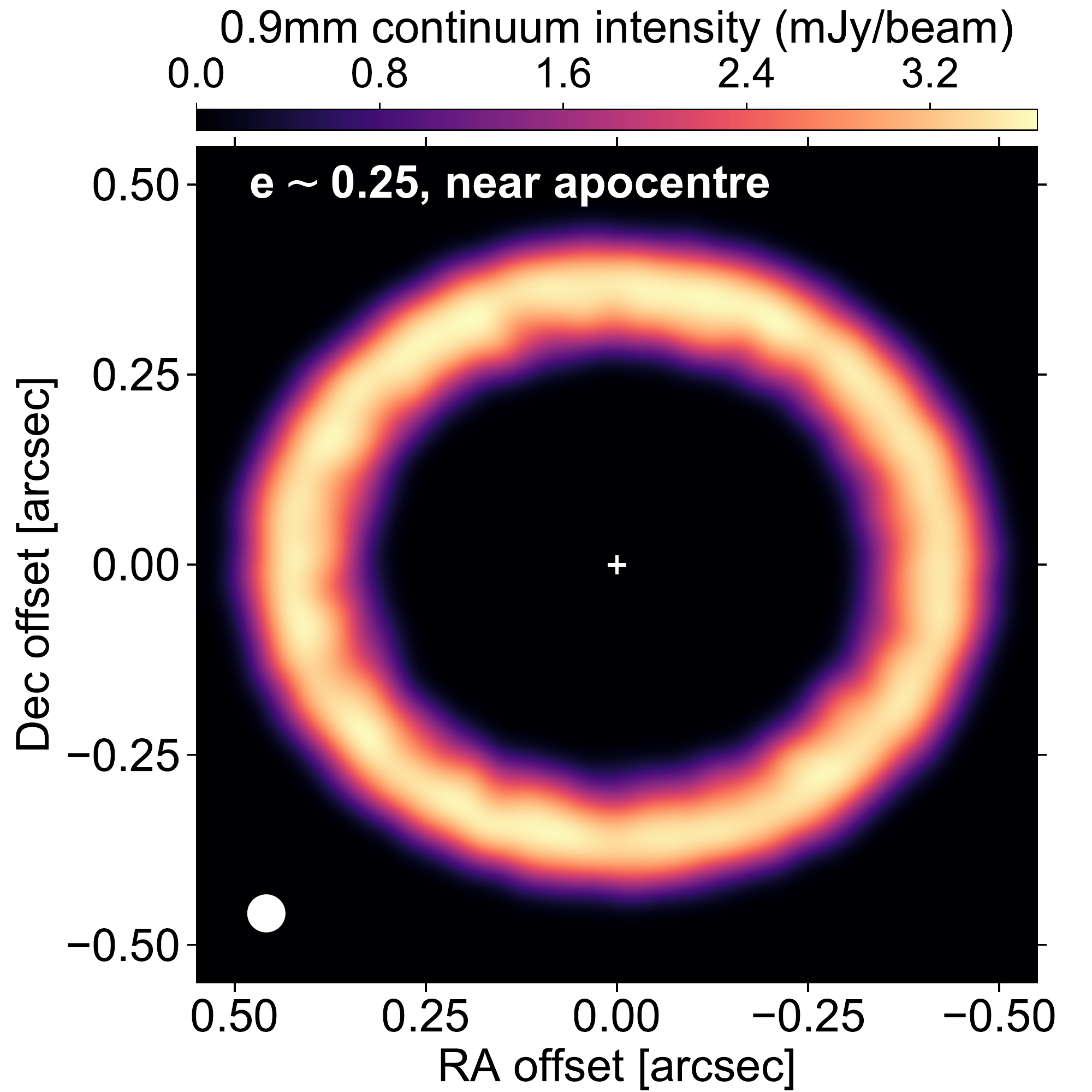}
}
\caption{Left: gas surface density (in black and white) and position of the dust particles (coloured dots) in the gas+dust restart simulation at the same time as in the right panel of Fig.~\ref{fig:fig2}, when the planet is close to the apocentre of its $e \approx 0.25$ eccentric orbit. Right: synthetic image of the dust's continuum emission at 0.9 mm.}
\label{fig:fig10}
\end{figure*}

In this section, we briefly address how the planet eccentricity impacts the $^{13}$CO and C$^{18}$O J=3$\rightarrow$2 lines, which are both centred around 0.91 mm. Fig.~\ref{fig:fig9} displays the integrated intensity maps for both lines, with and without noise in the channels. A four-lobed pattern is clearly seen in all maps without noise. It is actually the most salient feature in these maps, in contrast to those of $^{12}$CO J=3$\rightarrow$2, due to the low optical depth of the lines inside the cavity. We point out that the peak integrated intensity is higher by 6 to 10\% in the upper (northern) half of the four-lobed pattern compared to the lower half. This slight asymmetry is not visible in panels (b) and (c) of Fig.~\ref{fig:fig5} because of a different choice of colormap. Moreover, inside the cavity, the $^{13}$CO integrated intensity is clearly non-axisymmetric when the planet is eccentric. A similar behaviour can be seen for C$^{18}$O, but the reduced intensity inside the cavity makes it less visible.

Due to the overall low level of emission, with peak integrated intensities less than 6$-$8~mJy/beam km s$^{-1}$ outside the cavity, the aforementioned features become barely visible, if at all, when noise with a 1~mJy/beam rms is included in the channels. As far as $^{13}$CO is concerned, the main difference between panels (d) to (f) is the apparent size of the cavity, which is about twice as small when the planet is circular. This is due to the overall larger surface density of the gas inside the cavity when the planet is circular, as already seen in Fig.~\ref{fig:fig2}. The four-lobed pattern is only suggestive in panels (e) and (f). Regarding C$^{18}$O, panels (j) to (l) show no significant differences. 

Although not shown here, we have checked that the velocity maps of $^{13}$CO and C$^{18}$O J=3$\rightarrow$2 are very similar to those of $^{12}$CO J=3$\rightarrow$2 when noise is not included in the channels. In particular, the narrowing, stretching and twisting of the iso-velocity contours inside the cavity when the planet is eccentric are just as visible as in panels (b) and (c) of Fig.~\ref{fig:fig7}. However, addition of noise in the channels renders all these features indiscernible in the $^{13}$CO and C$^{18}$O velocity maps. From this section, we conclude that, for our disc model and assumed level of noise, the $^{13}$CO and C$^{18}$O J=3$\rightarrow$2 line emissions are not prime targets to diagnose the presence of eccentric Jupiters in the gas cavity of protoplanetary discs.

\subsection{Dust continuum emission in the sub-millimetre}
\label{sec:res_cont}

In contrast to the previous sections which focus on the gas emission, we now examine in this section and the following how the planet eccentricity affects the dust emission. We start in this section by presenting our results of dust radiative transfer calculations in the sub-millimetre continuum. 

First, the left panel of Fig.~\ref{fig:fig10} displays the location of the Lagrangian dust particles on top of the gas surface density in the restart gas+dust simulation at the same time as in the right panel of Fig.~\ref{fig:fig2}, that is when the planet is near the apocentre of its $e \approx 0.25$ eccentric orbit inside the gas cavity. We see that the dust particles are all located in a ring outside the cavity. Yet, in the restart simulation, dust particles are initially located between 18 and 27 au inside the cavity, and one may wonder why there is no dust particles left inside the cavity. This is because inside the cavity, the low surface density of the gas implies that dust particles have rather large Stokes numbers: between  $\sim 10^{-2}$ and a few tens. Particles therefore undergo efficient radial drift due to gas drag which, combined with efficient dust turbulent diffusion inside the cavity, rapidly brings particles to the outer edge of the cavity or sweeps them through the inner edge of the computational grid. About 25000 particles remain in the grid at the time shown in the left panel of Fig.~\ref{fig:fig10} (that is, 83\% of their initial number).

\begin{figure*}
\centering
\resizebox{\hsize}{!}
{
\includegraphics{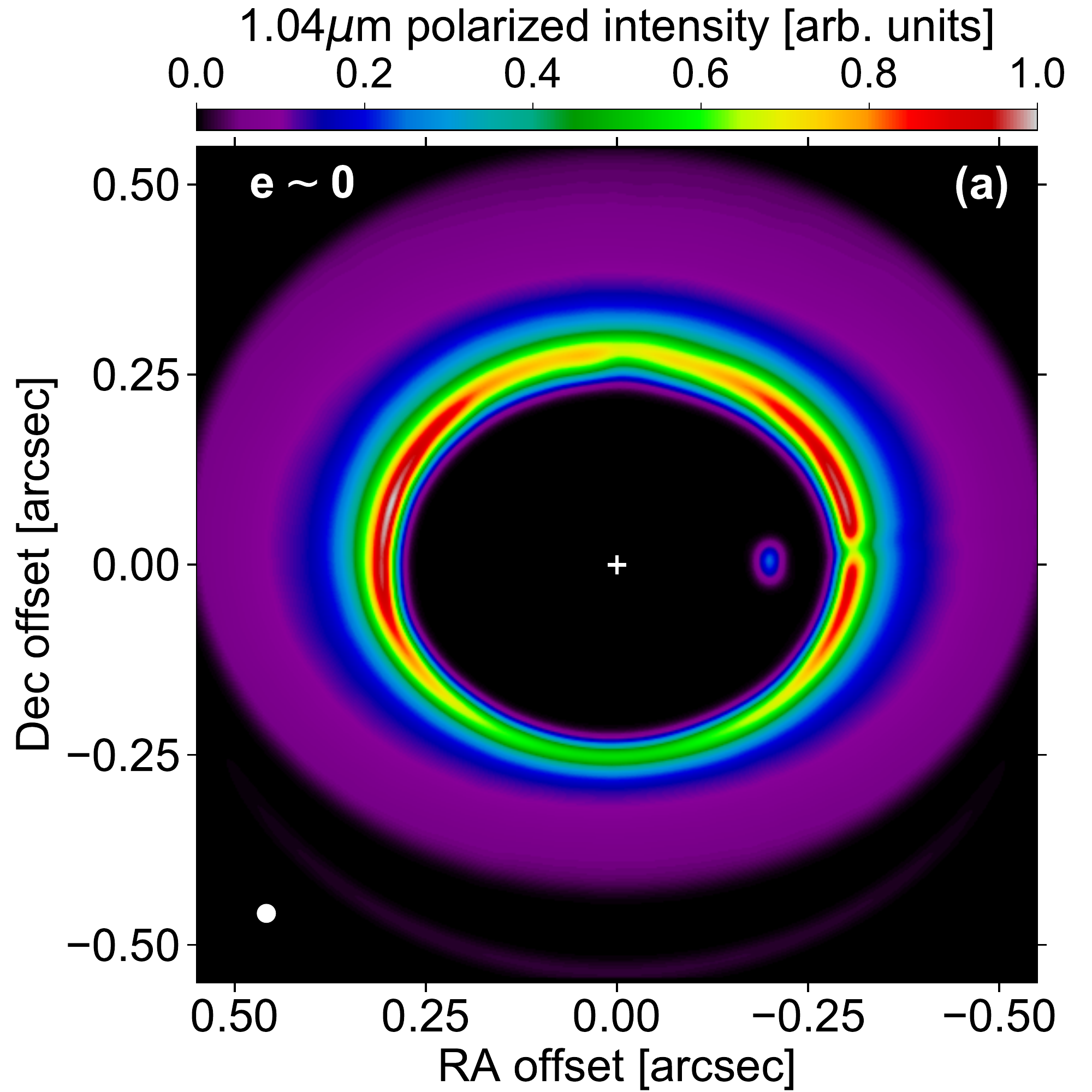}
\includegraphics{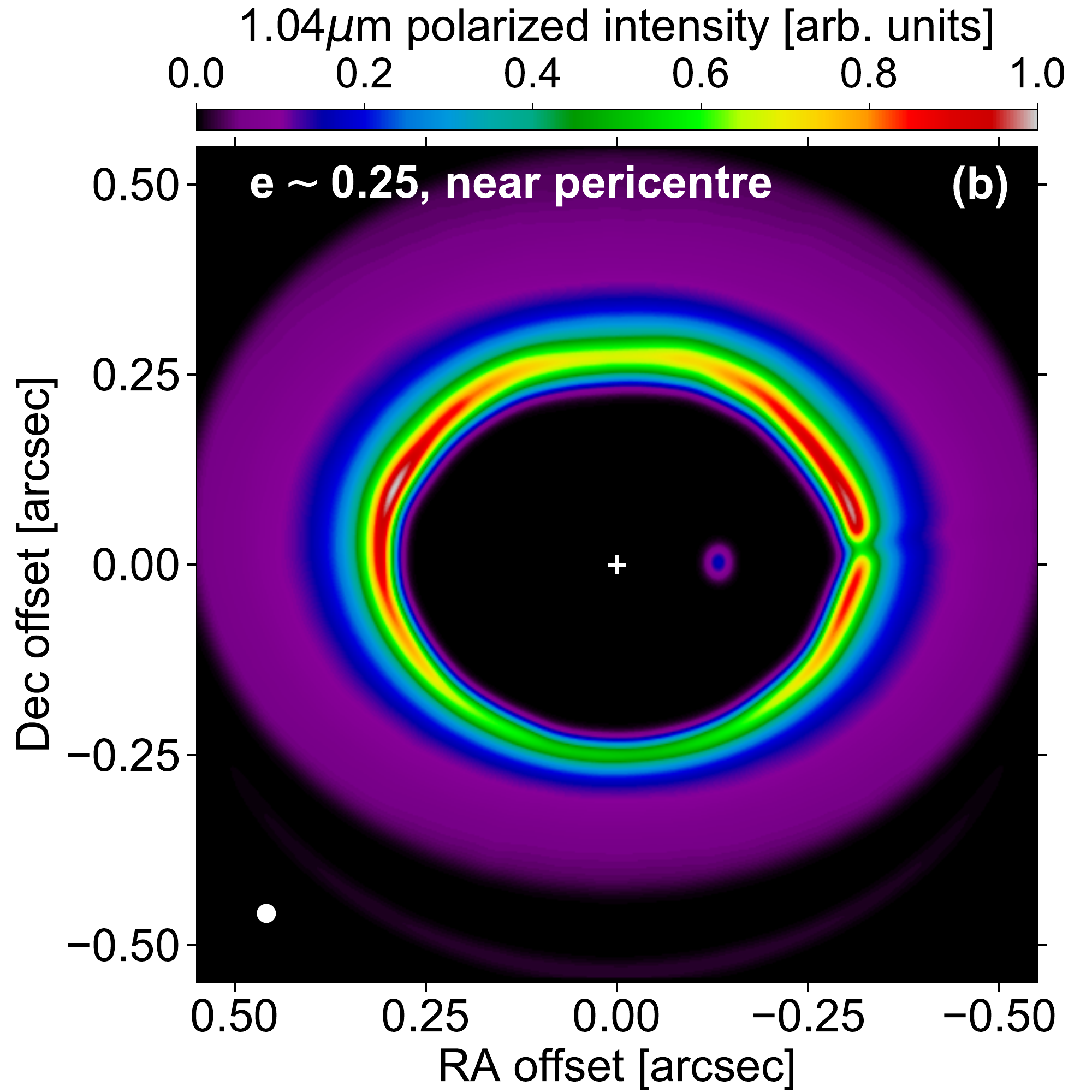}
\includegraphics{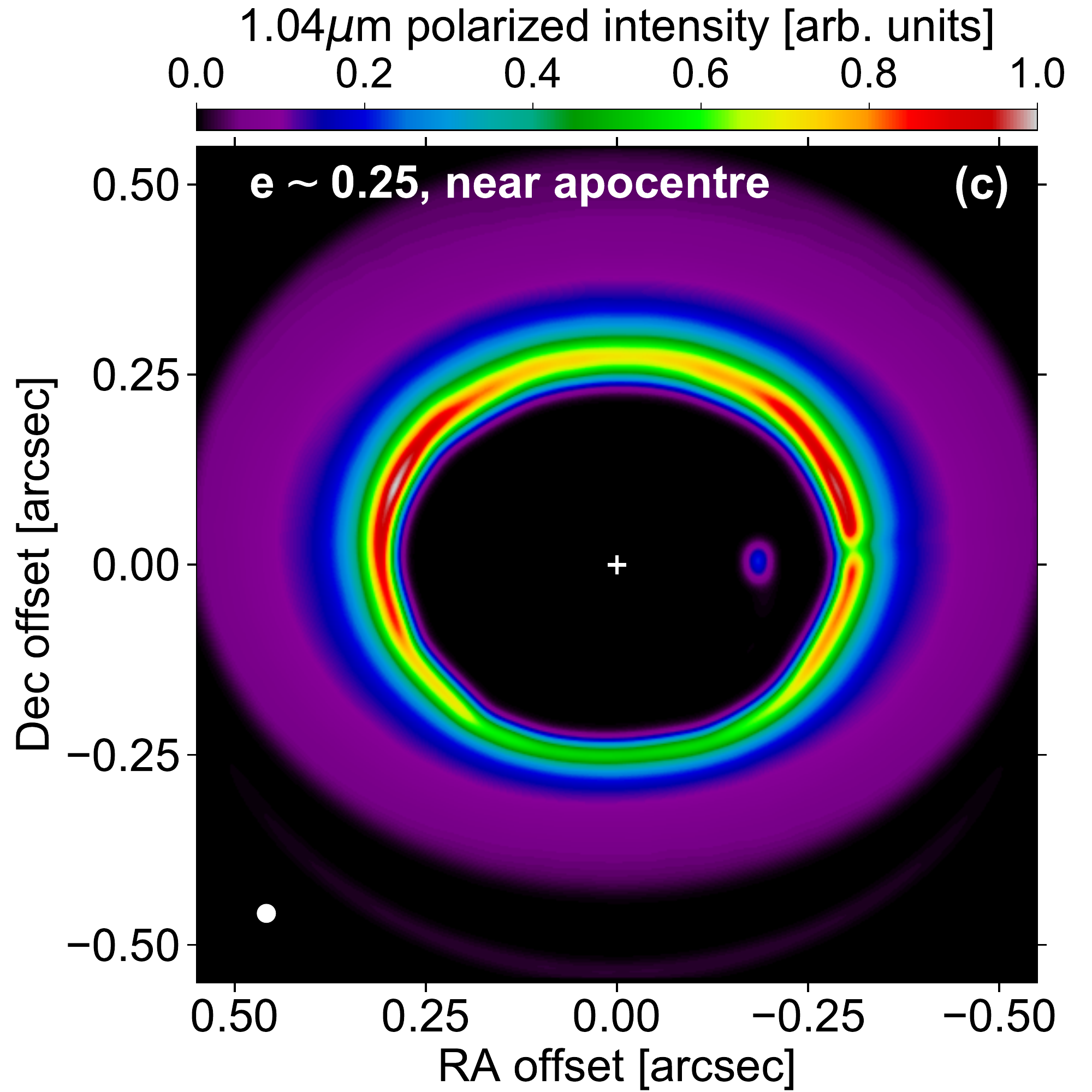}
}
\caption{Scattered light polarized intensity in the Y-band (at 1.04 $\micron$). In each image, the polarized intensity is scaled by the square of the deprojected distance from the star, and normalized such that the intensity of the strongest pixel equals unity. Once again, the white circle in the bottom-left corner of each panel shows the synthetic beam size (25 mas here) and the white cross marks the star position.}
\label{fig:fig11}
\end{figure*}

Outside the cavity, the Stokes number ranges from about $10^{-4}$ to $10^{-1}$ with increasing particle size from 10~$\micron$ to 1~cm. The larger particles are more sensitive to gas drag than to turbulent diffusion, and therefore occupy a smaller radial extent about the location of the pressure maximum outside the cavity at $\sim$ 43 au. Also, from the particles position in the left panel of Fig.~\ref{fig:fig10}, it is clear that the dust ring is circular and that the dust distribution should be axisymmetric. Just like the gas, the dust particles outside the cavity are too distant from the planet and probably still too coupled to the gas to develop significant eccentricity growth. It is therefore not surprising to see, in the right panel of Fig.~\ref{fig:fig10}, that the flux of continuum emission at 0.9~mm takes the form of a circular ring with a quasi uniform intensity distribution. It is also not surprising that the appearance of the continuum ring does not change with the orbital phase of the eccentric planet inside the cavity (which we have checked). 

Finally, we did not add noise in the synthetic image, because the intensity along the continuum ring, about 3~mJy/beam, is much larger than the typical rms noise level that can be achieved in ALMA continuum observations of discs at 0.9 mm and at this angular resolution. For instance, the ALMA band 7 image of the MWC 758 disc in \citealp{Dong18} has an rms noise level of about 20~$\umu$Jy/beam for an approximately 52 mas $\times$ 42 mas beam.

\subsection{Polarized scattered light in the Y-band}
\label{sec:res_pol}
We now present in this section our synthetic images of polarized scattered light at 1.04 $\micron$. Before describing the images, we briefly come back to the methodology detailed in the last paragraph of Section~\ref{sec:dustRTsetup}. 

Scattered light in the near-infrared is assumed to arise from small dust particles, typically up to a few microns, and it is usually safe to say that these small dust particles should be perfectly coupled to the gas, meaning that their spatial distribution should be the same as the gas. This approximation is valid so long as the Stokes number (St) of the dust particles remains very small. Question is: how small is small? The competition between drifting due to gas drag and mixing due to dust turbulent diffusion tells us that St needs to be much smaller than $\alpha_{\rm t}$, the dust turbulent diffusivity expressed as an equivalent alpha viscosity, by at least two orders of magnitude (see, e.g., Eq. 8 of \citealp{Birnstiel13}). 

The maximum particle size in our radiative transfer calculations of polarized scattered light is 0.3~$\micron$. Outside the cavity, this particle size corresponds to a Stokes number of a few $\times 10^{-6}$, which is about two orders of magnitude smaller than $\alpha_{\rm t} \approx \alpha$. Since the Stokes number is proportional to particle size, all dust particles in the size range for these calculations can thus be safely assumed to have a spatial distribution identical to the gas. However, inside the cavity, where the gas surface density is 2 to 4 orders of magnitude smaller than outside the cavity when the planet is eccentric, a 0.3~$\micron$ particle has a Stokes number that can reach a few percent, which is comparable to $\alpha$ inside the cavity. Only particles with St $\la$ a few $\times 10^{-4}$ can be safely regarded as perfectly coupled to the gas inside the cavity. To reflect this condition, we set to 0 the local mass volume density of the dust particles where their Stokes number exceeds $10^{-4}$ (which, again, only occurs inside the cavity).

This criterion can be further justified by the evolution of the Lagrangian particles in the gas+dust restart simulation presented in the previous section. We have checked that the smallest particles in the simulation ($\sim$ 10~$\micron$) escape the cavity in only $\sim$ $10^4$~years, either by drifting outwards towards the edge of the cavity or by zipping inward through the inner edge of the disc. Actually, this timescale is more probably an upper estimate since, at the time when the particles are inserted in the simulation, the planet is only mildly eccentric ($e\approx0.06$) and so is the gas, whose surface density does not take as small values as when the planet reaches its near maximum eccentricity. Furthermore, the effect of radiation on the dust particles is neglected in our simulation. Given the low gas density in the cavity, Poynting-Robertson drag and radiation pressure blow-out could make the smallest particles escape the cavity even faster. Assuming that the above timescale is inversely proportional to particle size, we see that a 0.3~$\micron$ particle would escape the cavity in less than a few $\times10^5$ years. 

We now describe our synthetic images, which are displayed in Fig.~\ref{fig:fig11}. The images do not include noise. As in most previous figures, they are shown from left to right at the same times as the gas density panels in Fig.~\ref{fig:fig2} to underline the impact of the planet eccentricity. Overall, we see that all three images are very similar. This is essentially due to the lack of scattered light arising from inside the cavity, which comes about because of the above condition that there should be no dust particles left inside the cavity with Stokes number $\ga10^{-4}$. As a result, only the denser circumplanetary material shows up in the cavity as a pale blue dot. Outside the cavity, the $R^2$-scaled polarized intensity is maximum near the edge of the cavity. It peaks near $\sim$ 3 and $\sim$ 9 o'clock in all three images, and the far side of the disc is somewhat brighter than the near side. This is the expected behaviour for an inclined ring when scattering is dominated by sub-micron sized grains (see, e.g., \S4.3 in \citealp{Keppler18}). Note the slight dimming at $\sim$ 3 o'clock in the images, which is due to a shadow cast by the circumplanetary material. We also point out that the back side of the disc is (slightly) visible in the bottom part of the images.

For testing purposes, we have increased the above threshold Stokes number to $10^{-3}$. Although the corresponding images are not shown here, we find that, when the planet still has a near-circular orbit, the cavity contributes to the polarized intensity image: the planet gap and its wakes become clearly visible, much like in the central parts of panel (a) in Fig.~\ref{fig:fig9}. When the planet is eccentric, however, the cavity is hardly discernible, but the polarized intensity at the edge of the cavity shows much less smooth variations as in panels (b) or (c) in Fig.~\ref{fig:fig11}. From the results of this section, we reach the conclusion that polarized scattered light may not be able to distinguish between a circular and an $e\sim0.25$ eccentric planet in the gas cavity of a protoplanetary disc.

\section{Concluding remarks}
\label{sec:conclusion}

\begin{figure*}
\centering
\resizebox{\hsize}{!}
{
\includegraphics{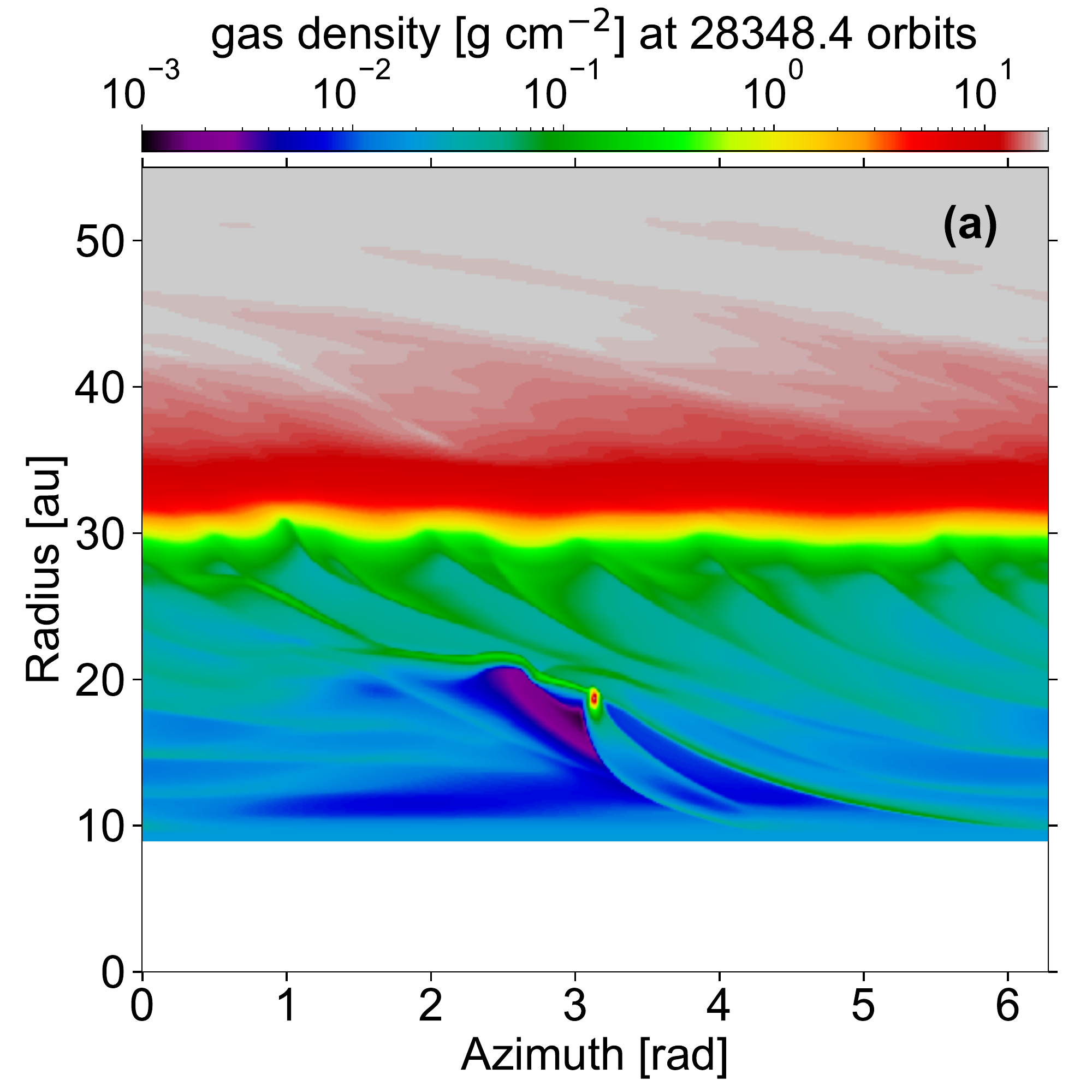}
\includegraphics{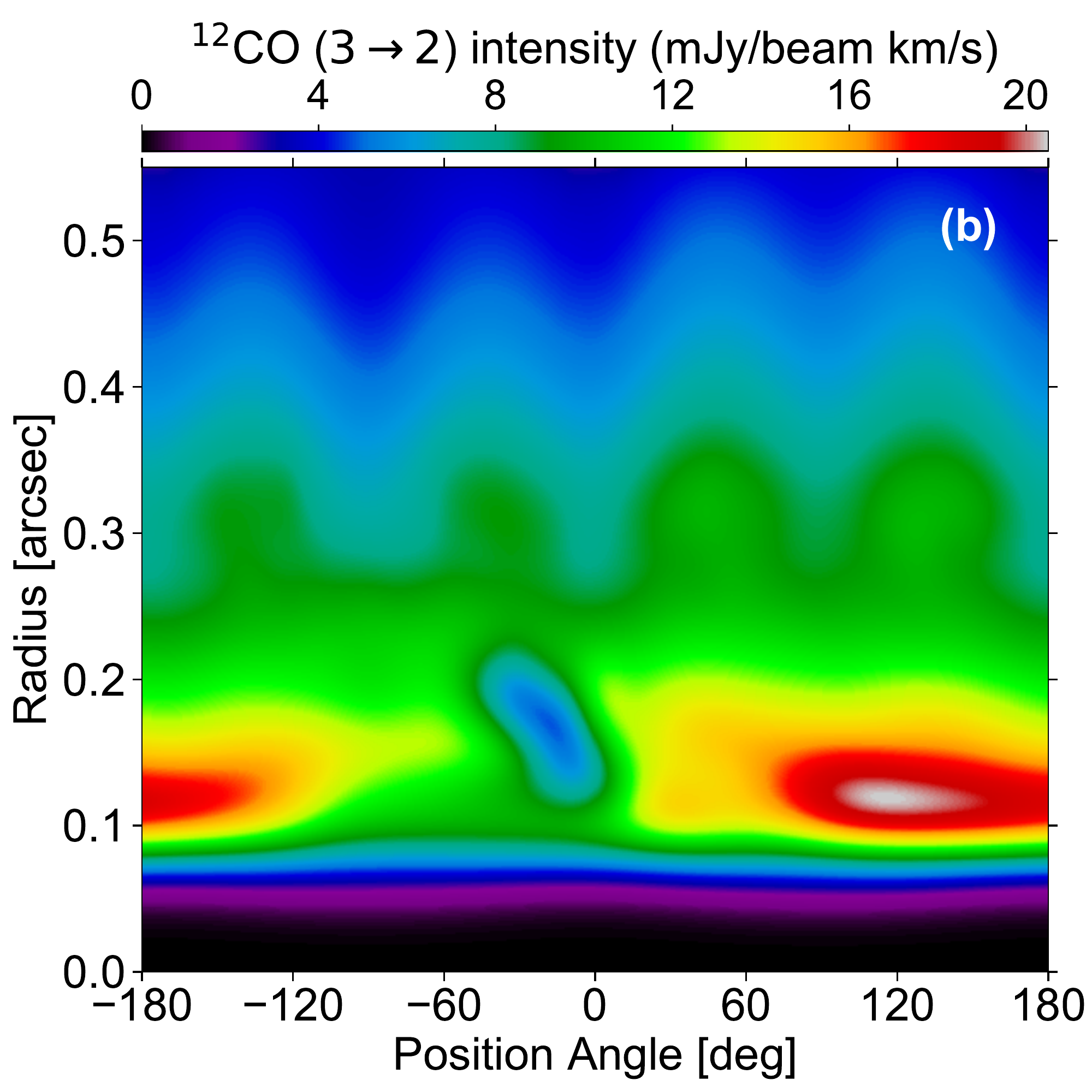}
\includegraphics{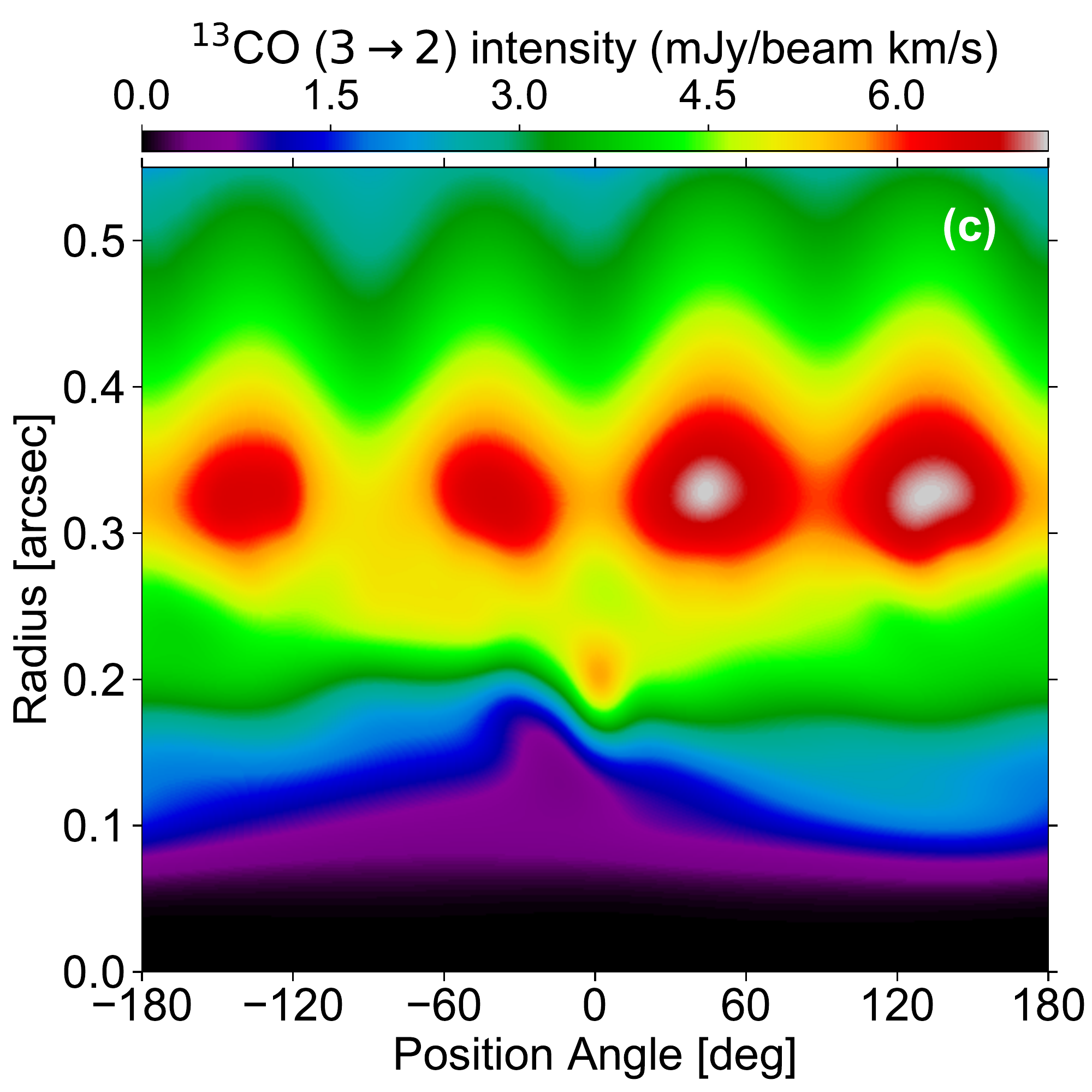}
}
\resizebox{\hsize}{!}
{
\includegraphics{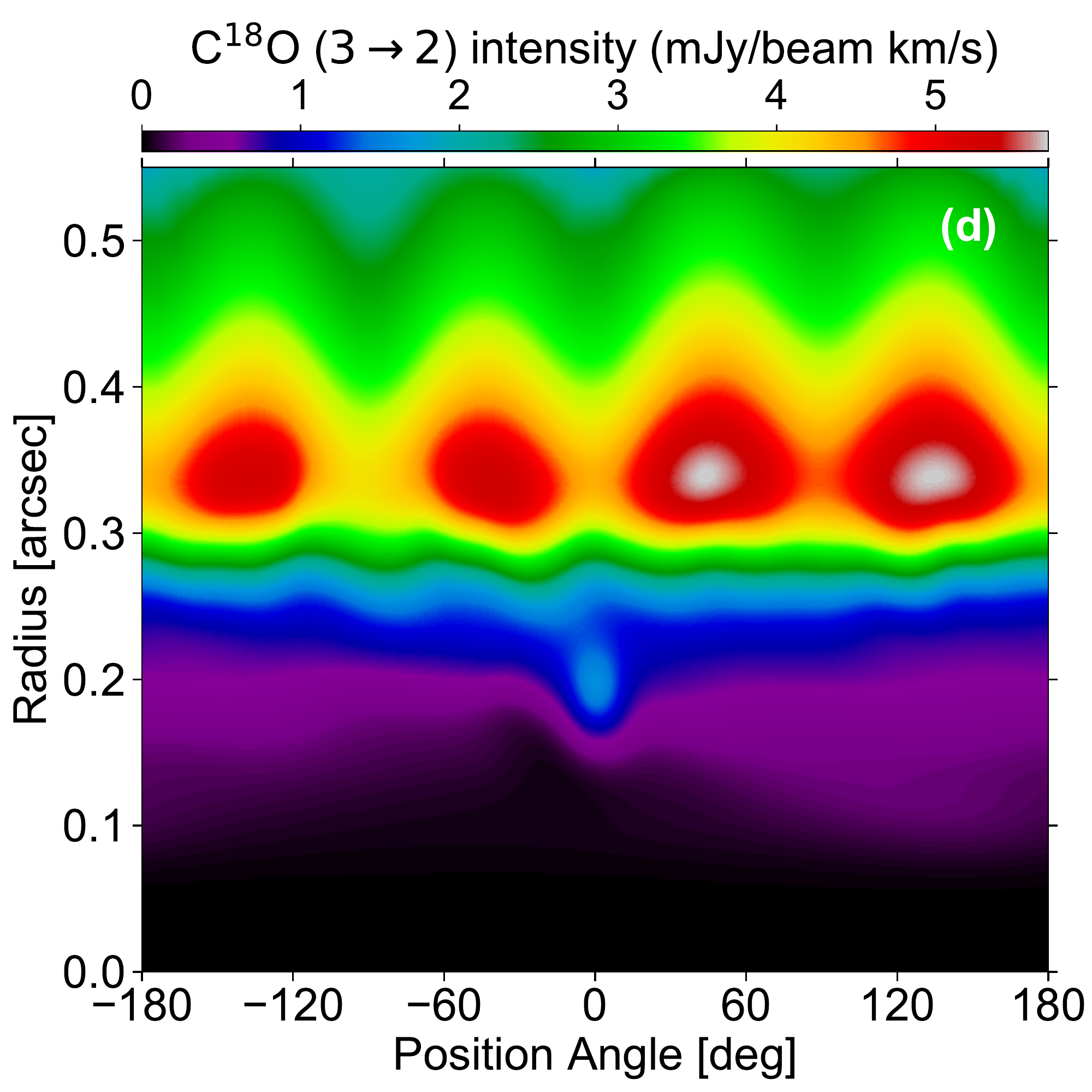}
\includegraphics{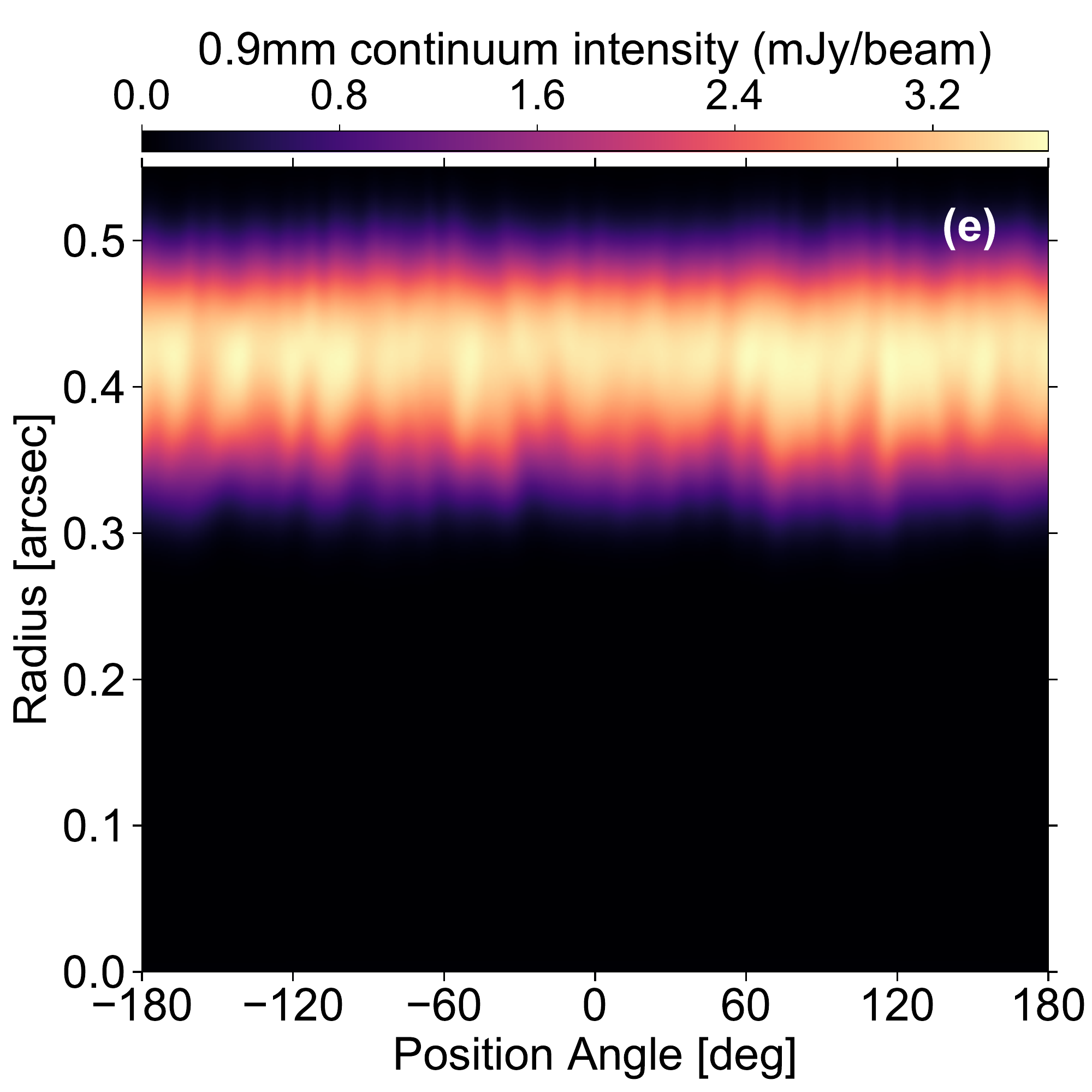}
\includegraphics{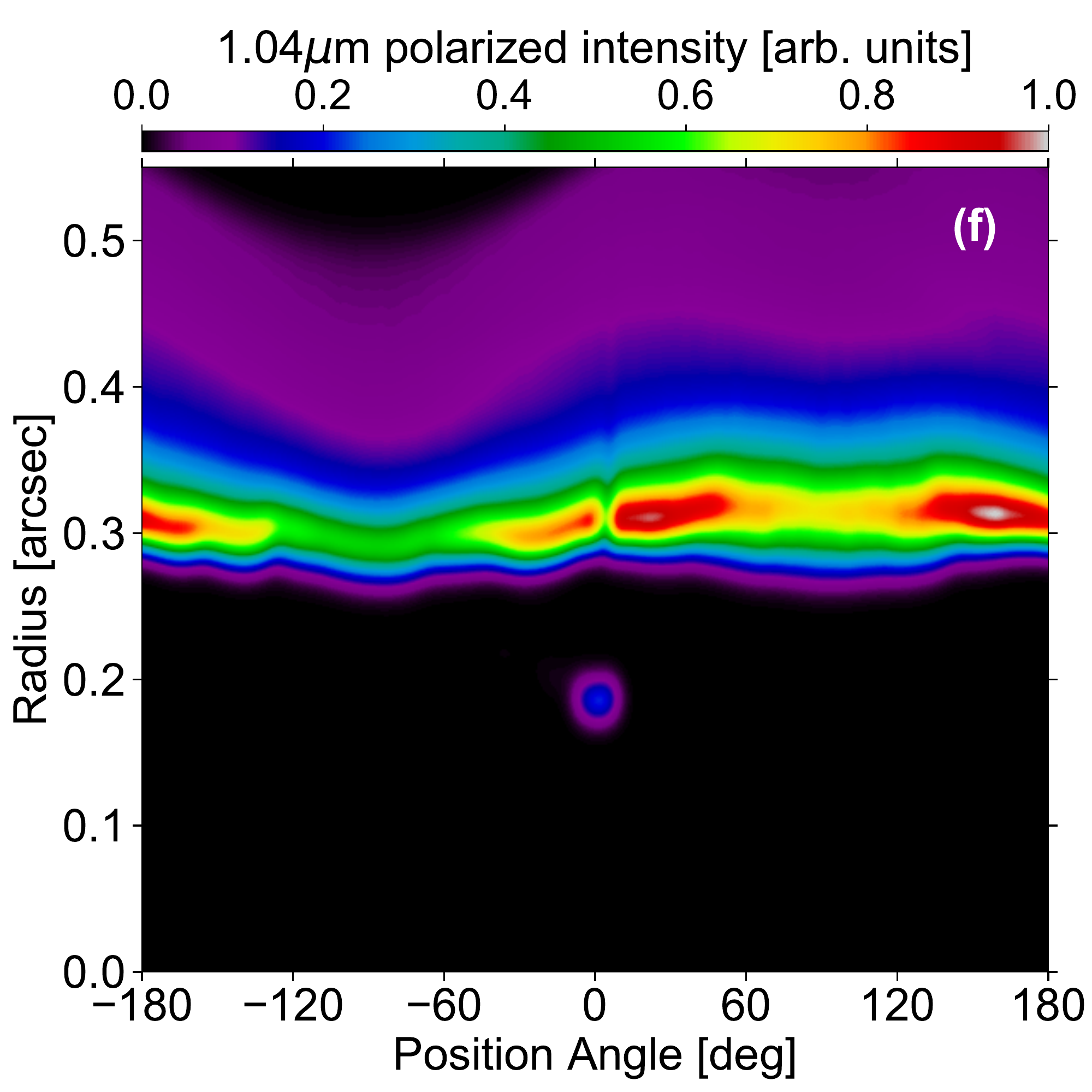}
}
\caption{Results summary when the planet is near apocentre on an $e \approx 0.25$ eccentric orbit inside the gas cavity. We compare the gas surface density in polar coordinates (panel a) with the deprojected maps of integrated intensity for the $^{12}$CO, $^{13}$CO and C$^{18}$O J=3$\rightarrow$2 line emissions (panels b to d), of the continuum emission at 0.9 mm (panel e) and of the polarized scattered light at 1.04~$\micron$ (panel f). No noise is included in the synthetic emission maps. The beam FWHM is 50 mas in all images except in that of polarized intensity, where it is 25 mas.}
\label{fig:fig12}
\end{figure*}

The exquisite sensitivity and angular resolution of ALMA and SPHERE disc observations have stimulated a substantial body of theoretical work predicting observable signatures of planets in the dust and gas emission of their disc \citep[e.g.,][]{Dong15,Zhang18,Perez18_seba,Pinte19,WFB20}. Our study differs from previous works in that it examines the case of an eccentric 2~Jupiter mass planet orbiting a 2~Solar-mass star, and which is located inside a gas cavity in its protoplanetary disc. It builds on the recent results of \citet{Debras21}, who have shown that planets in the Jupiter-mass range could reach eccentricities up to 0.3$-$0.4 after migrating into a low-density gas cavity, with eccentricity growth driven solely by disc-planet interactions. It is based on 2D hydrodynamical simulations including either gas or gas+dust, which are post-processed by 3D radiative transfer calculations. 

In our disc model, the planet reaches a maximum eccentricity $\sim$ 0.25 in a few Myr, which triggers strong asymmetries in the surface density of the gas inside the cavity. In our gas radiative transfer calculations, these asymmetries are enhanced by photodissociation and manifest themselves as large-scale asymmetries in integrated intensity maps of the $^{12}$CO J=3$\rightarrow$2 line emission in the sub-millimetre. We have shown that these asymmetries are detectable for a beam and a noise level similar to those currently achieved in ALMA disc gas observations. Although our synthetic maps were obtained with a 50 mas beam and a 1~mJy/beam rms noise level per channel map, we have checked that the asymmetries remain easily detectable up to a $0\farcs15$ beam (which is the solid angle subtended by $\sim$ half of the cavity) and a rms noise level increased to 4~mJy/beam per channel map. Without photodissociation, the asymmetries are less pronounced but remain detectable for our fiducial beam and noise level. Also, the planet eccentricity renders the gas eccentric inside the cavity, which causes detectable narrowing, stretching and twisting of the iso-velocity contours in velocity maps of $^{12}$CO J=3$\rightarrow$2. In contrast, for a noise level similar to that currently achieved with ALMA, we find that the planet eccentricity does not have clear detectable signatures in the $^{13}$CO and C$^{18}$O J=3$\rightarrow$2 line emissions inside the cavity. Still, when the planet is eccentric, we find that the apparent size of the cavity is roughly the same in $^{13}$CO and C$^{18}$O J=3$\rightarrow$2, whereas when the planet is still nearly circular the cavity appears much broader in C$^{18}$O than in $^{13}$CO (by a factor $\sim$ 2 in our disc model). Outside the cavity, the gas retains near circular orbits and the optically thick, vertically extended gas emission takes the form of a four-lobed pattern in CO integrated intensity maps for sufficient disc inclinations. In our synthetic images, this four-lobed pattern is clearly visible for disc inclinations $\ga30\degr$.

Due to the lack of large ($\ga10$~$\micron$) dust particles inside the cavity in our disc model, continuum emission is only visible at the edge of the cavity. Like the gas, the dust outside the cavity describes near-circular orbits and the continuum emission forms an axisymmetric ring around the star. Similarly, due to the expected lack of small (0.01 to 0.3~$\micron$) dust particles inside the cavity, polarized intensity images in the near-infrared do not feature scattered light arising from inside the cavity, with the exception of the circumplanetary material. Thus, these images may not allow to differentiate between a circular and an eccentric planet inside the cavity, at least for eccentricities $e \la 0.25$. A more massive planet could change this picture.

The aforementioned findings are summarised with the help of Fig.~\ref{fig:fig12}, which displays the gas surface density and the deprojected images of the gas and dust emission in polar coordinates, when the planet eccentricity $e \sim 0.25$ and the planet near apocentre. The figure also allows to compare the deprojected distance at which the emissions in both the gas and dust peak. The integrated intensity maps of the CO isotopologues do not include noise in the channels. 

Finally, several simplifications have been made in this work to keep the problem tractable. Most of them have already been discussed in Sections~\ref{sec:methods} and~\ref{sec:results}, and below we highlight a few hypotheses that we have made and which should be tested in future studies:
\begin{itemize}
\item[\sbt] First of all, our predictions are based on the results of 2D hydrodynamical simulations. As already stated in \citet{Debras21}, 3D simulations seem very relevant, not only to re-assess the level of eccentricity pumping that can be obtained once the planet has migrated into the cavity, but also to have a more accurate modelling of the vertical structure of the wakes triggered by the (eccentric) planet. In this regard, inclusion of an energy equation would seem relevant as well. Running such 3D simulations over the several tens of thousands of planet orbits required to reach a large eccentricity according to 2D simulations might prove to be numerically challenging.
\item[\sbt]  We have assumed that photodissociation causes the number density of CO gas species to instantaneously plummet where the column density of gas falls below a threshold value (see Eq.~\ref{pd_eq1}). However, when the planet is eccentric, the density of the gas inside the cavity evolves quite dynamically over the orbital phase of the planet. The typical timescale for CO gas species to react to photodissociation under these dynamical conditions should be investigated. Also, ideally, the number density of gas species should be computed with the help of chemical models using the hydrodynamical simulations data for the gas and dust as inputs.
\item[\sbt] Our gas radiative transfer calculations have focused on the J=3$\rightarrow$2 line for the three main CO isotopologues, since this transition usually provides a good compromise in terms of signal-to-noise ratio in sub-millimetre observations (see Section~\ref{sec:gasRTsetup}). Other rotational levels need to be explored. For instance, although not presented in this paper, we have computed integrated intensity maps of the $^{12}$CO J=6$\rightarrow$5 line at $\sim$ 0.43 mm. The images are overall similar to those of $^{12}$CO J=3$\rightarrow$2, except that the integrated intensity is about 4 times larger throughout the disc for the same 50 mas beam. In particular, the large-scale asymmetries in the gas cavity remain visible when assuming a rms noise level 6 times larger for $^{12}$CO J=6$\rightarrow$5 than for $^{12}$CO J=3$\rightarrow$2, which is about the same factor as in the synthetic images in Appendix D of \citet{Facchini18} or in the TW Hya observations of \citet{Calahan21}.
\item[\sbt] In the same vein, other molecules need to be explored, especially given the growing chemical inventory in several protoplanetary discs with gas cavities, like AB Aur \citep{RM20ABAUR} or PDS70 \citep{Facchini21}. Likewise, \citet{Regaly10} showed that a planet that is massive enough to pump the eccentricity of its surrounding gas could cause a detectable distortion in the line profile of the CO ro-vibrational fundamental emission at $\sim$ 4.7~$\micron$. It would be interesting to revisit this prediction for the cool gas in our rather large cavity, especially with CRIRES+ and METIS on the horizon.
\end{itemize}

In future work, it would be interesting to test some of the ideas and predictions of this paper to specific protoplanetary discs. The PDS 70 disc, which is host to two companions inside its cavity, including one with an eccentricity similar to that obtained in our disc model \citep{Wang21}, seems a relevant target. The AB Aur disc, which shows suggestive evidence for planetary companion(s) in its large cavity \citep[e.g.,][]{Fuente2017,Boccaletti20} is another one.

\section*{Acknowledgments}
Numerical simulations were performed on the CALMIP Supercomputing Centre of the University of Toulouse. We would like to thank Wing-Fai Thi for helpful discussions, and the referee, Ken Rice, for constructive comments. G.W.F acknowledges funding from the European Research Council (ERC) under the European Union's Horizon 2020 research \& innovation programme (grant agreement \#815559, MHDiscs). R.L.G. acknowledges support from a CNES fellowship. F.D acknowledges funding from the ERC under the H2020 research \& innovation programme (grant agreement \#740651, NewWorlds). A.C. acknowledges funding from the French National Research Agency (ANR) under contract number ANR-18-CE31-0019 (SPlaSH). This work is partly supported by ANR in the framework of the Investissements d'Avenir program (ANR-15-IDEX-02), through the funding of the {\it Origin of Life} project of the Grenoble-Alpes University. A.F. and P.R.M. thank the Spanish MICIU for funding support from AYA2016-75066-C2-2-P and PID2019-106235GB-I00.

\section*{Data availability}
The FARGO-ADSG code is available from \href{http://fargo.in2p3.fr/-FARGO-ADSG-}{http://fargo.in2p3.fr/-FARGO-ADSG-}. The code version including a Lagrangian treatment of the dust particles, Dusty FARGO-ADSG, can be made available for use on a collaborative basis upon request to C. Baruteau. The RADMC-3D code is available from \href{https://www.ita.uni-heidelberg.de/~dullemond/software/radmc-3d/}{https://www.ita.uni-heidelberg.de/~dullemond/software/radmc-3d/}. The python packages used for the post-processing of the simulations data and for the analysis of the radiative transfer calculations are available via their GitHub repositories: fargo2radmc3d (\href{https://github.com/charango/fargo2radmc3d}{https://github.com/charango/fargo2radmc3d}) and bettermoments (\href{https://github.com/richteague/bettermoments}{https://github.com/richteague/bettermoments}). The input files for generating our hydrodynamical simulations and radiative transfer calculations will be shared on reasonable request to the corresponding author.

\bibliographystyle{mnras}


\appendix

\section{$^{12}$CO J=3$\rightarrow$2 integrated intensity maps without photodissociation}
\label{sec:res_12CO_mom0_nopd}

\begin{figure*}
\centering
\resizebox{0.99\hsize}{!}
{
\includegraphics{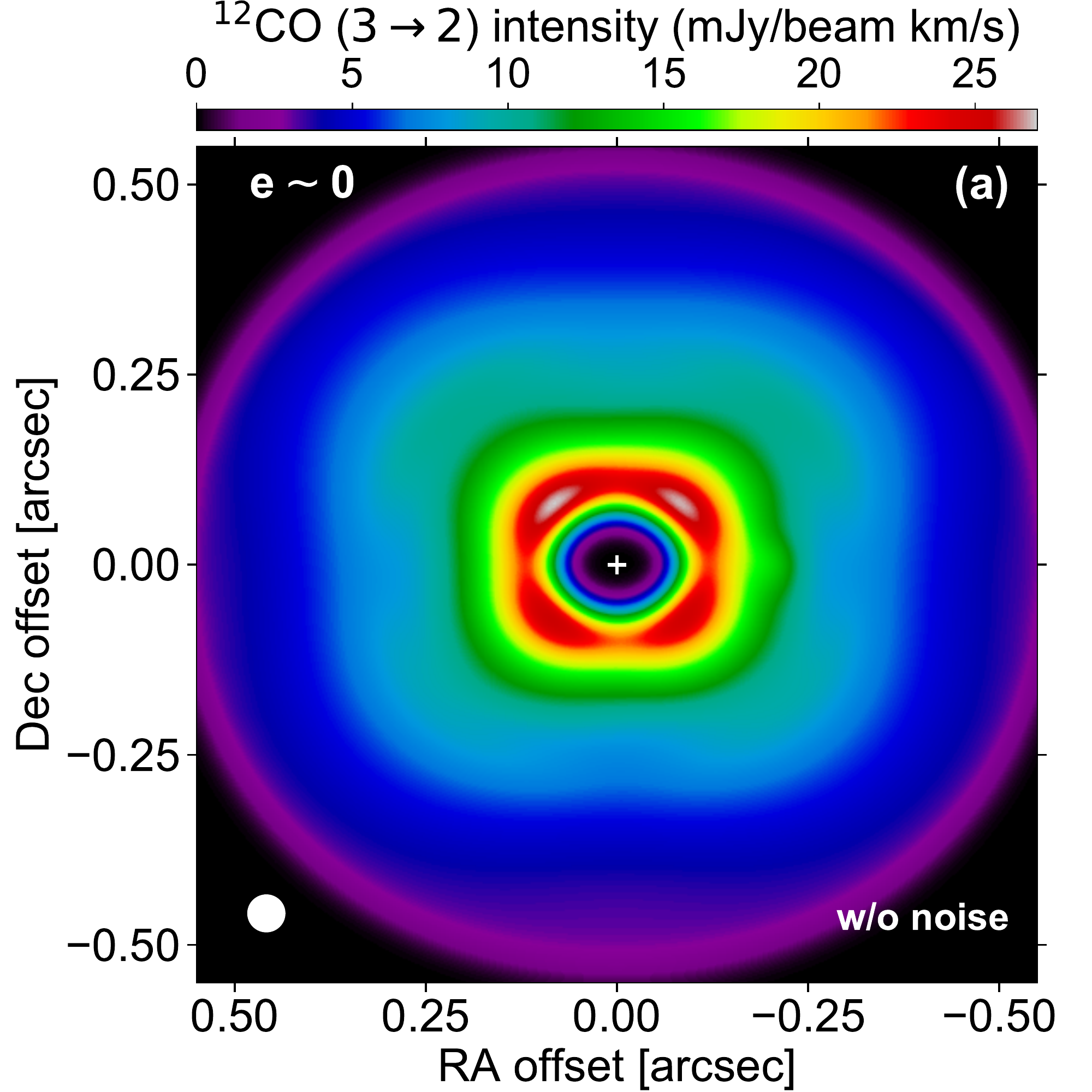}
\includegraphics{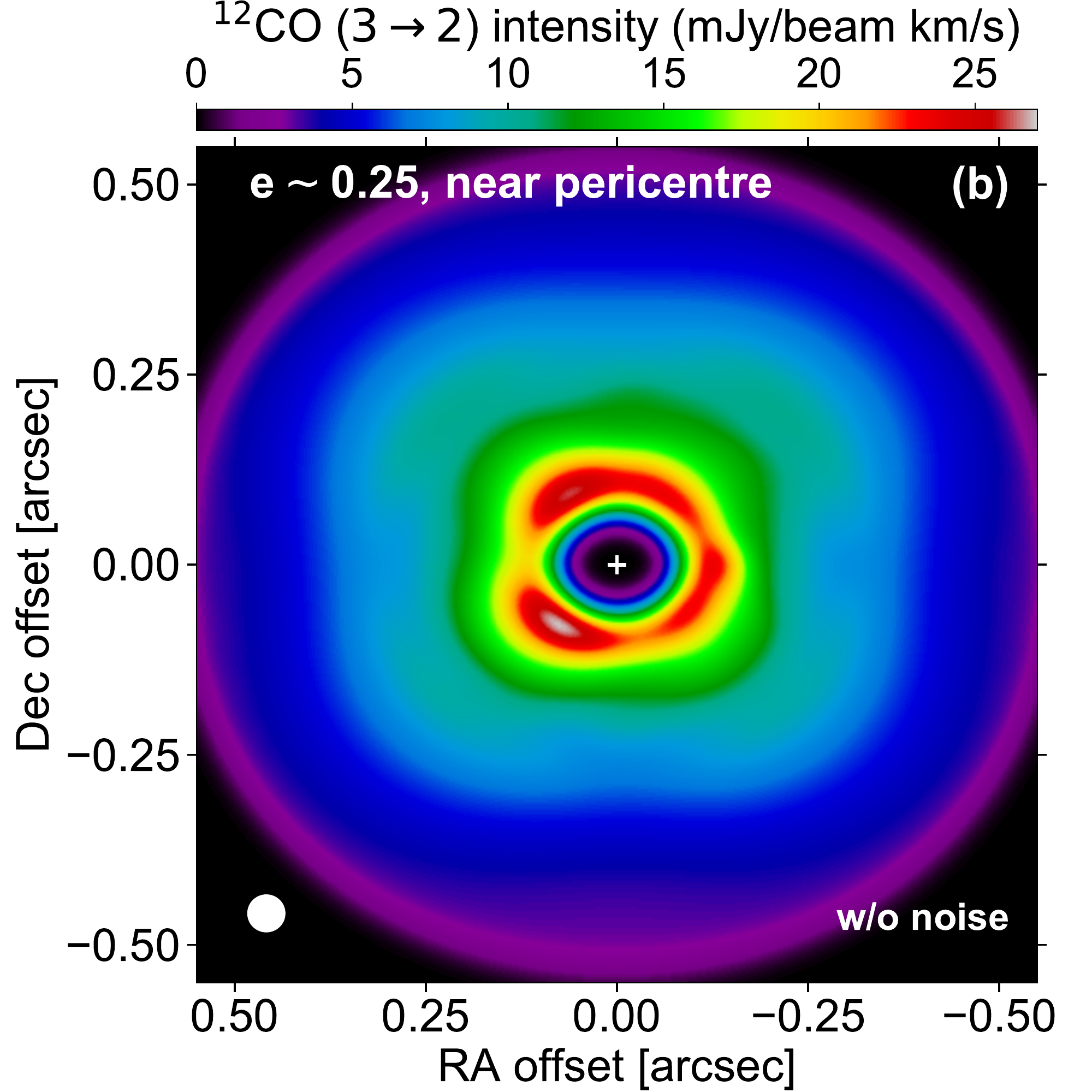}
\includegraphics{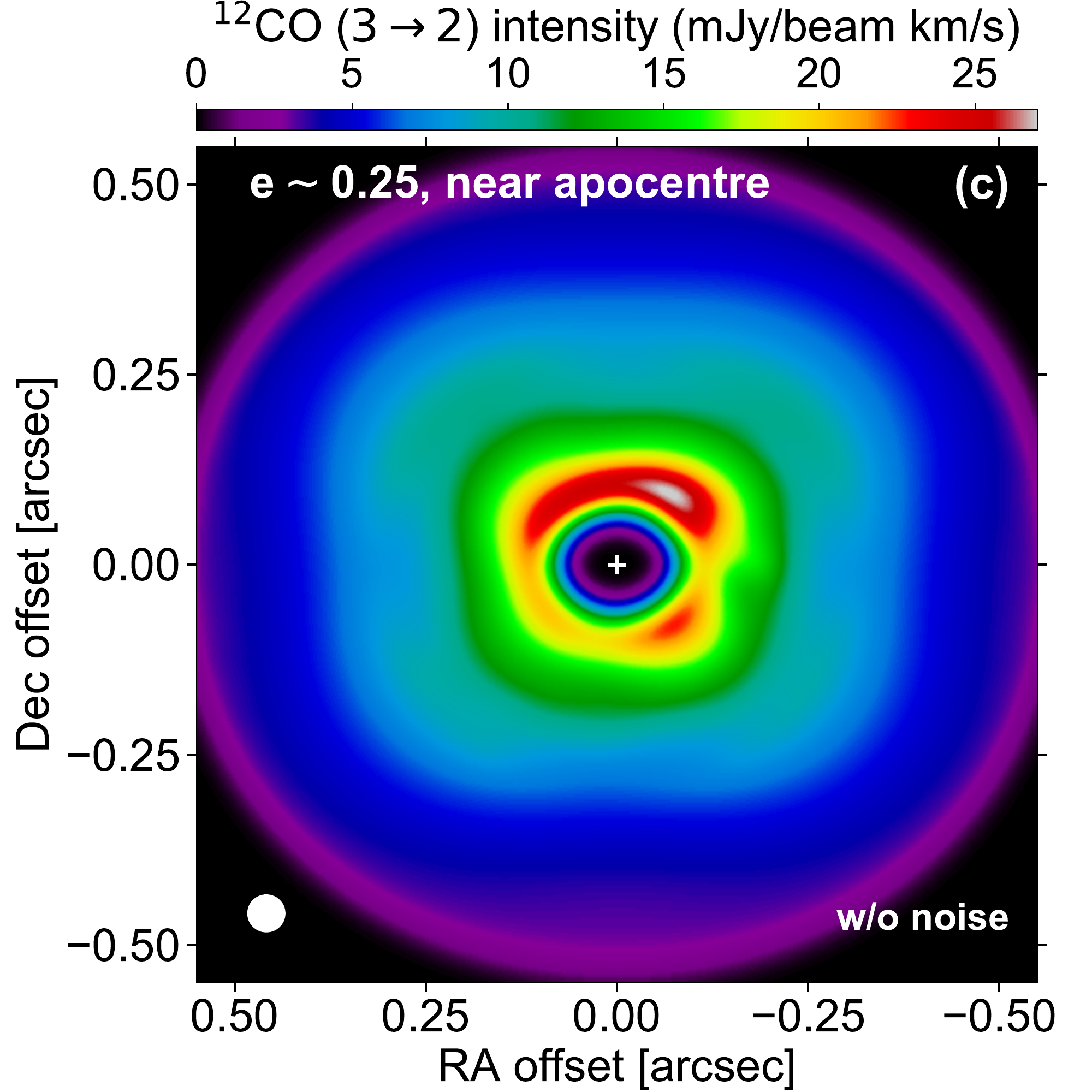}
}
\resizebox{0.99\hsize}{!}
{
\includegraphics{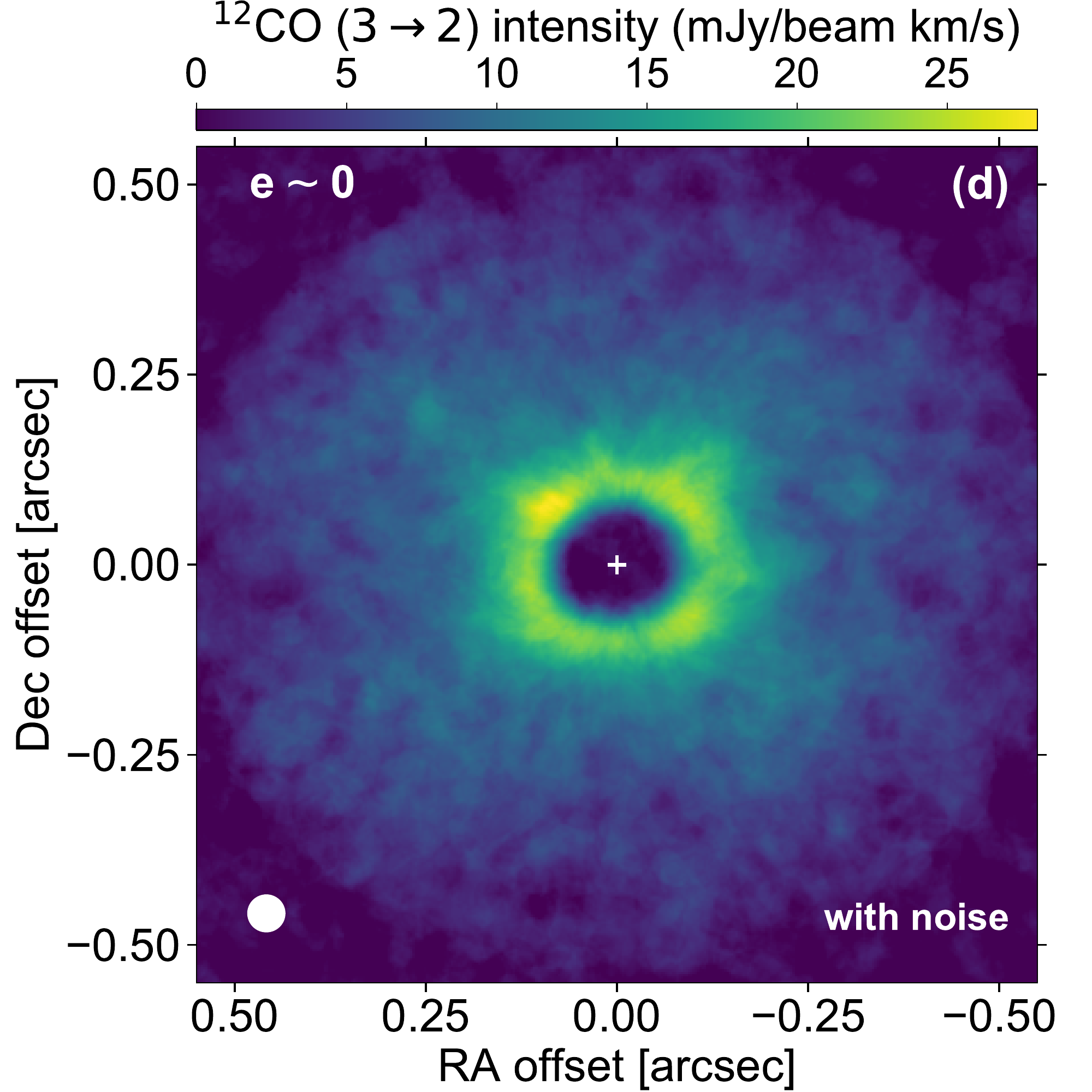}
\includegraphics{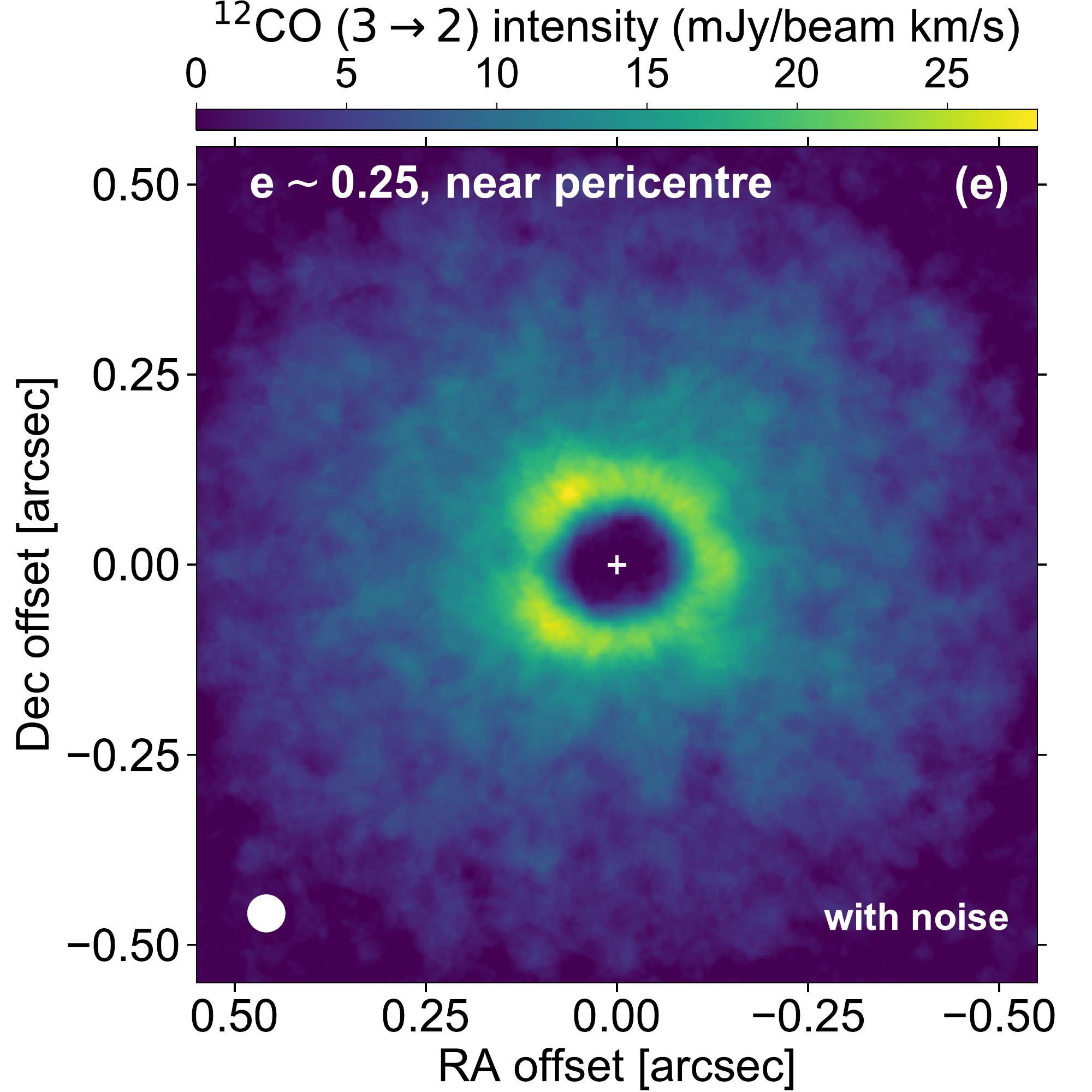}
\includegraphics{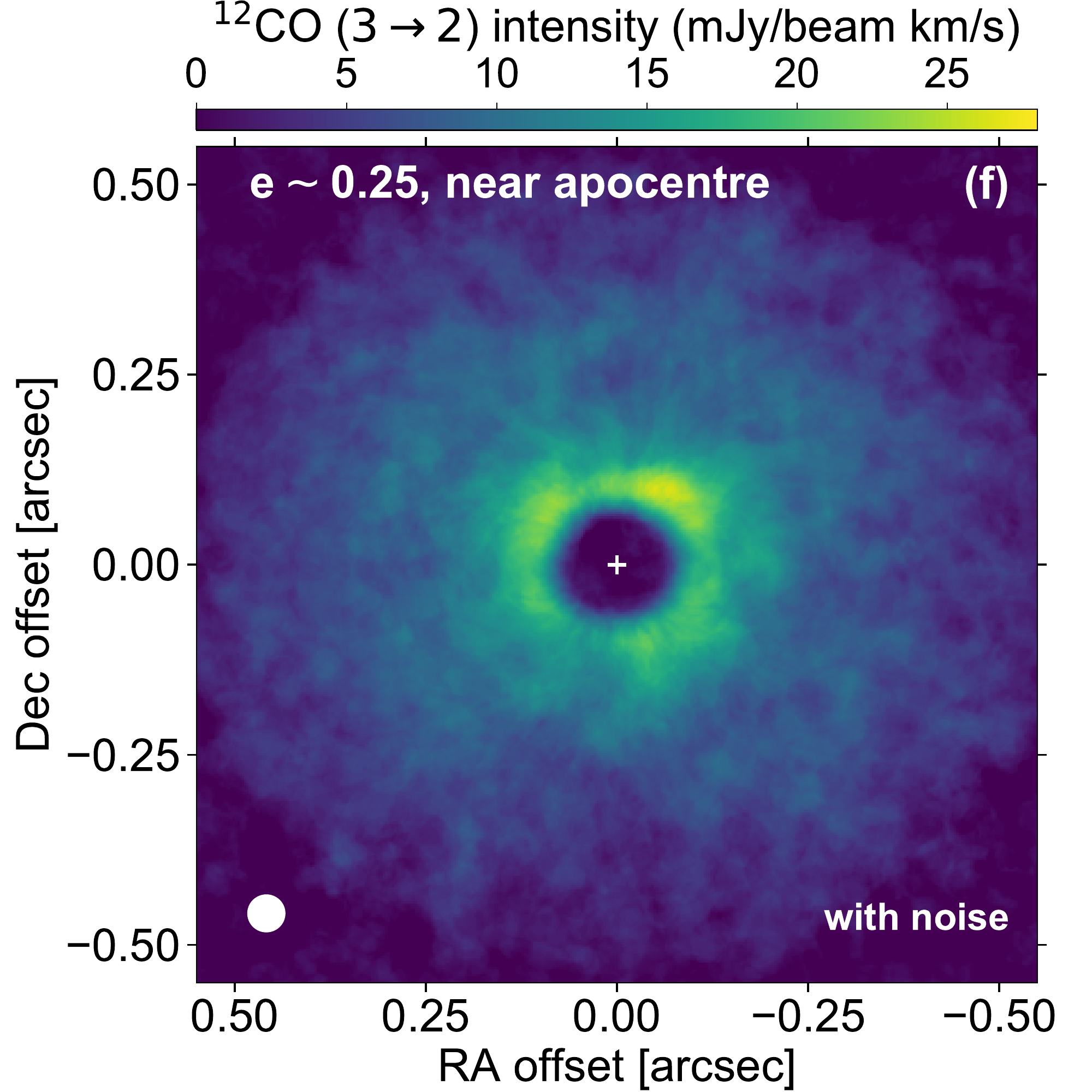}
}
\caption{$^{12}$CO J=3$\rightarrow$2 integrated intensity maps without photodissociation. In the lower panels, maps include white noise with standard deviation $\sigma =$ 1~mJy/beam per channel map.}
\label{fig:figa1}
\end{figure*}

We have seen in Section~\ref{sec:res_12CO_mom0} that the asymmetries in the gas density inside the cavity due to the planet eccentricity lead to large-scale asymmetries in integrated intensity maps of the $^{12}$CO J=3$\rightarrow$2 line. Photodissociation significantly enhances these asymmetries, and one may wonder whether they would still be detectable without photodissociation. This is what is illustrated with Fig.~\ref{fig:figa1}, where we display $^{12}$CO J=3$\rightarrow$2 integrated intensity maps with and without noise at the same three times as, e.g., the panels shown in Figs.~\ref{fig:fig2} to~\ref{fig:fig4}. While the contrast in the integrated intensity is definitely smaller without photodissociation (compare panels b and c with panels k and l of Fig.~\ref{fig:fig3}), it is still significantly above the effective noise level in our emission maps, and therefore detectable, at least in the more favourable case where the planet is near apocentre (panel f).

\section{$^{12}$CO J=3$\rightarrow$2 integrated intensity maps at different disc inclinations}
\label{sec:res_12CO_mom0_incl}

The $^{12}$CO J=3$\rightarrow$2 integrated intensity maps shown in Fig.~\ref{fig:fig3} display a four-lobed pattern of optically thick emission outside the cavity, which is due to the near overlapping along the disc minor and major axes of the emission from the near to the far sides of the disc (see Section~\ref{sec:res_12CO_4lobe}). When noise is added to the channel intensities, however, the four-lobed pattern no longer shows up. A similar conclusion is obtained for the $^{13}$CO and C$^{18}$O J=3$\rightarrow$2 lines. The aim of this section is to briefly explore how the disc inclination impacts these findings. 

\begin{figure*}
\centering
\resizebox{\hsize}{!}
{
\includegraphics{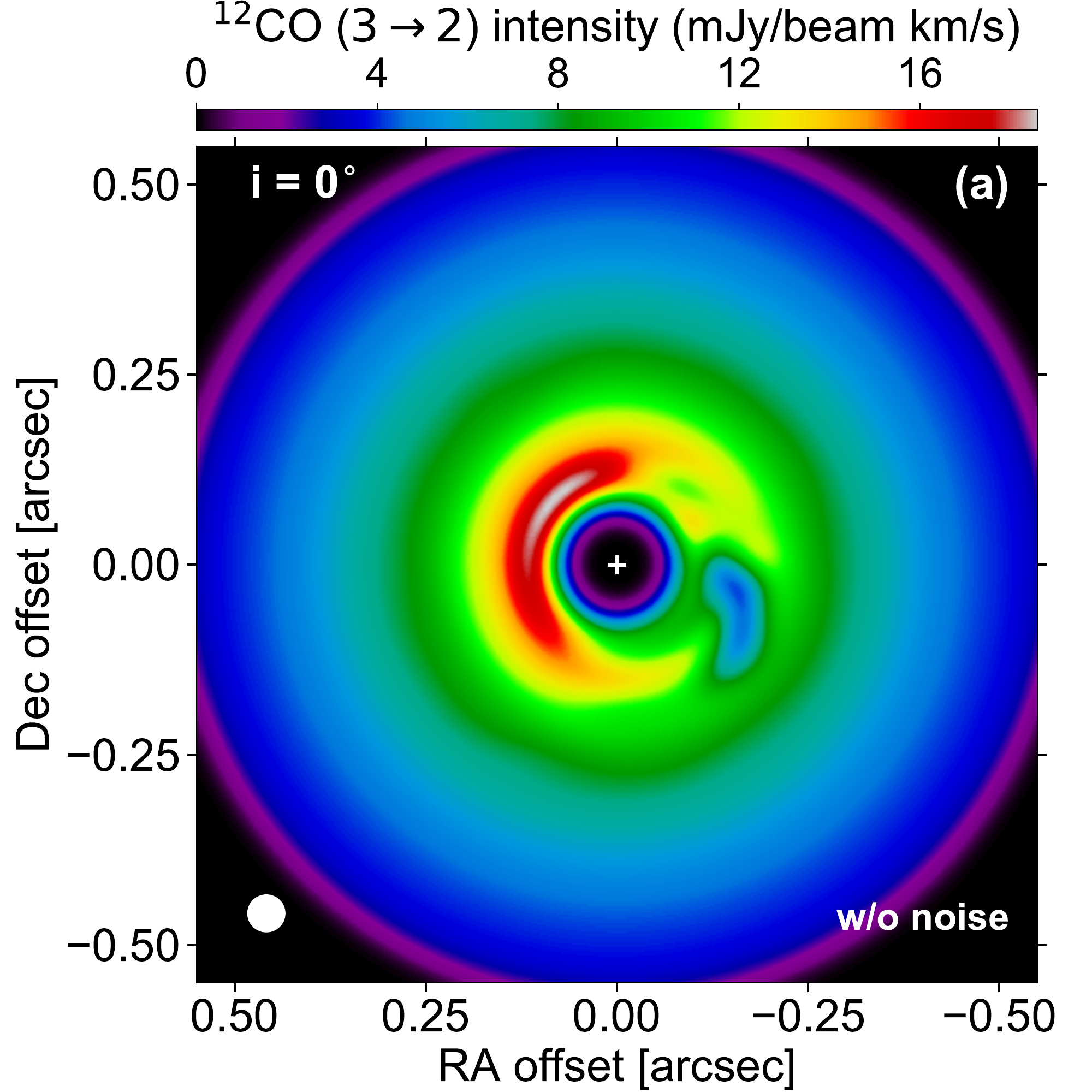}
\includegraphics{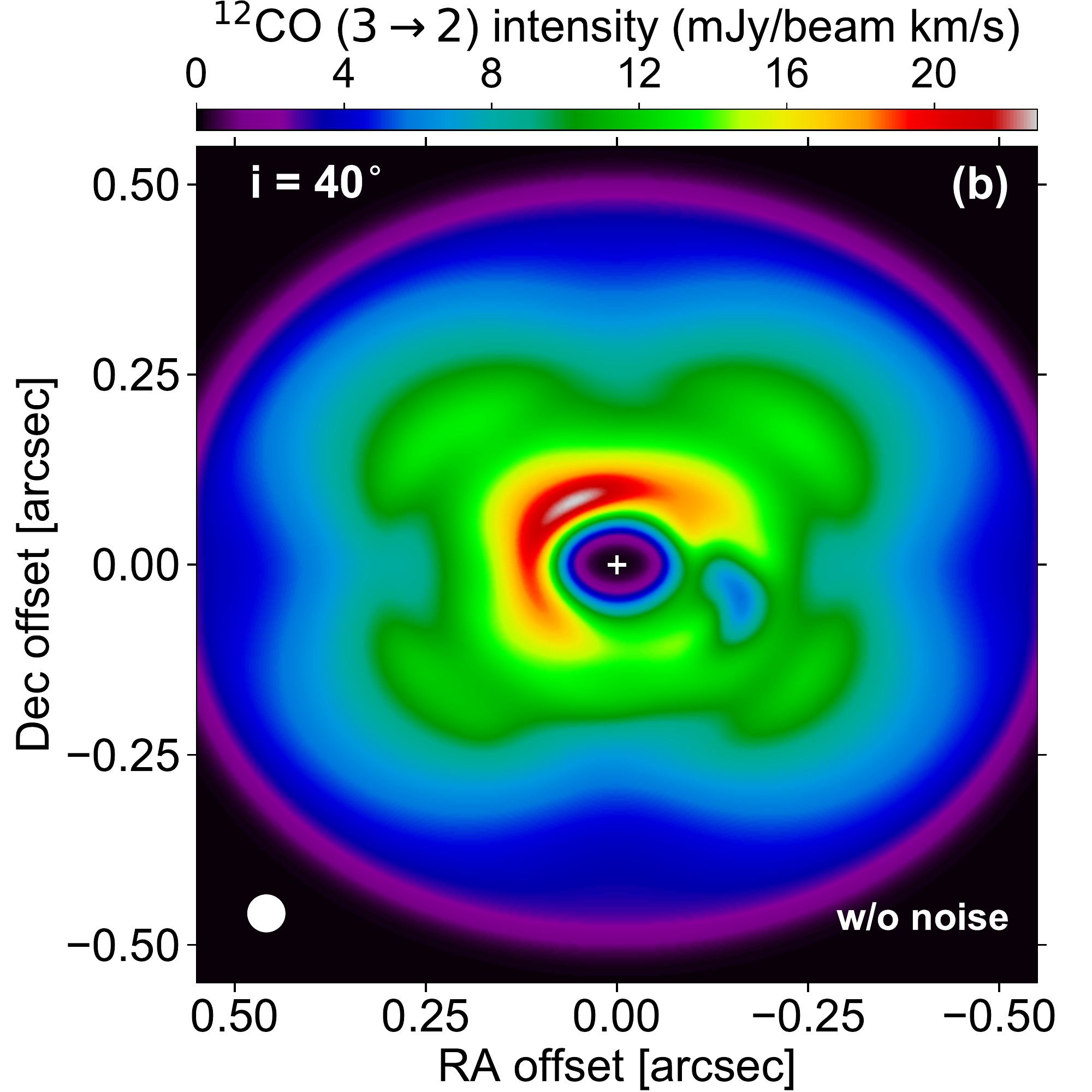}
\includegraphics{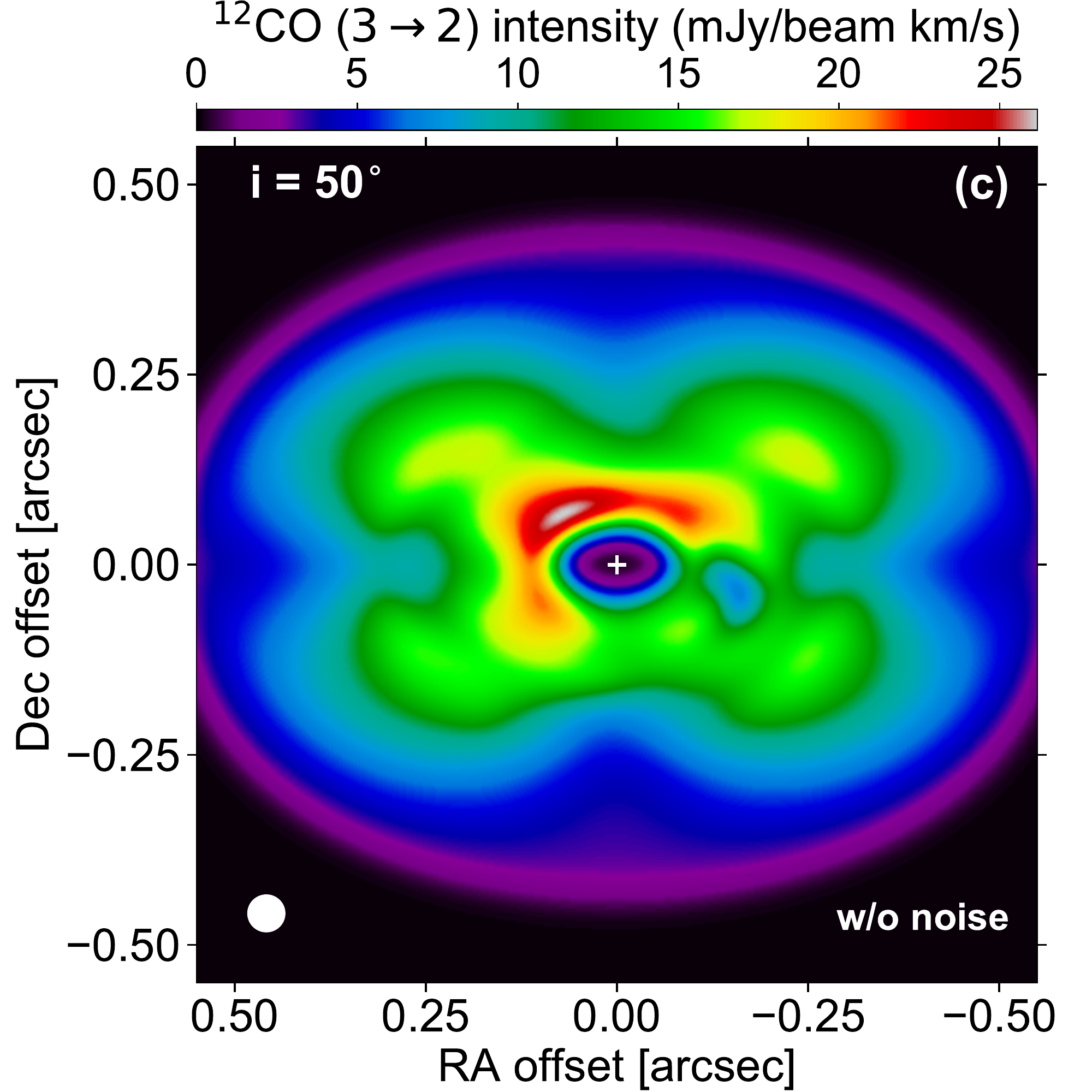}
\includegraphics{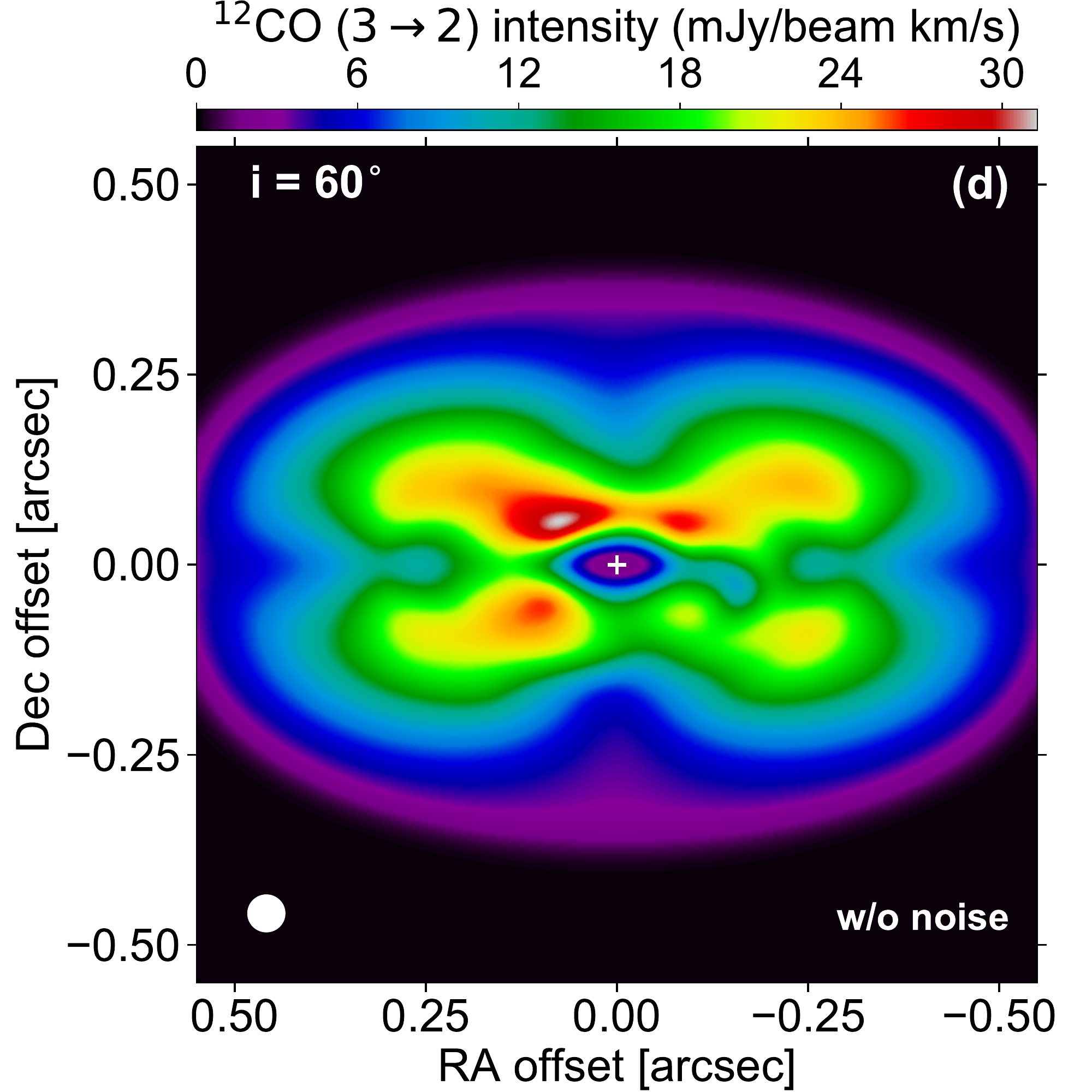}
}
\resizebox{\hsize}{!}
{
\includegraphics{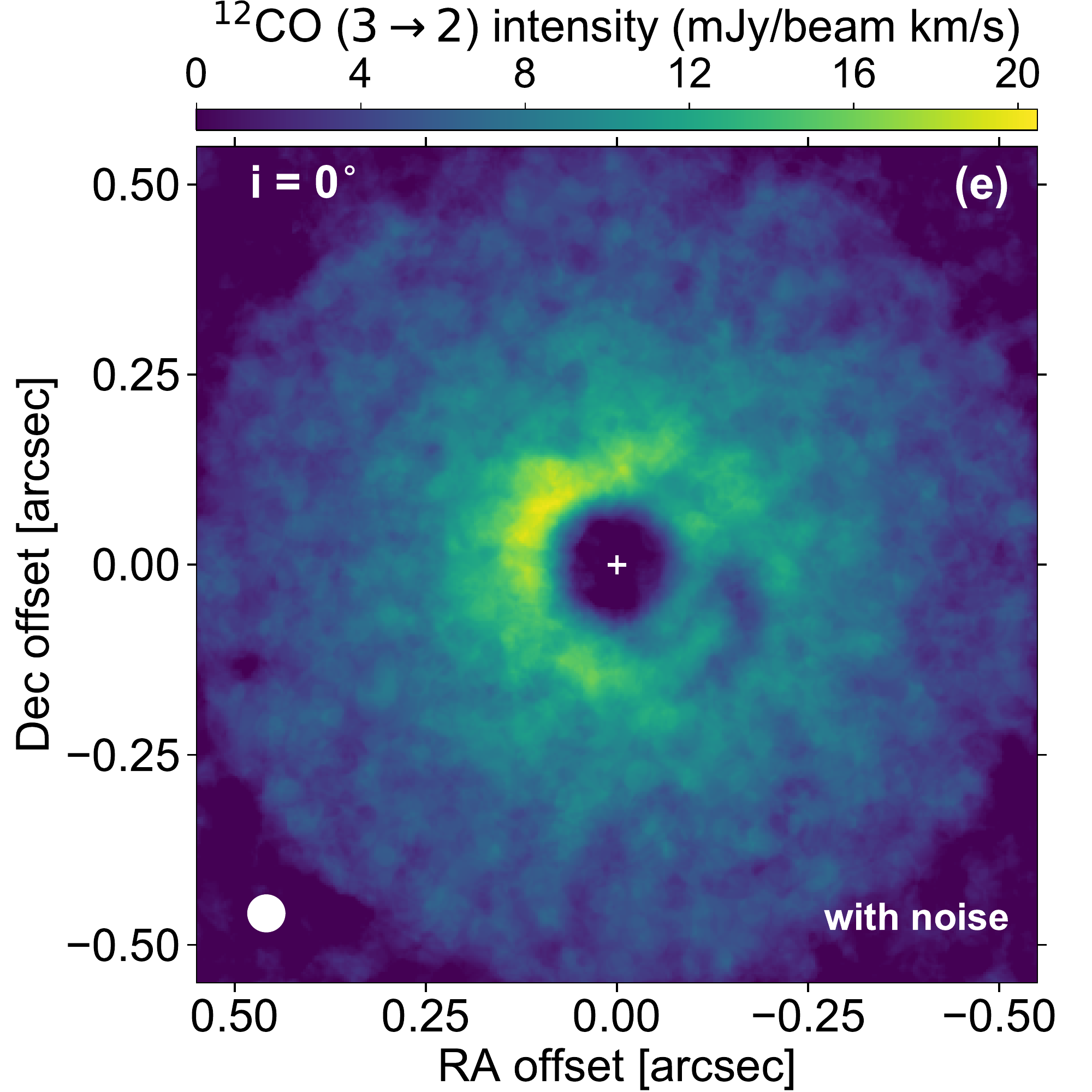}
\includegraphics{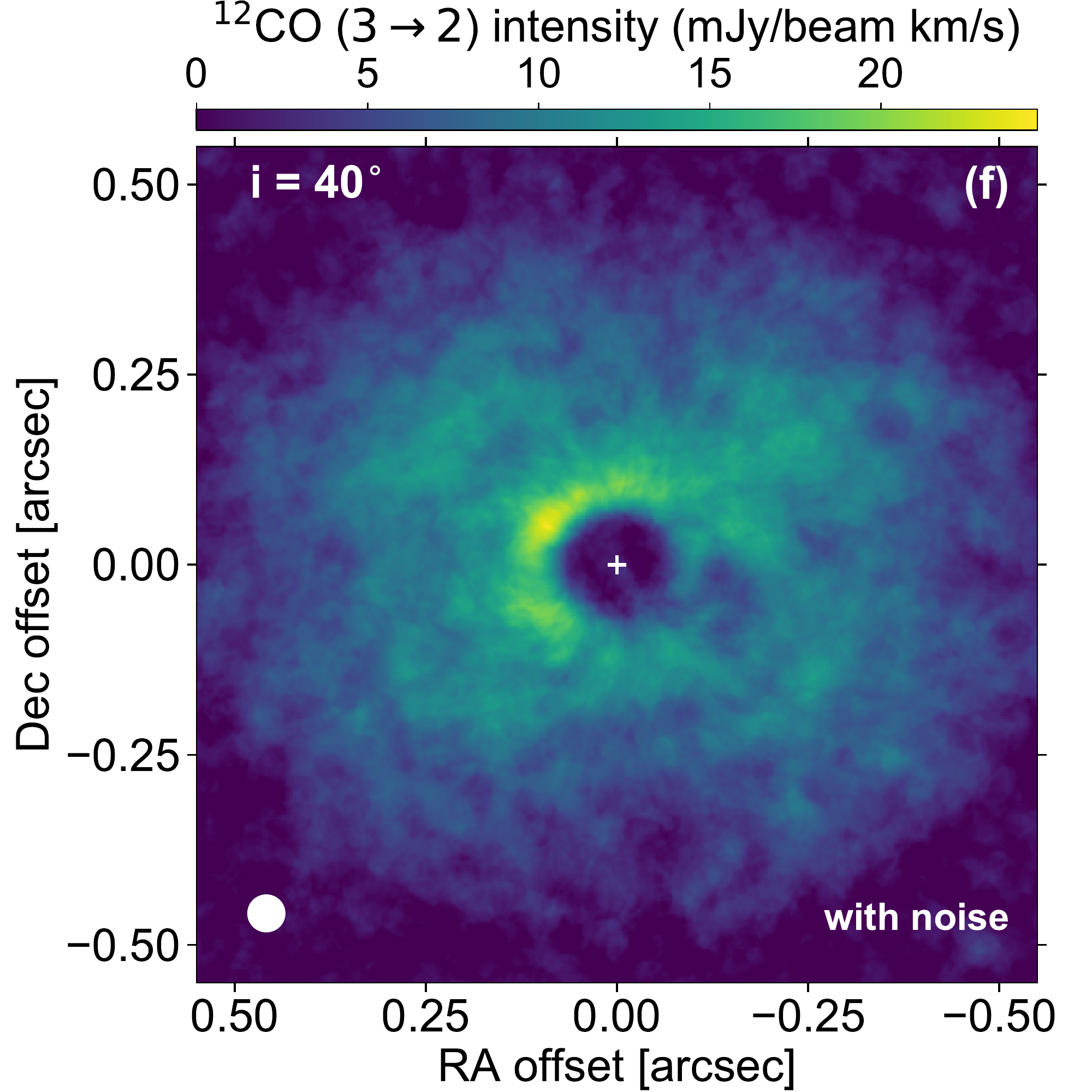}
\includegraphics{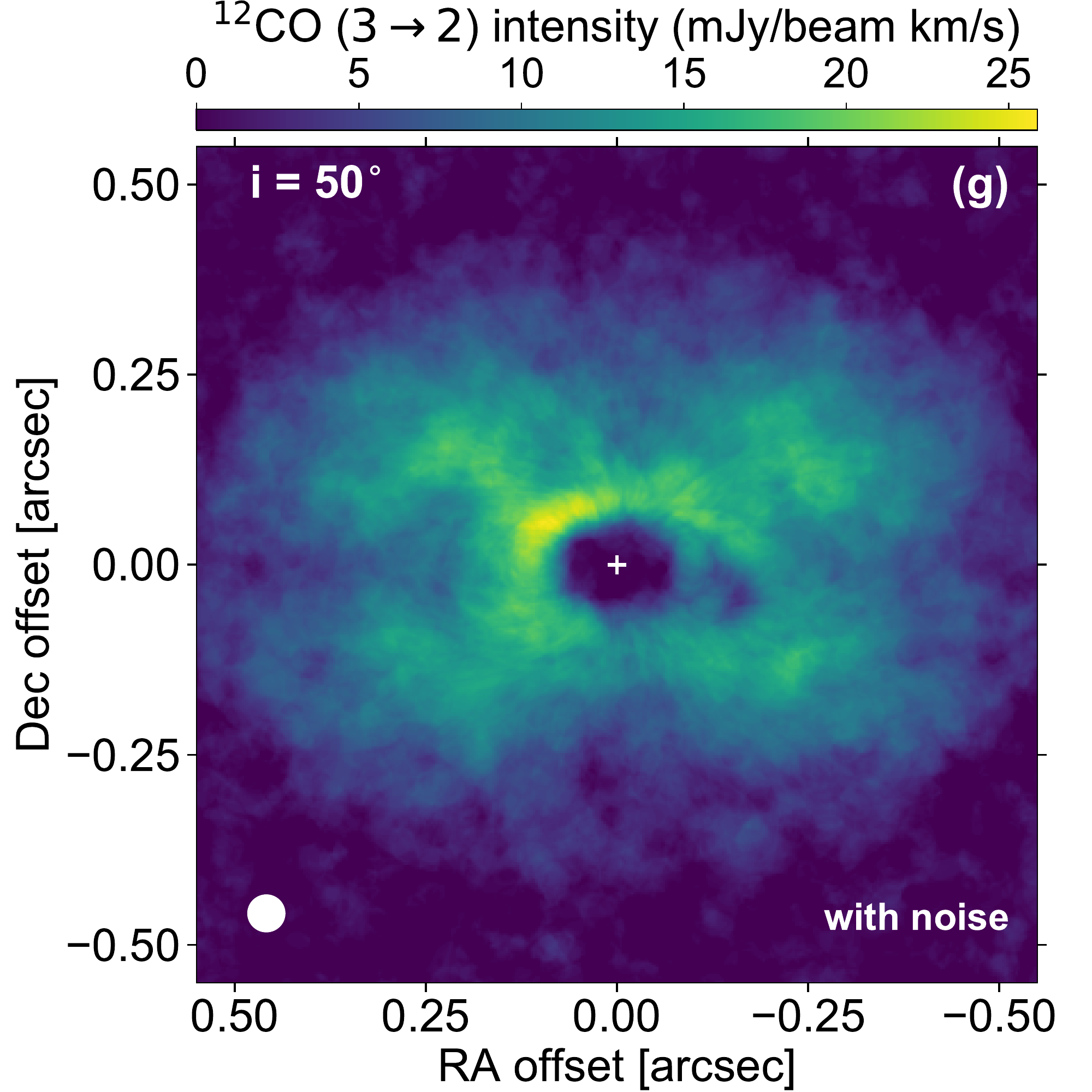}
\includegraphics{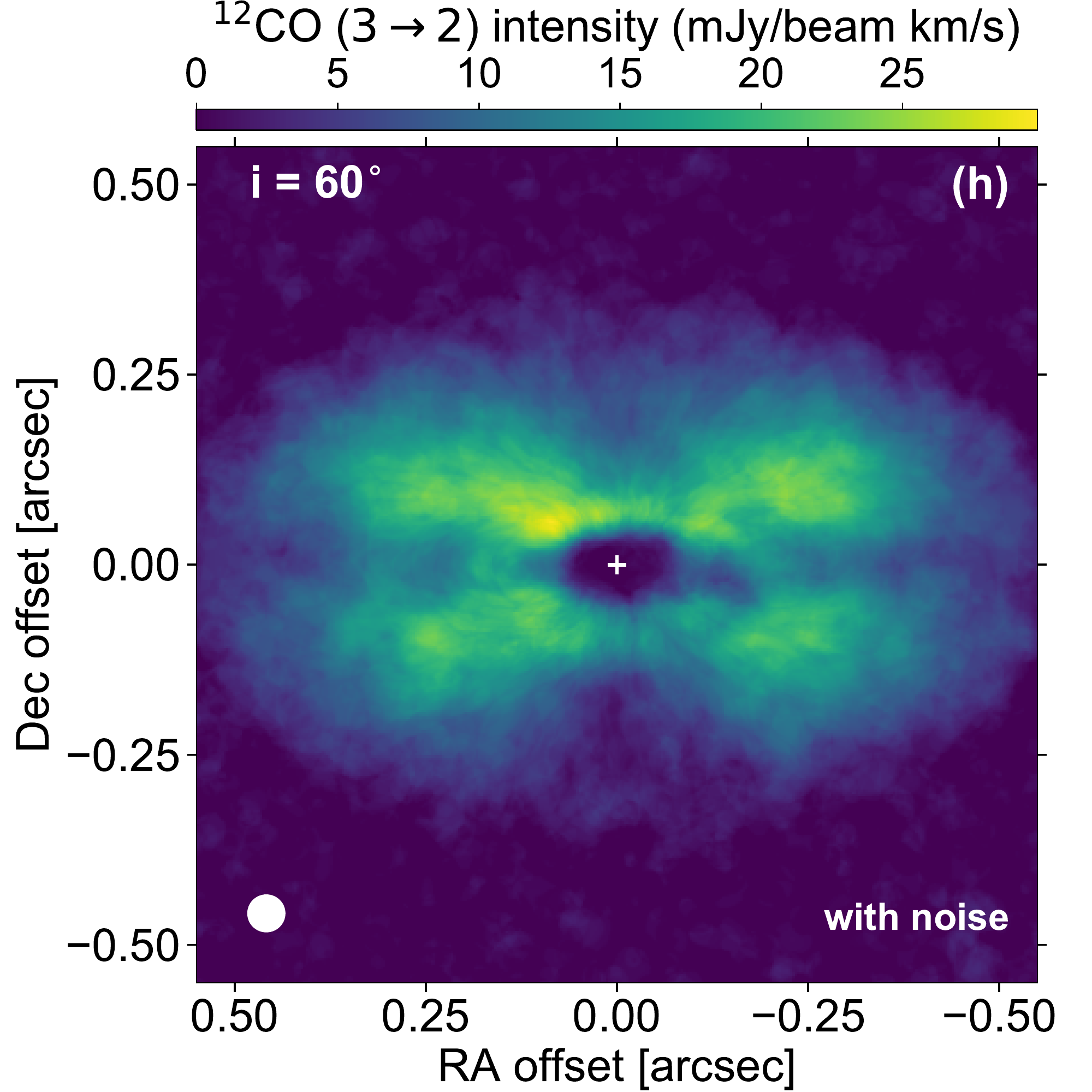}
}
\caption{$^{12}$CO J=3$\rightarrow$2 integrated intensity maps with photodissociation, obtained for different disc inclinations (the inclination $i$ is indicated in the upper-left corner in each panel). In the lower panels, maps include white noise with standard deviation $\sigma =$ 1~mJy/beam per channel map.}
\label{fig:figb1}
\end{figure*}

For this, we focus on the $^{12}$CO J=3$\rightarrow$2 line for which we have carried out additional gas radiative transfer calculations by varying the disc inclination up to 60$\degr$. Since the line-of-sight velocity increases with disc inclination, these calculations use channel maps now covering $\pm$ 11 km s$^{-1}$ around the systemic velocity. To keep the same spectral resolution, the number of channel maps is increased to 121. Fig.~\ref{fig:figb1} displays the $^{12}$CO J=3$\rightarrow$2 integrated intensity maps without and with noise in the channels when the $e\approx0.25$ eccentric planet is near apocentre. From left to right, the disc inclination is 0, 40$\degr$, 50$\degr$ and 60$\degr$. The beam FWHM is again 50 mas. The case with 30$\degr$ inclination can be found in panel (l) of Fig.~\ref{fig:fig3} (without noise) and in panel (c) of Fig.~\ref{fig:fig4} (with noise). The figure shows clearly that the four-lobed pattern outside the cavity becomes more prominent when increasing disc inclination, and that it is detectable for i $\ga50\degr$ given the assumed level of noise in the channels.

\bsp	
\label{lastpage}
\end{document}